\documentclass[review,12pt]{elsarticle}

\pdfoutput=1
\journal{Physics Reports}

\usepackage{comment}
\usepackage{graphicx}          
\usepackage{physics,slashed}
\usepackage{amsfonts,amsmath,amssymb}
\usepackage{mathrsfs}
\usepackage{bm}                

\usepackage{epsfig}            
\usepackage{subfigure}         
\usepackage{verbatim}
\usepackage{url}
\usepackage{array}
\usepackage{dcolumn}           
\usepackage{multirow}
\usepackage{cprotect}
\usepackage{framed}
\usepackage{setspace}
\usepackage{enumitem}
\setlist{nolistsep}
\usepackage{booktabs}
\usepackage{caption}
\usepackage{underscore}
\usepackage{xcolor}
\usepackage{hyperref}
\usepackage{rotating}
\hypersetup{
    colorlinks=true,
    linkcolor=blue!50!black,
    citecolor=red!70!black,
    urlcolor=red!70!black
}
\usepackage{microtype}
\usepackage{fontawesome5}

\definecolor{nicecol1}{rgb}{0.56,0.,1.}
\definecolor{nicecol2}{rgb}{1.,0.3,0.8}
\definecolor{nicecol3}{rgb}{0.,1.,0.}

\renewcommand\[{\begin{equation}}
\renewcommand\]{\end{equation}}

\begin{document}

\begin{frontmatter}

\title{Inside the Black Box of Big Bang Nucleosynthesis: A Comprehensive
Sensitivity Atlas for the Precision Era}

\author{Anne-Katherine Burns}
\ead{annekatherineburns@icc.ub.edu}
\affiliation{organization={Institut de Ci\`encies del Cosmos de la Universitat de Barcelona},
             city={Barcelona},
             country={Spain}}

\begin{abstract}
In this study we present a comprehensive sensitivity atlas for Big Bang Nucleosynthesis (BBN) in which we quantify the dependence of the primordial abundances of helium-4, deuterium, and lithium-7 as well as $N_{\rm{eff}}$ on variations in 14 fundamental particle physics and cosmological parameters and 63 thermonuclear reaction rates. We use the publicly available BBN code \faGithub \href{https://github.com/vallima/PRyMordial}{\,\texttt{PRyMordial}} to compute each sensitivity using two nuclear reaction rate compilations and two weak-rate normalization schemes, and provide a model independent reference applicable to Beyond the Standard Model (BSM) models in which MeV scale physics is modified. In addition, we rank each parameter's contribution to the theoretical uncertainty budget. We compare our predictions against the latest observational determinations of the primordial abundances, including a recent LBT measurement of the helium-4 abundance \cite{Aver:2026dxv} which roughly halves the observational uncertainty relative to previous determinations. We present these results both fixing $\Delta N_{\rm eff}$ at its Standard Model (SM) value, and allowing it to be a free parameter using the latest uncertainty from the combined CMB+BAO+BBN 2026 value \cite{Goldstein:2026iuu}. When $\Delta N_{\rm eff}$ is allowed to be a free parameter, it dominates the theoretical uncertainty of the helium-4 abundance, highlighting the importance of upcoming observations from the Simons Observatory \cite{SimonsObservatory:2025wwn}. As illustrative applications, we examine the deuterium tension and the Lithium Problem in light of our sensitivity analysis. The full set of numerical results and figures is publicly available on GitHub \faGithub \href{https://github.com/Anne-KatherineBurns/bbn-sensitivity-atlas}{\,\texttt{bbn-sensitivity-atlas}}.
\end{abstract}

\begin{keyword}
Big Bang nucleosynthesis \sep primordial abundances \sep helium-4 abundance
\sep deuterium abundance \sep lithium problem \sep thermonuclear reaction rates
\sep weak interaction rates \sep neutron lifetime
\sep physics beyond the Standard Model \sep variation of fundamental constants
\sep early-universe cosmology
\end{keyword}

\end{frontmatter}

\tableofcontents

\section{Introduction}

The prediction of the light-element abundances produced during Big Bang Nucleosynthesis (BBN)~\cite{Weinberg:1977ji} is one of the most quantitatively successful achievements of modern cosmology. The foundational theory work of the subject was conducted in 1948 by Alpher, Bethe, and Gamow~\cite{Alpher:1948ve}, who first proposed that the light elements were synthesized in a hot, dense, expanding early universe, followed by the more detailed treatment of Alpher, Follin, and Herman~\cite{Alpher:1953zz} which incorporated the role of the weak interactions in setting the neutron-to-proton ratio. While the original idea that the abundances of all chemical elements could be accounted for through primordial nucleosynthesis was ultimately superseded by stellar nucleosynthesis for heavier nuclei~\cite{Burbidge:1957vc}, the framework was placed on firm physical footing in the 1960's by Hoyle and Tayler~\cite{Hoyle:1964zz}, Peebles~\cite{Peebles:1966rol, Peebles:1966zz}, and Wagoner, Fowler, and Hoyle~\cite{Wagoner:1966pv}, whose calculations established that the helium-4 mass fraction is robustly predicted to be near 0.25 and that the lighter nuclei D, $^{3}$He, $^{7}$Li, etc. are produced at trace levels sensitive to the baryon-to-photon ratio. The discovery of the cosmic microwave background~\cite{1965ApJ...142..419P,Dicke:1965zz} provided independent confirmation of the hot Big Bang picture in which these predictions depended. In the decades since, BBN has matured through precise measurements of the primordial abundances~\cite{Spite:1982dd,1976A&A....50..461A}, increasingly accurate determinations of the relevant nuclear cross sections~\cite{Ando06,Mossa:2020gjc,Iliadis16,Xu:2013fha,Serpico:2004gx,Descouvemont04,Cyburt:2004cq,Fields:2019pfx}, refined treatments of weak-rate and radiative corrections~\cite{Sirlin:1967zza,Abers:1968zz,Dicus:1982bz,Ivanov:2012qe,Brown:2000cp,Pitrou:2018cgg}, and the development of dedicated public codes~\cite{Burns:2023sgx,Arbey:2018zfh,Giovanetti:2024zce,Pitrou:2018cgg}.

Today, the primordial abundances of helium-4 ($Y_p$) and deuterium ($D/H$) have been measured to percent-level precision, and corresponding theoretical predictions are comparably precise~\cite{Aver:2026dxv, PDG2025, Yanagisawa:2025mgx}. The theoretical uncertainty is roughly an order of magnitude smaller than the observational error for helium-4, while for deuterium the two are comparable in size~\cite{Aver:2026dxv,Pitrou:2020etk}. Today, BBN plays an important role in modern cosmology, serving both as a consistency check of Standard Model (SM) physics, nuclear physics inputs, and cosmological parameters, and as a constraint on beyond the SM scenarios that alter the radiation content, neutrino properties, or expansion rate, or that introduces late-time particle decays or energy injection \cite{Cyburt:2015mya,Kawasaki:2004qu}. These constraints are complementary to those from the Cosmic Microwave Background (CMB) and Large Scale Structure (LSS): BBN probes MeV-scale physics at early times, whereas the CMB and LSS constrain related parameters at lower energies and later epochs.

Among the parameters connecting these probes, the effective number of relativistic neutrino species, $N_{\rm eff}$, is particularly impactful. In the Standard Model, $N_{\rm eff}$ = 3.044, accounting for three active neutrino species with small corrections from non-instantaneous neutrino decoupling from the thermal bath. Deviations from this value, parametrized by $\Delta N_{\rm eff}$, arise in a variety of BSM scenarios involving light degrees of freedom that contribute to the total radiation energy density during or after neutrino decoupling. Because $N_{\rm eff}$ directly affects the expansion rate of the universe during BBN, even modest deviations propagate strongly into the predicted helium-4 abundance. The recent combined CMB+BAO+BBN determination of $N_{\rm eff} = 2.990 \pm 0.070$~\cite{Goldstein:2026iuu} has nearly halved the previous uncertainty, making it timely to assess how this improved constraint reshapes the BBN error budget.

Because the helium-4 and deuterium abundances have been so precisely measured, small shifts in their predicted values are meaningful. As observational uncertainties improve, theoretical uncertainties and modeling choices may become a limiting factor and must be quantified clearly. Currently, there exists extensive literature exploring the impact of varying physics inputs on BBN~\cite{Cyburt:2015mya, Barenboim:2025vrc, Burns:2022hkq, Burns:2024ods, Jung:2025dyo, Meissner:2023voo,Giovanetti:2024eff, Barenboim:2026ocd,Froustey:2024mgf,Domcke:2025jiy,Yang:2026gdy}. However, it is not straightforward to compare the sensitivity of the predicted abundances to different parameters across existing studies, since those analyses rely on different compilations of empirical nuclear reaction-rate data, different implementations of QED plasma and weak-rate corrections, different likelihood choices and datasets, and different numerical codes. 

The literature contains a number of important reviews of BBN, each emphasizing different aspects of the subject. Early comprehensive treatments by Boesgaard and Steigman~\cite{1985ARA&A..23..319B} and by Malaney and Mathews~\cite{1993PhR...229..145M} established the framework of the standard BBN calculation and surveyed its potential as a probe of physics beyond the Standard Model in the era before high-redshift deuterium measurements. Sarkar~\cite{Sarkar:1995dd} subsequently provided an extensive pedagogical and quantitative review focused on the use of BBN as a constraint on new physics, particularly light degrees of freedom and unstable massive relics. Schramm and Turner~\cite{Schramm:1997vs} marked the transition to the precision era by surveying the implications of the first reliable QSO deuterium measurements and anticipating the role of upcoming CMB anisotropy data. Iocco, Mangano, Miele, Pisanti, and Serpico~\cite{Iocco:2008va} reviewed the connection between BBN and precision cosmology with an emphasis on the underlying nuclear-physics inputs, while Pospelov and Pradler~\cite{Pospelov:2010hj} gave a focused review of BBN as a probe of new physics, organizing the literature into broad categories of beyond-Standard-Model scenarios. The status of standard BBN in light of Planck-era CMB data was comprehensively reviewed by Cyburt, Fields, Olive, and Yeh \cite{Cyburt:2015mya}, which remains the most widely cited modern reference; Pitrou, Coc, Uzan, and Vangioni~\cite{Pitrou:2018cgg} presented a parallel review focused on theoretical precision, in particular improvements to the predicted helium-4 abundance accompanying the development of the PRIMAT code. Most recently, Cooke~\cite{Cooke:2024nqz} provided a pedagogical overview of both the theoretical framework and the observational techniques used to determine the primordial abundances. 

Our work differs from these reviews in scope and emphasis. Rather than surveying the field broadly or focusing on a specific class of BSM scenarios, we present a single-code sensitivity atlas that studies 77 parameters independently within a uniform framework, repeated across two weak-rate normalization schemes and two nuclear rate compilations. For each observable, we provide an explicit ranking of contributions to the theoretical uncertainty, both for SM BBN and with $\Delta N_{\rm eff}$ allowed to vary using the most recent CMB+BAO+BBN constraint~\cite{Goldstein:2026iuu}. We present these results in light of the most recent data, particularly highlighting the determination of the helium-4 abundance from the LBT~\cite{Aver:2026dxv}. The resulting atlas is intended as a model-independent reference that complements existing reviews by making transparent which inputs control the current theoretical uncertainty and the responses of each observable to early-universe modifications of any of the inputs we consider.

All calculations are performed using the code \faGithub \href{https://github.com/vallima/PRyMordial}{\,\texttt{PRyMordial}} \cite{Burns:2023sgx}, with input observations of helium-4, deuterium, lithium-7, and $N_{\rm{eff}}$ taken from the latest determinations. While the Lithium-Problem persists, we include it in our analysis in order to quantify the way in which parameter variation affects the predicted value. We include these results not as proposed solutions, but rather as a starting point for future work.

\faGithub \href{https://github.com/vallima/PRyMordial}{\,\texttt{PRyMordial}} \cite{Burns:2023sgx} is among the latest in a long line of public BBN codes whose development has paralleled the precision improvements outlined above. The first such code was developed by Wagoner \cite{Wagoner69, 1973ApJ...179..343W} and served as the basis for essentially all subsequent calculations; it was modernized and made widely available by Kawano \cite{Kawano:1988vh,1992STIN9225163K} as the NUC123 program, which became the de facto standard for two decades. More recent public codes include PArthENoPE \cite{Pisanti:2007hk, Gariazzo:2021iiu, Consiglio:2017pot}, which traces its lineage to the Wagoner-Kawano framework while incorporating updated nuclear rates and weak-rate corrections; AlterBBN \cite{Arbey:2011nf, Arbey:2018zfh}, designed specifically to facilitate the study of non-standard cosmological scenarios; PRIMAT \cite{Pitrou:2018cgg}, which introduced a number of precision improvements to the calculation of the helium-4 abundance including a careful treatment of weak-rate radiative corrections; and most recently LINX \cite{Giovanetti:2024zce}, a differentiable code written in JAX designed to enable fast joint CMB+BBN parameter inference. \href{https://github.com/vallima/PRyMordial}{\,\texttt{PRyMordial}} \cite{Burns:2023sgx} differs from these other codes in several respects. It computes the background thermodynamics from first principles by directly solving the coupled Boltzmann equations rather than interpolating tabulated results, includes non-instantaneous neutrino decoupling effects natively, and provides built-in support for both the PRIMAT and NACRE-II nuclear reaction rate compilations as well as two different weak-rate normalization schemes. These features make \href{https://github.com/vallima/PRyMordial}{\,\texttt{PRyMordial}} \cite{Burns:2023sgx} particularly well-suited to the present sensitivity analysis, since they allow each of the 77 input parameters to be varied independently and the response of each observable to be tracked across multiple choices of nuclear inputs and weak-rate normalization within a single, uniform numerical framework.

Tables \ref{tab:sens_taun_nacre_side_by_side}, \ref{tab:sens_fund_side_by_side}, \ref{tab:sens_taun_side_by_side_neff}, and \ref{tab:sens_tau_n_rates_first12_side_by_side} present sensitivity matrices for the BBN observables considered here. These tables clearly identify which parameters dominate the local response of each observable, allowing rapid estimates of the induced changes in the predicted abundances. These tables can be easily used as a diagnostic tool in searches for new physics. In scenarios where one or more parameters take values in the early universe that differ from their present-day determinations, the response functions presented here provide a simple first estimate of how the primordial abundances are shifted, and therefore which directions in parameter space are most promising to explore. Although our sensitivities are computed locally about fiducial SM values, they offer a useful guide for identifying early-universe modifications that could motivate a full non-linear analysis. 

In addition, we present ranked uncertainty budgets clearly showing current limitations on BBN predictions. We construct these budgets by propagating the uncertainties in the input quantities to the predicted abundances. For each observable, Tables \ref{tab:budget_comparison_taun}, \ref{tab:budget_comparison_fund}, \ref{tab:unc_neff_taun}, and \ref{tab:unc_neff_fund} provide an ordered list of the parameters that most strongly contribute to the theoretical error for each observable. This allows us to identify which future measurements could most effectively sharpen BBN predictions. 

\begin{figure}[t]
    \centering
    \includegraphics[width=0.75\linewidth]{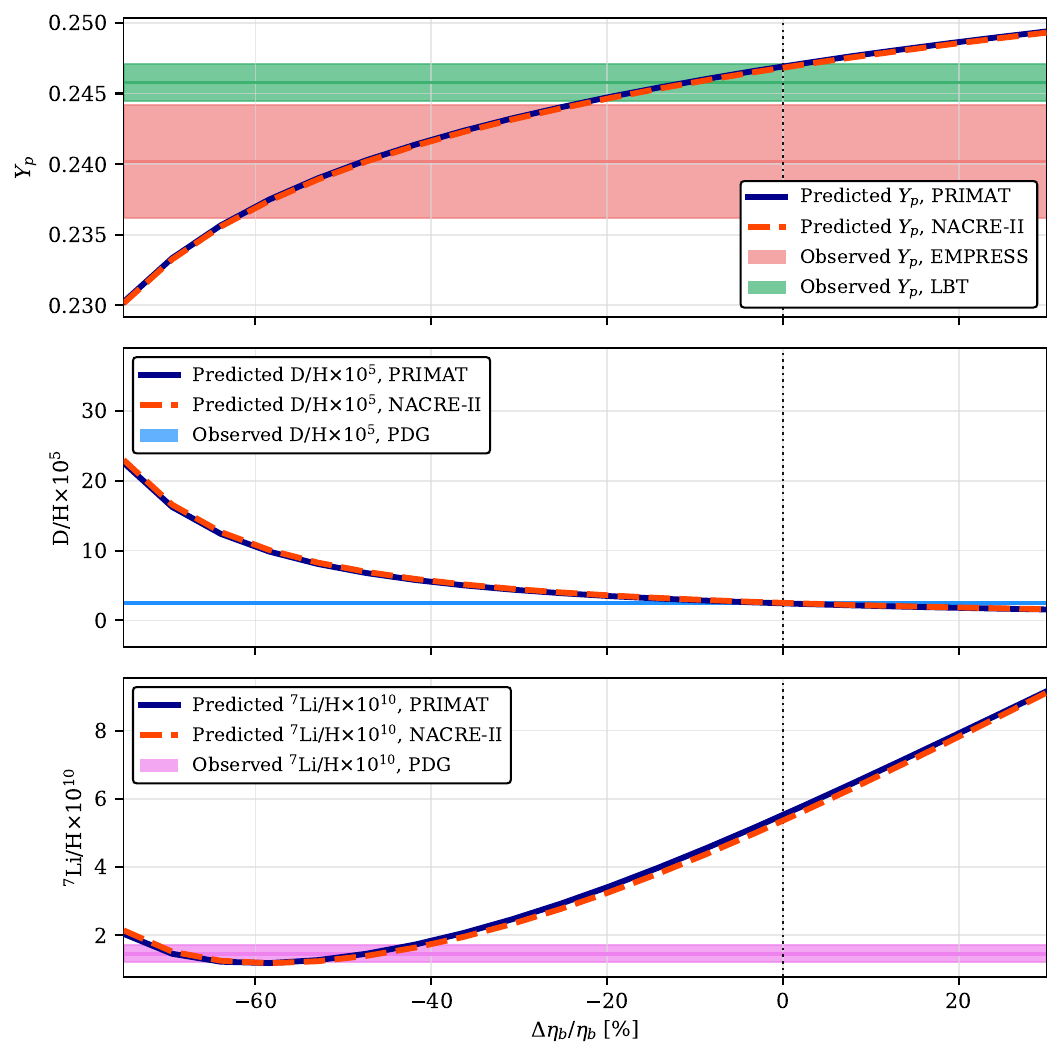}
    \caption{Predicted primordial abundances of $Y_p$ (top), D/H$\,\times 10^{5}$ (middle), and $^{7}$Li/H$\,\times 10^{10}$ (bottom) as functions of the fractional deviation $\Delta \eta_b / \eta_b$ of the baryon-to-photon ratio from its fiducial value, computed with the PRIMAT (solid, dark blue) and NACRE-II (dashed, orange-red) nuclear reaction rate compilations using the $\tau_n$-normalized weak rates. Horizontal bands indicate the $1\sigma$ observational determinations: EMPRESS~\cite{Yanagisawa:2025mgx} and LBT~\cite{Aver:2026dxv} for $Y_p$, PDG~\cite{PDG2025} for D/H, and PDG~\cite{PDG2025} for $^{7}$Li/H. The vertical dotted line marks the fiducial value $\Delta \eta_b / \eta_b = 0$.}
    \label{fig:eta0b_intro}
\end{figure}

Figure~\ref{fig:eta0b_intro} illustrates the style of analysis carried out throughout this work, using the baryon-to-photon ratio, $\eta_b$, as a familiar example. The predicted abundances of $Y_p$, D/H, and $^7$Li/H are shown as functions of the fractional deviation $\Delta\eta_b/\eta_b$ from its fiducial value, computed independently with the PRIMAT and NACRE-II nuclear reaction rate compilations. Observational determinations are overlaid as horizontal bands. The predicted $^7$Li/H lies well above the observed value across the full scanned range, a manifestation of the cosmological Lithium Problem described in Section~\ref{sec:Li7}. Throughout the rest of this work we vary the physical baryon density, $\Omega_b h^2$, rather than $\eta_b$. The two are linearly related through the present-day CMB photon number density and therefore encode identical physics, but $\Omega_b h^2$ is the quantity directly constrained by the CMB and is the natural choice when combining BBN with CMB likelihoods. Analogous scans for each of the 77 parameters considered in this work form the basis of the sensitivity and uncertainty analyses that follow.

All of these results will be presented for two different compilations of nuclear reaction rates, allowing us to identify compilation-dependent systematics. This shows us which sensitivities are robust, and which are dependent on the chosen nuclear reaction rate set. 

These results are particularly timely in light of the newly published helium data from the Large Binocular Telescope (LBT)~\cite{Aver:2026dxv}, discussed in detail in Section \ref{sec:LBT}. As observational data continue to improve, a standardized framework for interpreting parameter sensitivity and theoretical uncertainty will become increasingly important.

This paper is organized as follows. In Section \ref{sec:obs}, we review the observational inputs adopted in this work, including the current determinations of the primordial abundances of helium-4, deuterium, and lithium-7, as well as the latest constraint on $N_{\rm{eff}}$. Section \ref{sec:method} describes the theoretical framework and numerical methodology implemented in \faGithub \href{https://github.com/vallima/PRyMordial}{\,\texttt{PRyMordial}}. In Section \ref{sec:PVar}, we present the sensitivity of the primordial abundances to variations in the fundamental physics and cosmological parameters, while Section \ref{sec:p_var_nuc} summarizes the corresponding sensitivity to the thermonuclear reaction rates. Section \ref{sec:sens} presents the tabulated logarithmic sensitivity coefficients. Sections \ref{sec:unc} and \ref{sec:unc_neff} contain the propagated uncertainty budgets for SM BBN and for varying $N_{\rm{eff}}$, respectively. In Section \ref{sec:app}, we provide several illustrative applications of the atlas, including possible resolutions of the deuterium tension and the cosmological Lithium Problem. We conclude in Section \ref{sec:conclusion}.

\section{Cosmological Observations}
\label{sec:obs}

\subsection{A New Measurement of $Y_p$ from the LBT}
\label{sec:LBT}

The primordial abundance of helium-4 is inferred from emission-line spectroscopy of young, metal-poor galaxies.  Recently, a new determination of the primordial helium-4 abundance, $Y_p$, was published using data from the Large Binocular Telescope (LBT)~\cite{Aver:2026dxv}. The value was determined via the analysis of 54 metal-poor extragalactic HII regions. 41 targets were determined to have sufficiently low systematic errors and were used in the analysis. Notably, 15 low-metallicity targets with high signal-to-noise ratios show no significant trend of $Y_p$ with O/H, permitting a direct determination of the primordial abundance from their weighted average without the traditional linear extrapolation to zero metallicity. The resulting value of $Y_p$ was determined to be, 

\begin{equation}
    Y_p = 0.2458 \pm 0.0013  \quad 
    \text{\cite{Aver:2026dxv}.}
\end{equation}

The precision of this measurement, at 0.5\%, represents a factor of $\sim2$ improvement over previous determinations~\cite{PDG2025} and allows us to put precise constraints on New Physics affecting the universe at MeV scales. In addition, the result is consistent with the SM theory prediction for $Y_p$, $0.2467 \pm 0.0002$ at the $\sim1\sigma$ level. 

\subsection{EMPRESS Result for $Y_p$}

Recently the EMPRESS collaboration has reported a new independent determination for the primordial helium-4 abundance. Their analysis is
based on Subaru near-infrared spectroscopy of 29 galaxies, including 14 extremely metal-poor galaxies (EMPGs) with less than one-tenth the metal content of the Sun, targeting the He\,\textsc{i} $\lambda10830$\,\AA{}
line to break the density--temperature degeneracy. After selection cuts, the qualifying galaxies are added to a literature sample of 58 galaxies, yielding a final sample of 68. Their fiducial value,

\begin{equation}
Y_p = (0.2402\pm0.0040) \quad \text{\cite{Yanagisawa:2025mgx},}
\end{equation}

is $\sim 1\sigma$ lower than most previous estimates but consistent with recent EMPG-based determinations and with the CMB constraint from ACT. Combined with the primordial deuterium abundance, this implies
$N_\mathrm{eff} = 2.54^{+0.25}_{-0.20}$, in mild $1$--$2\sigma$ tension
with the Standard Model value and Planck. This analysis and the one by the LBT collaboration differ in several respects, including the method of extrapolation to zero metallicity, the homogeneity of the observational dataset, and the treatment of unphysical parameter values in the MCMC modeling. We defer a detailed comparison of these datasets to future work.

\subsection{Observation of Deuterium} 

Deuterium is observed via its isotope-shifted high-order Lyman series absorption features in low-metallicity gas clouds. Because BBN is the only known significant source of deuterium in the universe, which is subsequently destroyed in stellar processing, observed values serve as lower bounds on the primordial abundance. 

The PDG world average for the observed abundance of deuterium is the weighted average of the 12 most precise and recent measurements and is,

\begin{equation}
D / H \times 10^5 = (2.508 \pm 0.029) \quad \text{\cite{PDG2025}.}
\end{equation}

This value of the deuterium abundance is in tension at the level of almost 2$\sigma$ with the SM predicted value, $D/H \times 10^5 = 2.439 \pm 0.037$~\cite{PDG2025}. This mild tension, together with the $Y_p$ measurements discussed above, motivates the parameter sensitivity study pursued in this work.

\subsection{Observation of Lithium-7}
\label{sec:Li7}

The primordial abundance of lithium-7 is measured in the atmospheres of metal-poor stars. The observed value has a significantly higher relative uncertainty as compared to the uncertainties on the observed value of the primordial helium-4 and deuterium abundances, and is in significant tension with the SM predicted value, $(5.464 \pm 0.220) \times 10^{-10}$. This discrepancy is widely known as the ``Lithium Problem" to which many solutions have been proposed spanning nuclear physics, stellar physics, and new physics~\cite{Fields:2011zzb, Miranda:2025wcp, Makki:2024sjq, Ali:2022moz, Koren:2022axd, Franchino-Vinas:2021nsf, Hayakawa:2021jxf, Deal:2021kjs, Hou:2021ynb, Anchordoqui:2020djl, Clara:2020efx, Flambaum:2018ohm, Hou:2017uap, Salvati:2016jng, Yamazaki:2014fja,
Cumberbatch:2007me}. In particular, Ref.~\cite{Fields:2022mpw} argues that  the discrepancy may be resolved by stellar lithium depletion, which, if correct, could bring the inferred primordial abundance into agreement with the SM prediction.

The recommended PDG value for the observed primordial abundance of lithium-7 is:

\begin{equation}
^7Li / H \times 10^{10} = (1.45 \pm 0.25) \quad \text{\cite{PDG2025}.}
\end{equation}

\subsection{Observation of $N_{\rm{eff}}$: A new value determined using LBT data} 

The value of $N_{\rm{eff}}$, the effective number of neutrino species, can be determined from the CMB independently of BBN, providing a complementary probe of the radiation content of the universe at recombination. However, the most precise current value comes from a combination of the primordial helium abundance  from the Large Binocular Telescope $Y_p$ Project, CMB data from ACT Data Release 6, Planck, and the South Pole Telescope, and baryon acoustic oscillation (BAO) data from the Dark Energy Spectroscopic Instrument. While this error bar likely sits at the optimistic end of plausible estimates, the central value is shown to be robust within their analysis, being insensitive to whether or not optical depth constraints inferred from CMB polarization data at large scales are included. However, it should be noted that alternative assumptions about the primordial abundance likelihoods or analysis details can yield somewhat different central values and more conservative uncertainties~\cite{ACT:2025tim,SPT-3G:2025bzu,Yanagisawa:2025mgx}. The new value, assuming a fixed total neutrino mass of 0.06 eV, is:

\begin{equation}
N_{\rm eff} = (2.990 \pm 0.070) \quad \text{\cite{Goldstein:2026iuu}.}
\end{equation}

The error bars on this measurement have been reduced by nearly a factor of two relative to previous determinations \cite{ACT:2025tim}. The result is in excellent agreement with the Standard Model prediction, $N_{\rm eff} = 3.044$.

\section{Theoretical Framework and Numerical Methods}
\label{sec:method}

In this study we use the publicly available BBN code, \faGithub \href{https://github.com/vallima/PRyMordial}{\,\texttt{PRyMordial}} which computes the primordial abundances of the light elements along with $N_{\rm eff}$ from first principles. The calculation of the light element abundances in \faGithub \href{https://github.com/vallima/PRyMordial}{\,\texttt{PRyMordial}}  proceeds in three stages. 

\begin{enumerate}
    \item \textbf{Background thermodynamics.} 
    The neutrino and photon temperatures as functions 
    of time, the scale factor as a function of photon 
    temperature, and $N_\text{eff}$ are computed by 
    solving coupled Boltzmann equations governing 
    the energy densities of the relativistic species.

    \item \textbf{Proton--neutron interconversion rates.} 
    Starting from the Born approximation, and 
    incorporating thermal corrections, radiative 
    corrections, and finite nucleon mass effects, 
    the $n \leftrightarrow p$ rates are obtained as 
    functions of photon temperature.

    \item \textbf{Nuclear reaction network.} 
    The rates from the previous stages are fed into 
    a nuclear reaction network that evolves the 
    abundances of D, T, ${}^3$He, ${}^4$He, ${}^7$Li, 
    and ${}^7$Be to their freeze-out values.
\end{enumerate}

\begin{figure}
    \centering
    \includegraphics[width=1.0\linewidth, trim=0 1600 0 0, clip]{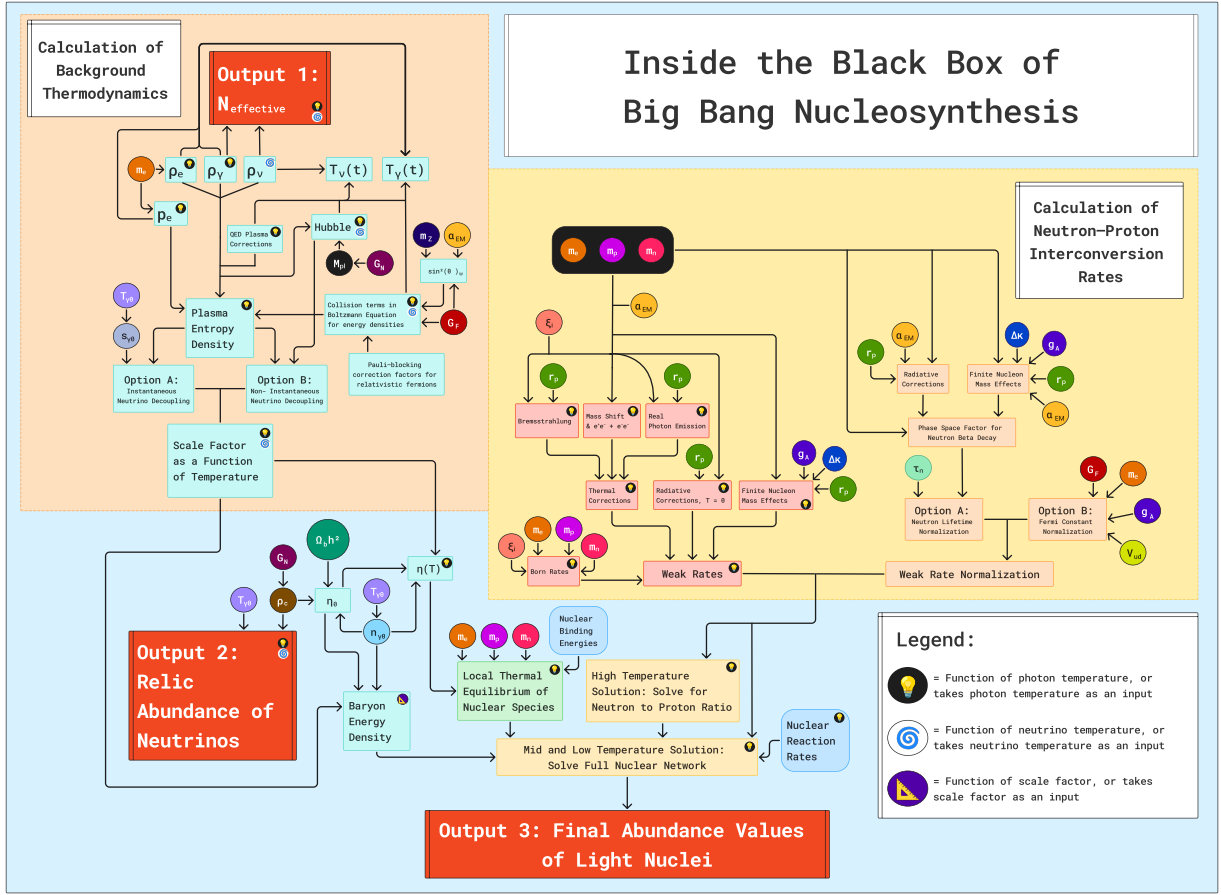}
    \caption{Schematic overview of the BBN calculation as implemented in \faGithub \href{https://github.com/vallima/PRyMordial}{\,\texttt{PRyMordial}}. The three computational stages: background thermodynamics , neutron--proton interconversion rates and the nuclear reaction network are shown along with their three outputs: $N_\text{eff}$, the relic neutrino abundance, and the final light-element abundances. Each colored circle represents a fundamental input parameter varied in this analysis. The symbols described in the legend indicates whether a given quantity depends on the photon temperature, neutrino temperature, or scale factor. Arrows trace how each input propagates through the calculation.}
    \label{fig:PRyM}
\end{figure}

Figure \ref{fig:PRyM} is a schematic diagram of the calculation done by the \faGithub \href{https://github.com/vallima/PRyMordial}{\,\texttt{PRyMordial}}~\cite{Burns:2023sgx}. Each colorful circle represents an input parameter whose impact on the predicted abundances we quantify in this study. Using this diagram, we can see all of the ways in which each parameter participates in the final calculation. For example, the fine structure constant, $\alpha_{EM}$, enters the calculation in three areas: in the collision terms in the Boltzmann Equations describing the energy densities of neutrinos, in the corrections to the weak rates, and in the weak rate normalization calculation. 

Four physical inputs are most consequential in the determination of the light element abundances: (a) the neutron-to-proton ratio, (b) the expansion rate of the universe,  (c) the baryon abundance, $\Omega_b h^2$, and (d) the thermonuclear reaction rates themselves. The sensitivity of each observable to these inputs is quantified in Section~\ref{sec:PVar}.

\paragraph{n/p} The neutron-to-proton ratio at the onset of BBN is controlled by the proton-neutron interconversion rates and the timing of their freeze-out. In short, a higher neutron-to-proton ratio leads to greater production of the light elements. $Y_p$ is particularly sensitive to this ratio, as the majority of the neutrons present prior to BBN ultimately end up bound in helium-4.

The neutron to proton interconversion rates are calculated via the Born Rates to which corrections are added to achieve percent-level accuracy or better. The Born Rates are a simplification of the full neutron to proton interconversion rates in which the masses of the nucleons are taken to be infinite, denoted by the superscript $\infty$, and thus their kinetic energy may be neglected ~\cite{Seckel:1993dc,Lopez:1997ki}. The Born Rates are given by the following equations: 

\begin{eqnarray}
    \label{eq:GammaBorn}
    \Gamma_{\rm n \to p}^{\infty} & = & \widetilde{G}_{\rm F}^2 \int_{0}^{\infty} dE_e \,E_e \, \sqrt{E_e^2-m_e^2} \, (E_\nu^-)^2 \left[ f_{\nu}(E_\nu^-)f_{e}(-E_e)+
    f_{\nu}(-E_\nu^-)f_{e}(E_e)\right] \ , \\
    \Gamma_{\rm p \to n}^{\infty} & = & \widetilde{G}_{\rm F}^2 \int_{0}^{\infty} dE_e \,E_e \, \sqrt{E_e^2-m_e^2} \, (E_\nu^+)^2 \left[ f_{\nu}(E_\nu^+)f_{e}(-E_e)+
    f_{\nu}(-E_\nu^+)f_{e}(E_e)\right] \ , \nonumber
\end{eqnarray}

in which $E_\nu^\pm = E_{e} \pm \mathcal{Q}$ with $\mathcal{Q} \equiv m_n - m_p$ and $\widetilde{G}_{\rm F}^2 = N / m_e^5$~\cite{Burns:2023sgx}. Here, $N$ is the normalization factor given by either Equation~\ref{eq:weakratenorm} or Equation~\ref{eq:weakratenorm_taun}, depending on the parameterization employed, discussed in further detail in Section \ref{sec:PVar}.

In our analysis we include the following corrections to the Born Rates: QED radiative corrections ~\cite{Sirlin:1967zza,Abers:1968zz,Dicus:1982bz,Ivanov:2012qe}, finite nucleon-mass effects and weak magnetism ~\cite{Seckel:1993dc,Lopez:1997ki}, and finite-temperature effects~\cite{Dicus:1982bz,Brown:2000cp}. Section 2.2 of~\cite{Burns:2023sgx} and Appendix~B of~\cite{Pitrou:2018cgg} give more information on the way in which all of the corrections are calculated.

\paragraph{Expansion rate of the Universe} The second key input is the expansion rate of the universe which is governed by the Friedmann equation,
\begin{equation}
    H^2 = \frac{8\pi}{3 M_{\rm Pl}^2} 
    \left[\rho_\gamma + \rho_{e^\pm} + \rho_\nu + 
    \rho_{\rm NP}\right] \,,
    \label{eq:Friedmann}
\end{equation}
where $\rho_{\rm NP}$ represents the energy density of any additional species beyond the SM.

A faster expansion causes the weak $n \leftrightarrow p$ rates to freeze out at a higher temperature, increasing the neutron-to-proton ratio and thus $Y_p$. In addition, earlier weak rate freeze out leaves less time for nuclear processing to proceed beyond helium-4. As a result, a faster expansion rate generally leads to higher final abundances of both helium-4 and deuterium.

\paragraph{Baryon Abundance} The third is the baryon energy density, $\rho_B(a)$ or equivalently, the baryon-to-photon ratio $\eta_b$, which sets the overall density of nucleons available for fusion into nuclei, 

\begin{equation}
    n_B(a) = \eta_b \, \frac{2\zeta(3)}{\pi^2} T_0^3 \, 
    a^{-3} = n_{B0} \, a^{-3} \,, \qquad 
    \rho_B(a) = m_a \, n_{B0} \, a^{-3} \,,
    \label{eq:rhoB}
\end{equation}
where $n_{B0}$ is the baryon number density today, $T_0$ is the photon temperature today, and $m_a$ is the atomic mass unit.

A higher baryon density increases the rate of all two-body nuclear reactions. The final deuterium abundance is particularly sensitive to this value as higher values of $\rho_B$ lead to more efficient burning of deuterium into heavier nuclei. The value of $\eta_b$ is precisely determined by CMB anisotropy measurements~\cite{Planck:2018vyg} via $\Omega_b h^2$, making deuterium a sensitive probe of any non-standard physics that causes a shift in the baryon density.

\paragraph{Thermo-nuclear Reaction Rates } Finally, the thermo-nuclear reaction rates directly determine the final abundance predictions of each light element. To determine the most influential reactions to the final prediction, we define a conservative sensitivity threshold: we consider a reaction to be significant if a 1\% change in its rate produces a shift of at least 0.001\% in any of the final light element abundances. For comparison, the most precise observable, $Y_p$, is currently measured to 0.5\% precision. Our threshold lies well below this, ensuring that no reaction capable of producing observable shifts when varied across its full uncertainty range is discarded. Applying this criterion, we find that only 12 of the 63 reactions in the network meet this threshold. These reactions are listed in Table~\ref{table:12}. For each of these 12 reactions, we use two different reaction rates in our analysis. A complete list of the nuclear reaction rates implemented in the code can be found in~\cite{Burns:2023sgx} and the complete set out results, including those for the additional 51 reactions can be found on our Github \faGithub \href{https://github.com/Anne-KatherineBurns/bbn-sensitivity-atlas}{\,\texttt{bbn-sensitivity-atlas}}.

The first set of rates are referred to as the PRIMAT rates as they are the same rates as those used by the PRIMAT code~\cite{Pitrou:2018cgg}. These rates are tabulated from a catalog of nuclear cross sections computed using numerical tools such as the TALYS code~\cite{Coc:2011az} or determined through statistical analyses within $R$-matrix theory~\cite{Descouvemont04,Longland:2010gs,Iliadis16,Gomez17}.   

The second set of rates, referred to as the NACRE-II rates, come from the NACRE~II database~\cite{Xu:2013fha}. This database  includes experimental results for the reaction rates of charged-particle induced reactions on target nuclei and adopts the  potential model to describe nuclear cross sections outside of the measured energy range~\cite{Angulo:1999zz}. Two of the rates in this database have been updated since it was compiled. In this study, we use the latest results for d(p, $\gamma$)$^3$He~\cite{Mossa:2020gjc} and $^7$Be(n, p)$^7$Li~\cite{Fields:2019pfx}.

The choice of reaction rates is particularly important for the determination of the primordial deuterium abundance. When the NACRE-II rates are used, the SM deuterium prediction is in agreement with the observed value. However, when the PRIMAT rates are used, the SM deuterium prediction is in tension with the observed value at the level of 2$\sigma$. The impact of changing the reaction rates on the primordial abundance values is discussed in further detail in Section \ref{sec:p_var_nuc}.

\begin{table}[t]
\centering
\vspace*{0pt}
\setlength{\heavyrulewidth}{0.08em}
\setlength{\lightrulewidth}{0.05em}
\setlength{\cmidrulewidth}{0.04em}
\setlength{\tabcolsep}{4pt}
\small
\begin{tabular}{l @{\hspace{1.5em}}  || @{\hspace{1.5em}} c @{\hspace{1.5em}} || @{\hspace{1.5em}}  c}

\textbf{Nuclear Reaction} & \textcolor{blue}{PRIMAT Ref.} & \textcolor{red}{NACRE-II Ref.} \\

$n+p \to \mathrm{D}+\gamma$                                        & \textcolor{blue}{\cite{Ando06}}          & \textcolor{red}{\cite{Ando06}} \\
$\mathrm{D}+p \to {}^{3}\mathrm{He}+\gamma$                        & \textcolor{blue}{\cite{Mossa:2020gjc}}   & \textcolor{red}{\cite{Mossa:2020gjc}} \\
$\mathrm{D}+\mathrm{D} \to {}^{3}\mathrm{He}+n$                    & \textcolor{blue}{\cite{Iliadis16}}       & \textcolor{red}{\cite{Xu:2013fha}} \\
$\mathrm{D}+\mathrm{D} \to {}^{3}\mathrm{H}+p$                     & \textcolor{blue}{\cite{Iliadis16}}       & \textcolor{red}{\cite{Xu:2013fha}} \\
${}^{3}\mathrm{H}+p \to {}^{4}\mathrm{He}+\gamma$                  & \textcolor{blue}{\cite{Serpico:2004gx}}  & \textcolor{red}{\cite{Serpico:2004gx}} \\
${}^{3}\mathrm{H}+\mathrm{D} \to {}^{4}\mathrm{He}+n$              & \textcolor{blue}{\cite{Descouvemont04}}  & \textcolor{red}{\cite{Xu:2013fha}} \\
${}^{3}\mathrm{H}+{}^{4}\mathrm{He} \to {}^{7}\mathrm{Li}+\gamma$  & \textcolor{blue}{\cite{Descouvemont04}}  & \textcolor{red}{\cite{Xu:2013fha}} \\
${}^{3}\mathrm{He}+n \to {}^{3}\mathrm{H}+p$                       & \textcolor{blue}{\cite{Descouvemont04}}  & \textcolor{red}{\cite{Cyburt:2004cq}} \\
${}^{3}\mathrm{He}+\mathrm{D} \to {}^{4}\mathrm{He}+p$             & \textcolor{blue}{\cite{Descouvemont04}}  & \textcolor{red}{\cite{Xu:2013fha}} \\
${}^{3}\mathrm{He}+{}^{4}\mathrm{He} \to {}^{7}\mathrm{Be}+\gamma$ & \textcolor{blue}{\cite{Iliadis16}}       & \textcolor{red}{\cite{Xu:2013fha}} \\
${}^{7}\mathrm{Be}+n \to {}^{7}\mathrm{Li}+p$                      & \textcolor{blue}{\cite{Descouvemont04}}  & \textcolor{red}{\cite{Fields:2019pfx}} \\
${}^{7}\mathrm{Li}+p \to {}^{4}\mathrm{He}+{}^{4}\mathrm{He}$      & \textcolor{blue}{\cite{Descouvemont04}}  & \textcolor{red}{\cite{Xu:2013fha}} \\

\end{tabular}
\caption{The key nuclear reactions, with corresponding references. The center column gives the references used for the \textcolor{blue}{PRIMAT} set of reaction rates, and the right column gives reactions corresponding to the \textcolor{red}{NACRE-II} set of reaction rates.}
\label{table:12}
\end{table}

Our approach offers several advantages over previous sensitivity studies. By computing numerical derivatives of each observable with respect to every input parameter individually, we construct a first-order sensitivity analysis with respect to each input parameter individually. Additionally, the use of two independent nuclear rate compilations, PRIMAT and NACRE-II, allows us to disentangle the uncertainty arising from nuclear physics inputs from that due to fundamental constants and cosmological parameters, and to quantify the systematic spread introduced by the choice of reaction rate database. Furthermore, the precision of the \faGithub \href{https://github.com/vallima/PRyMordial}{\,\texttt{PRyMordial}}  code which incorporates QED radiative corrections, finite nucleon-mass effects, weak magnetism, finite-temperature corrections, and non-instantaneous neutrino decoupling, ensures that the theoretical predictions are accurate at a level well below the input parameter uncertainties we are propagating. This combination of a high-precision BBN solver with a systematic, parameter-by-parameter exploration enables us to identify not only which inputs dominate the current error budget, but also where future improvements in experimental or observational precision would have the greatest impact on tightening the BBN predictions.

\section{Sensitivity Analysis: Fundamental Physics Parameters}
\label{sec:PVar}

We study the effect of varying 14 particle and astrophysics parameters and 63 nuclear reaction rates on the final primordial values of helium-4, deuterium, and lithium-7 as well as $N_{\rm eff}$. For each of these studies we compute the resulting abundance variation twice, once using the PRIMAT rates and once using the NACRE-II rates. In addition, for each study we chose one of two parameterizations for the normalization of the neutron-proton interconversion rates. The first, which we call the fundamental parameterization, has the following form:

\begin{equation}
N_{\textrm{Fundamental}} = (2\pi^3)^{-1} (G_F V_{ud})^2 (1+3g_A^2)  m_e^5.  
\label{eq:weakratenorm}
\end{equation}

The other, which we call the neutron lifetime parameterization is, 

\begin{equation}
N_{\tau_n} = 1 / (\tau_n F_n)
\label{eq:weakratenorm_taun}
\end{equation}

where $F_n$ is the neutron decay phase-space integral at zero temperature~\cite{Wilkinson:1982hu} with electroweak radiative corrections~\cite{Marciano:2005ec}. $F_n$ is a function of the neutron-proton mass difference, the nucleon axial coupling, and the weak magnetism constant.

For all parameters not entering into either of these two equations, we only report results using the neutron lifetime parameterization as in these cases the abundance variation is not affected by the choice of normalization. Full results for both choices of normalization are available on Github: \faGithub \href{https://github.com/Anne-KatherineBurns/bbn-sensitivity-atlas}{\,\texttt{bbn-sensitivity-atlas}}. For the Fermi constant, $G_F$, the nucleon axial coupling, $g_{A}$, the electron mass, $m_e$, the weak magnetism constant, $\Delta \kappa$, and the neutron-proton mass difference, $Q$ we report our results for both choices of weak rate normalization, as all of the parameters enter in one or the other of the normalization calculations, and in other places in the code so the predicted sensitivity depends on which normalization scheme is adopted. Finally, we report the result of varying the up-down element of the Cabibbo–Kobayashi–Maskawa (CKM) matrix, $V_{ud}$ only using the fundamental parameterization and the result of varying the neutron lifetime, $\tau_n$ only using the neutron lifetime parameterization as these two parameters exclusively enter in this part of the calculation.

Each parameter is varied independently about its Standard Model fiducial value, with all other parameters held fixed. The variation ranges are chosen to be broad enough to reveal the functional form of the response (typically $\pm 10-20\%$) while remaining within the domain where the observable varies smoothly. The sensitivity coefficients $d$ln$Y/d$ln$p$ are evaluated at the fiducial point and are therefore local, linear response functions which do not depend on the choice of variation range provided the response is approximately linear, as confirmed by the $R^2$ values reported in Section~\ref{sec:sens}. Correlations between parameters are neglected by construction: this choice produces a model-independent atlas, since the correlations between parameters are theory-dependent and vary across BSM scenarios. Users applying these sensitivities to a specific model in which parameters are correlated should propagate the full covariance matrix using the individual response functions provided here.

We find that $Y_p$ is most sensitive to the proton-neutron mass difference, $Q$, and parameters used to calculate the weak rate normalization, while D/H is dominated by $Q$ as well, and additionally the gravitational constant, $G_N$, and the baryon abundance, $\Omega_bh^2$. The full sensitivity matrices are presented in Tables \ref{tab:sens_taun_nacre_side_by_side}, \ref{tab:sens_fund_side_by_side}, \ref{tab:sens_taun_side_by_side_neff}, and \ref{tab:sens_tau_n_rates_first12_side_by_side}. In the following subsections, the result of varying each parameter is discussed, and plots are shown for the most impactful parameters.

\subsection{Neutron lifetime, $\tau_{n}$}

The neutron lifetime, $\tau_n$ only enters the calculation in one place: in the normalization of the charged-current weak rates governing $n\leftrightarrow p$ interconversion shown in Equation \ref{eq:weakratenorm_taun}. This parameter is particularly interesting to study due to the fact that its experimental determination remains limited by the beam--bottle discrepancy and associated uncertainties~\cite{PDG2025,UCNt:2021pcg,Yue:2013qrc}.

\begin{figure}[t]
    \centering
    \begin{minipage}{0.49\linewidth}
        \centering
        \includegraphics[width=\linewidth]{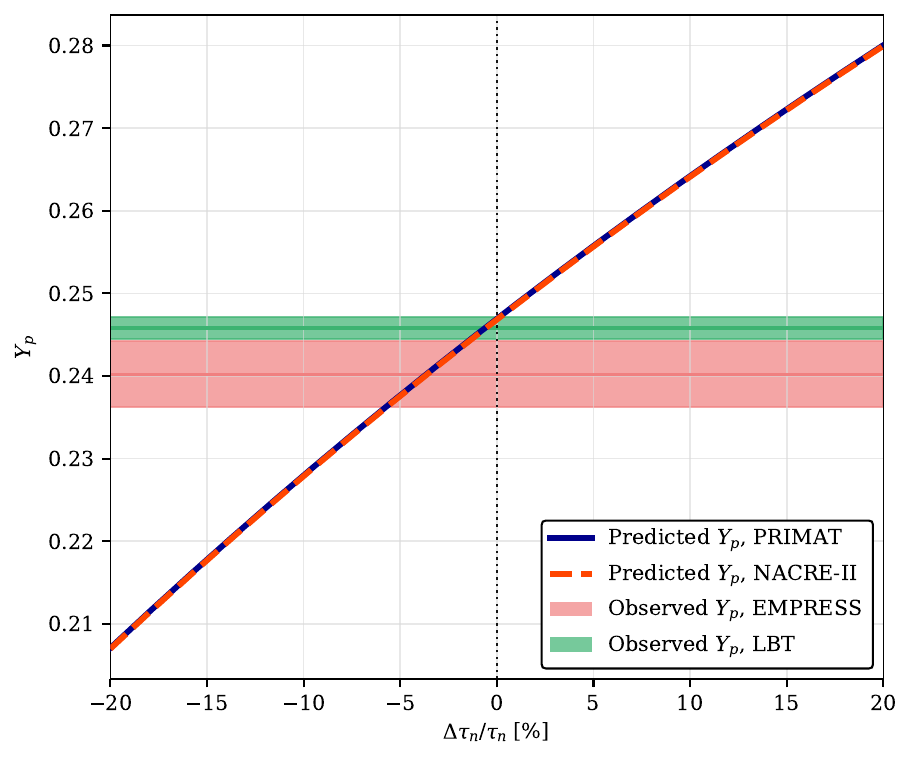}
        \caption{Predicted $Y_p$ as a function of $\tau_n$/$\tau_n^{\rm SM}$ calculated using PRIMAT (blue) and NACRE-II (red) reaction rates.}
        \label{fig:tau_Yp}
    \end{minipage}\hfill
    \begin{minipage}{0.49\linewidth}
        \centering
        \includegraphics[width=\linewidth]{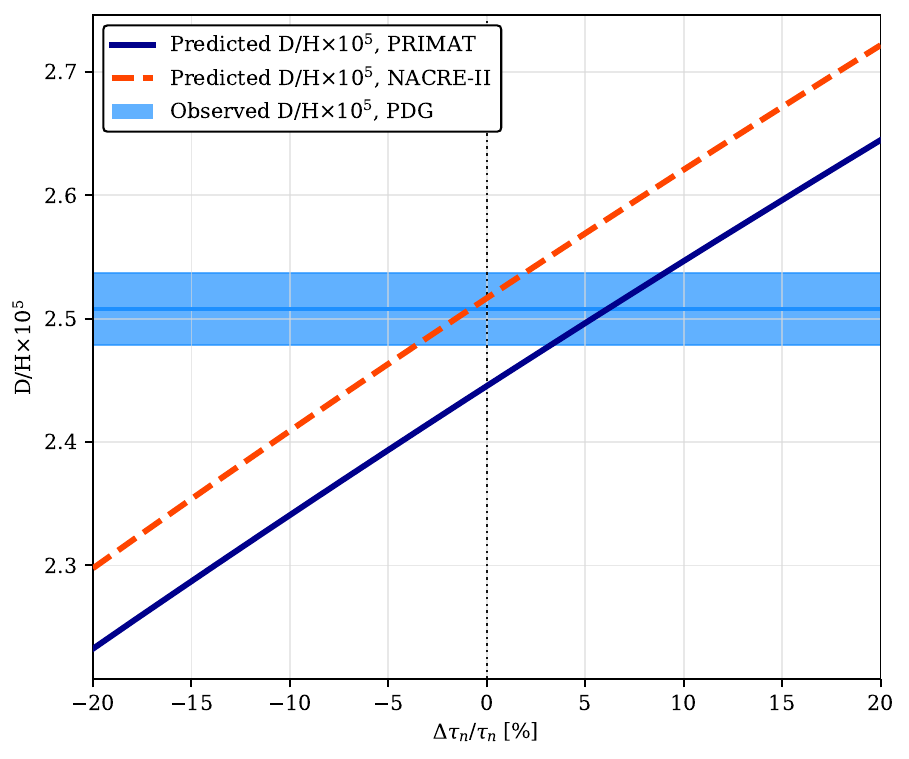}
        \caption{Predicted $D/H \times 10^5$ as a function of $\tau_n$/$\tau_n^{\rm SM}$ calculated using PRIMAT (blue) and NACRE-II (red) reaction rates.}
        \label{fig:tau_D}
    \end{minipage}
\end{figure}

Changing the normalization factor of the weak rates shifts the weak freeze-out temperature thereby impacting the ratio of the number of neutrons to number of protons before BBN, which then significantly impacts the final abundance values of helium-4 and deuterium, and modestly impacts the final abundance value of lithium-7. Here, increasing $\tau_n$, decreases the weak rate normalization factor which in turn, causes weak rate freeze out to occur earlier, increasing the n/p ratio at the onset of nucleosynthesis. As a result, $Y_p$ increases as more neutrons are available to be captured into helium-4, $D/H$ increases as the higher initial neutron abundance produces more deuterium than can be burned through the nuclear network, and $^7Li/H$ increases modestly due to enhanced $^7Be$ production. The changes in helium-4 and deuterium are shown in Figures \ref{fig:tau_Yp} and \ref{fig:tau_D}. Changing $\tau_n$ has no impact on the final value of $N_{\rm{eff}}$.

\subsection{Neutron-proton mass difference, Q}

The neutron-proton mass difference $Q = m_n - m_p$ affects BBN through two competing mechanisms whose relative importance depends on the weak rate normalization scheme. First, larger values of Q correspond to the neutron having more available phase space in $n \rightarrow p$ channels, increasing those rates. Conversely, larger values of Q correspond to the inverse direction being more endothermic, so $p \rightarrow n$ rates are suppressed. These effects are consistent with the equilibrium expectation: near equilibrium, the neutron-to-proton ratio is predominantly set by the Boltzmann factor for which Q is the dominant piece when the ratio of the electron-neutrino chemical potential to the neutron temperature, $\xi_{\nu e}$ is small or zero. 

\begin{equation}
    (n_n / n_p)|_{eq} \simeq e^{-Q/T - \xi_{\nu e}}
\label{eq:np_ratio}
\end{equation}

Both of these effects cause the number of neutrons to decrease when Q is increased, which cause the final abundance values of all of the light elements to be lower.  

\begin{figure}[t]
    \centering
    \begin{minipage}{0.49\linewidth}
        \centering
        \includegraphics[width=\linewidth]{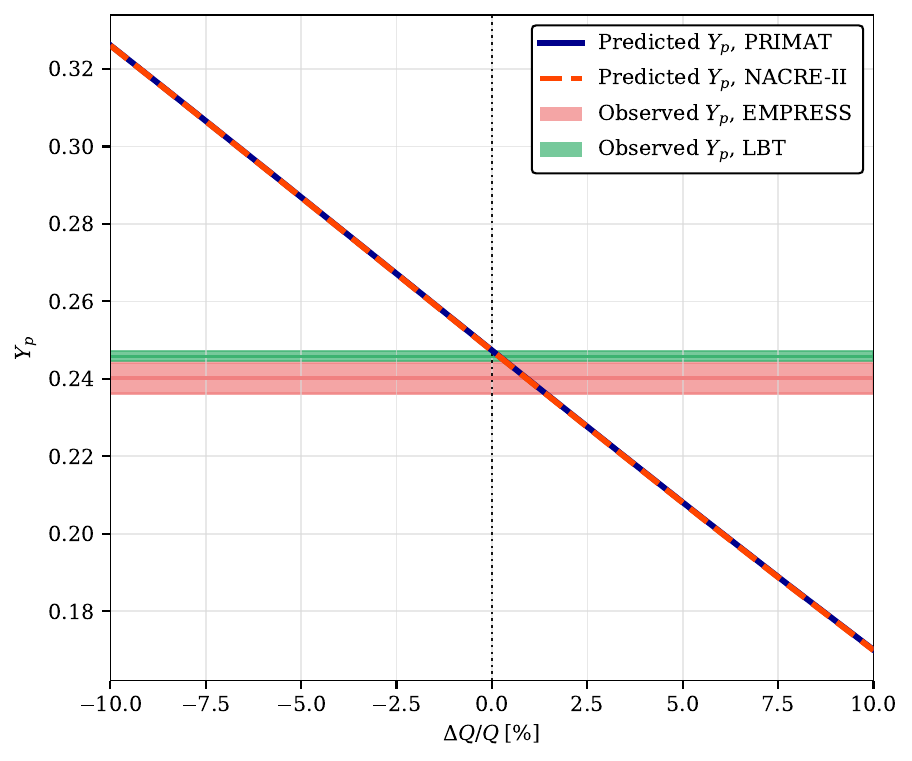}
        \caption{Predicted $Y_p$ as a function of $Q$/$Q^{\rm SM}$ calculated using PRIMAT (blue) and NACRE-II (red) reaction rates using the fundamental weak rate normalization.}
        \label{fig:Q_fund_Yp}
    \end{minipage}\hfill
    \begin{minipage}{0.49\linewidth}
        \centering
        \includegraphics[width=\linewidth]{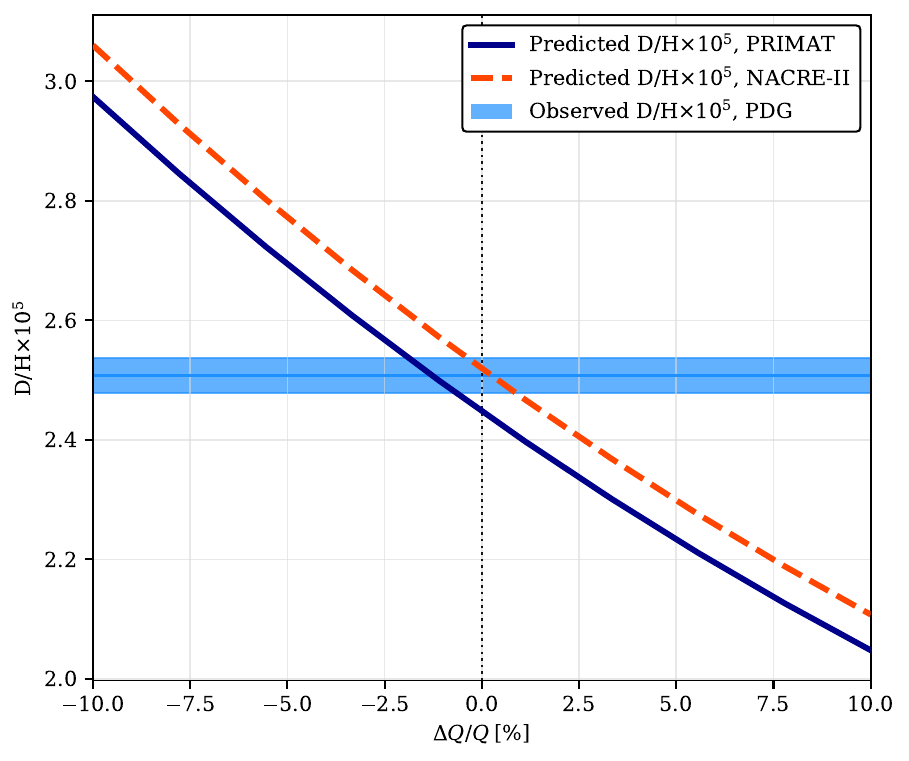}
        \caption{Predicted $D/H \times 10^5$ as a function of $Q$/$Q^{\rm SM}$ calculated using PRIMAT (blue) and NACRE-II (red) reaction rates using the fundamental weak rate normalization.}
        \label{fig:Q_fund_D}
    \end{minipage}
    \begin{minipage}{0.49\linewidth}
        \centering
        \includegraphics[width=\linewidth]{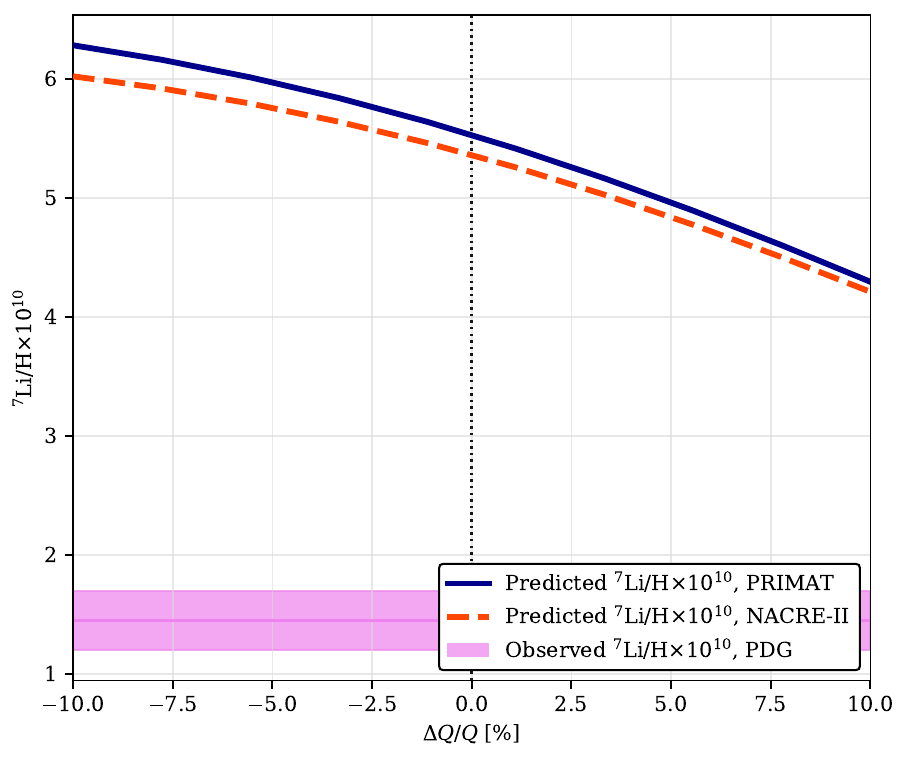}
        \caption{Predicted $^7Li/H \times 10^{10}$ as a function of $Q$/$Q^{\rm SM}$ calculated using PRIMAT (blue) and NACRE-II (red) reaction rates using the fundamental weak rate normalization.}
        \label{fig:Q_fund_Li7}
    \end{minipage}
\end{figure}

However, the final abundances' net sensitivity to Q depends critically on the weak rate normalization scheme used. When the fundamental weak rate normalization is used, Q does not play a role in the normalization and the final abundance values of all of the light elements decrease with increasing Q due to the effects outlined above as shown in Figures \ref{fig:Q_fund_Yp}, \ref{fig:Q_fund_D}, and \ref{fig:Q_fund_Li7}.

\begin{figure}[t]
    \centering
    \begin{minipage}{0.49\linewidth}
        \centering
        \includegraphics[width=\linewidth]{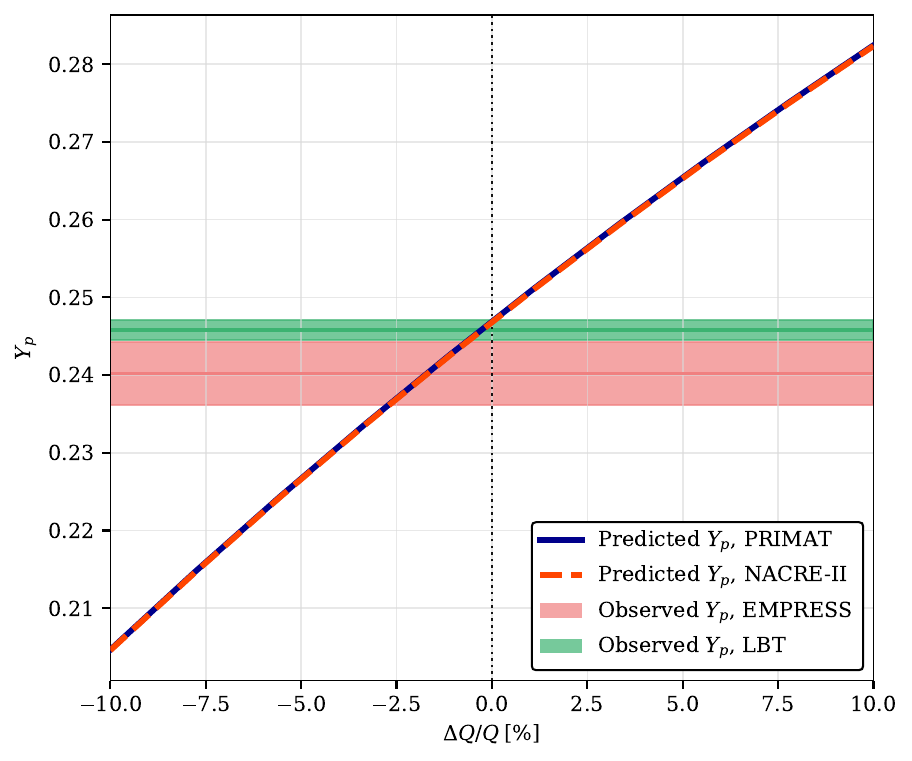}
        \caption{Predicted $Y_p$ as a function of $Q$/$Q^{\rm SM}$ calculated using PRIMAT (blue) and NACRE-II (red) reaction rates using the $\tau_n$ weak rate normalization.}
        \label{fig:Q_tau_Yp}
    \end{minipage}\hfill
    \begin{minipage}{0.49\linewidth}
        \centering
        \includegraphics[width=\linewidth]{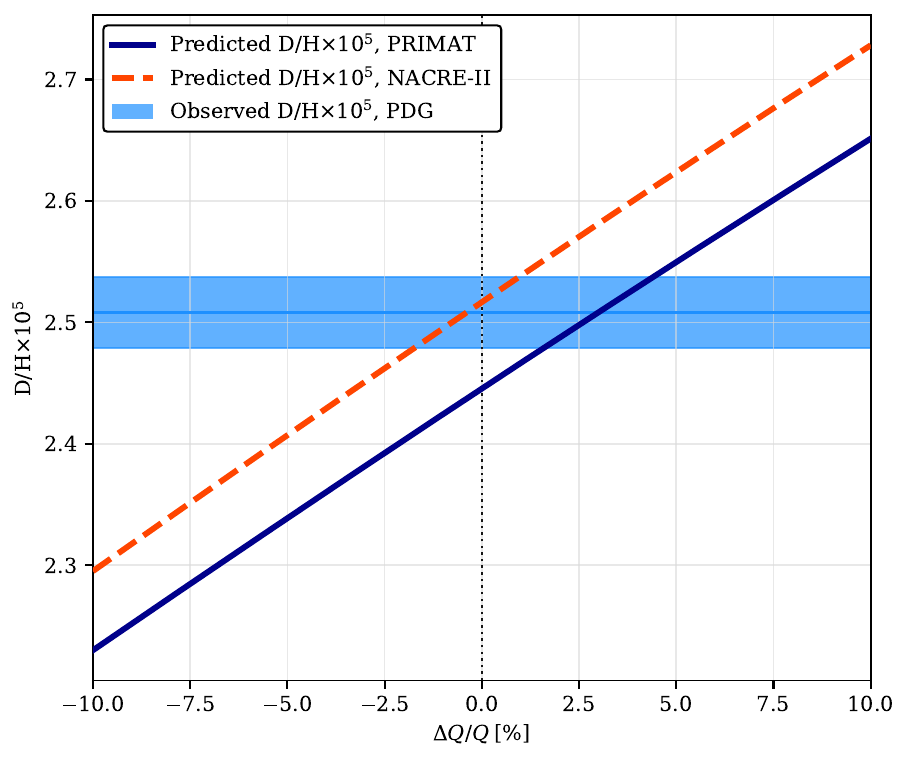}
        \caption{Predicted $D/H \times 10^5$  as a function of $Q$/$Q^{\rm SM}$ calculated using PRIMAT (blue) and NACRE-II (red) reaction rates using the $\tau_n$ weak rate normalization.}
        \label{fig:Q_tau_D}
    \end{minipage}
    \begin{minipage}{0.49\linewidth}
        \centering
        \includegraphics[width=\linewidth]{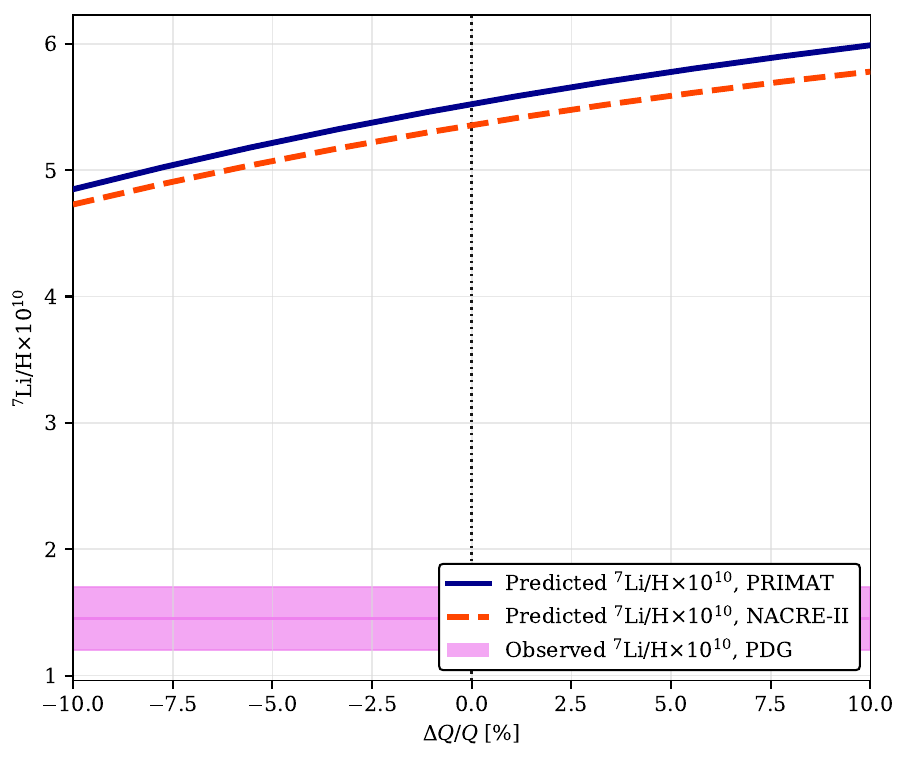}
        \caption{Predicted $^7Li/H \times 10^{10}$  as a function of $Q$/$Q^{\rm SM}$ calculated using PRIMAT (blue) and NACRE-II (red) reaction rates using the $\tau_n$ weak rate normalization.}
        \label{fig:Q_tau_Li7}
    \end{minipage}
\end{figure}

When the neutron lifetime normalization is used, by contrast, the prefactor on the rates takes the form in Equation \ref{eq:weakratenorm_taun}, where the neutron decay phase-space integral $F_n$ scales like $\sim Q^5$. Therefore, increasing Q steeply increases $F_n$, decreasing the weak rate normalization factor. This, in turn, causes weak rate freeze out to occur earlier, increasing the number of neutrons at the onset of nucleosynthesis. This effect is the dominant one when neutron lifetime weak rate normalization is used, and  thus the final abundance value of each of the light elements increases with increased Q, as shown in Figures \ref{fig:Q_tau_Yp}, \ref{fig:Q_tau_D}, and \ref{fig:Q_tau_Li7}.

Secondarily, masses of protons and neutrons, and thus Q, play a role in the calculation of the normalization of the nuclear reaction rates, as this is calculated using the measured values of atomic mass excess values for neutral atoms from NUBASE2020~\cite{Kondev_2021}. This effect is subdominant and does not alter the trends described above.

Notably, $Q$ is a parameter for which the predicted sensitivity is qualitatively normalization-dependent: the fundamental scheme predicts all abundances decrease with increasing $Q$, highlighting why both weak rate normalization schemes must be examined. 

\subsection{Fine structure constant, $\alpha_{EM}$}
\label{sec:alphaem}

The fine structure constant, $\alpha_{EM}$, plays two roles in the calculation of the light element abundances and $N_{\rm{eff}}$. The first is in the calculation of the proton-neutron interconversion rates. Specifically, $\alpha_{EM}$ is present in the standard electroweak correction factors which multiply tree-level neutron–proton weak rates. The first of these is the Fermi-Coulomb factor, which accounts for the Coulomb interaction between the outgoing electron or positron and the daughter proton, enhancing or suppressing the rate depending on the sign of the charge and the lepton velocity. 
\begin{equation}
F(b)
=
\left(1+\frac{\Gamma}{2}\right)\,4\,
\left(\frac{2\,r_p\,b}{\lambda_C}\right)^{2\Gamma}\,
\frac{1}{\Gamma(3+2\Gamma)^2}\,
\frac{e^{\left(\pi \alpha_{EM}/b\right)}}{(1-b^2)^{\Gamma}}\,
\left|\Gamma\!\left((\Gamma+1)+i\frac{\alpha_{EM}}{b}\right)\right|^2
\,,
\end{equation}

where $\Gamma = \sqrt{(1-\alpha_{EM}^2)}-1$, $\lambda_C$ is the reduced Compton wavelength of the electron, equal to $\hbar c/m_e$, and $b$ is the lepton velocity. 

The second factor is a multiplicative radiative correction to the weak rate, using the formulation from Czarnecki et al.~\cite{Czarnecki:2004cw}. Equation 15 in their paper shows the complete equation. 

In addition, $\alpha_{EM}$ plays a role in the finite-temperature radiative corrections, finite nucleon mass effects, Bremsstrahlung, mass shift, and real photon emission corrections to the weak rates, shown in Equation 107 of Brown and Sawyer's paper~\cite{Brown:2000cp}. 

Finally, $\alpha_{EM}$ is used in the calculation of the interactions between electrons, positrons and neutrinos. Here, it enters the neutrino–electron/positron collision terms indirectly through the tree-level determination of sin$^2 \theta_W$, which fixes the neutral-current couplings appearing in the prefactor of the collision terms of the Boltzmann equations for the neutrinos. Due to the fact that sin$^2 \theta_W$ must be real, the largest allowed value of $\alpha_{EM}$ that we test is 0.0109149, which corresponds to about a 49.6\% increase as compared to the SM value. 

\begin{figure}[t]
    \centering
    \begin{minipage}{0.49\linewidth}
        \centering
        \includegraphics[width=\linewidth]{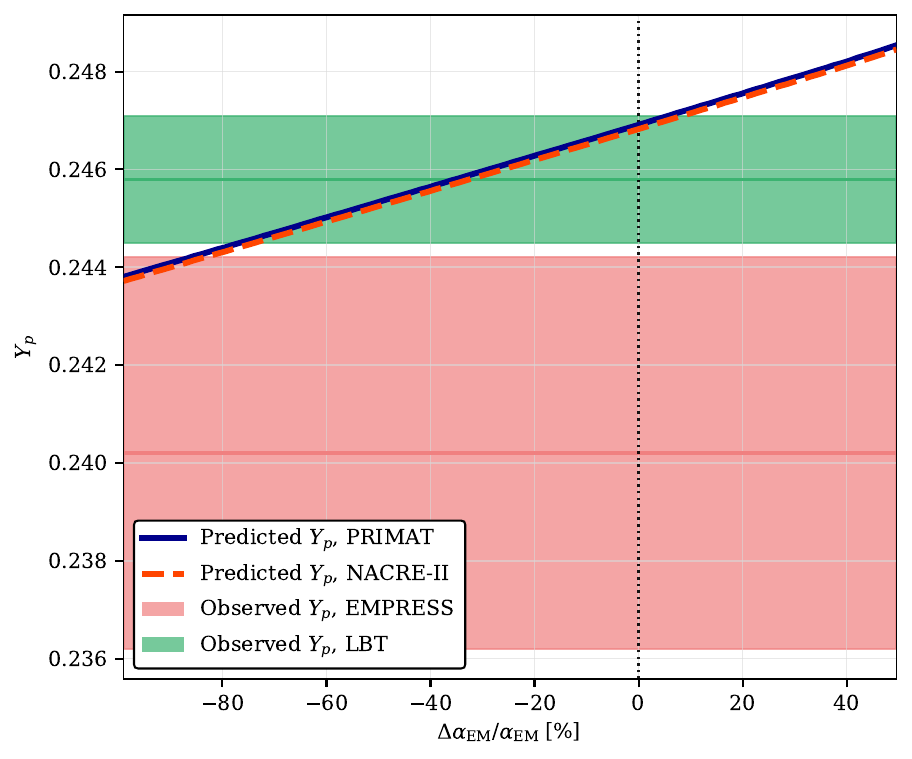}
        \caption{Predicted $Y_p$ as a function of $\alpha_{EM}$/$\alpha_{EM}^{\rm SM}$ calculated using PRIMAT (blue) and NACRE-II (red) reaction rates.}
        \label{fig:alpha_Yp}
    \end{minipage}\hfill
    \begin{minipage}{0.49\linewidth}
        \centering
        \includegraphics[width=\linewidth]{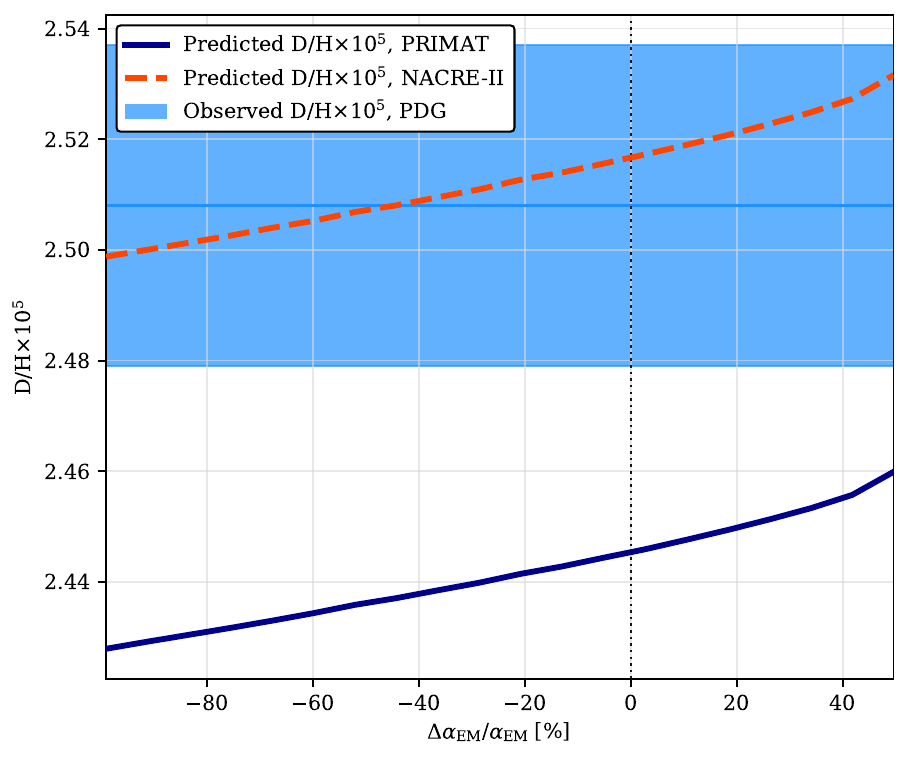}
        \caption{Predicted $D/H \times 10^5$ as a function of $\alpha_{EM}$/$\alpha_{EM}^{\rm SM}$ calculated using PRIMAT (blue) and NACRE-II (red) reaction rates.}
        \label{fig:alpha_D}
    \end{minipage}
\end{figure}

Changing $\alpha_{EM}$ has a significant impact on the final abundance value of helium-4 and a modest impact on the final abundance value of deuterium, primarily due to changes in the weak rates and thus, the neutron to proton ratio before BBN. Varying $\alpha_{EM}$ has a negligible impact on the lithium-7 abundance. These results are shown in Figures \ref{fig:alpha_Yp} and \ref{fig:alpha_D}.

\begin{figure}[t] 
    \centering 
    \includegraphics[width=.6\linewidth]{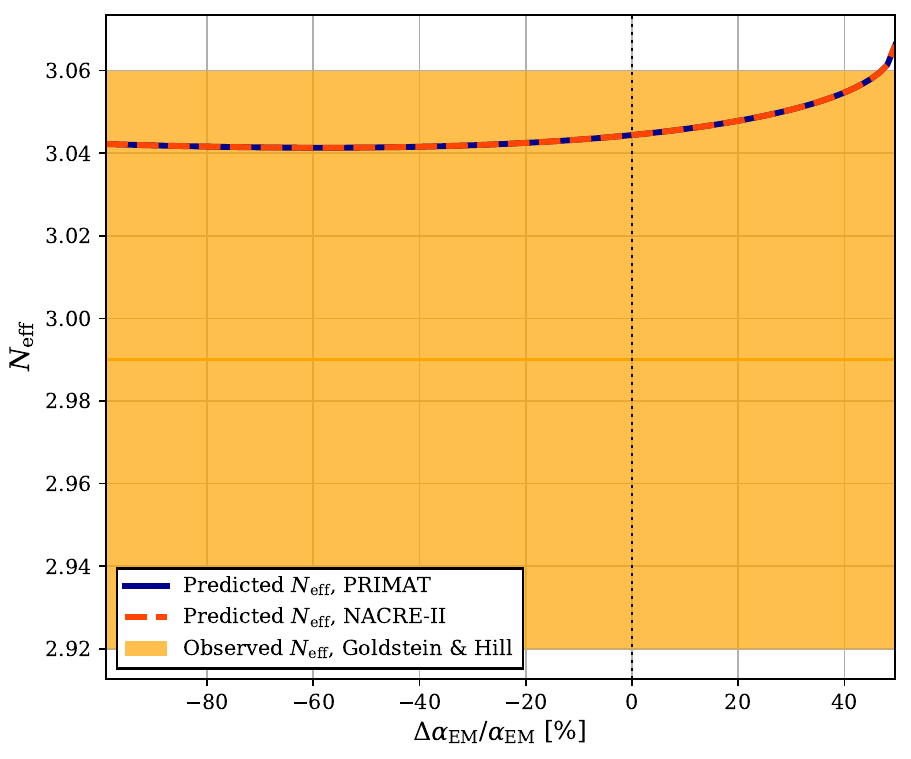} 
    \caption{Predicted $N_{\rm{eff}}$ as a function of $\alpha_{EM}$/$\alpha_{EM}^{\rm SM}$ calculated using PRIMAT (blue) and NACRE-II (red) reaction rates.} 
    \label{fig:alpha_Neff} 
\end{figure}

Changing $\alpha_{EM}$ has a small impact on the final value of $N_{\rm{eff}}$, as changing the collision terms in the Boltzmann equations of the neutrinos impacts the time of neutrino decoupling, and thus, the final neutrino temperature. This is because if neutrinos decouple later, more entropy is transferred to them from electron-positron annihilation, and their final temperature is higher. If they decouple earlier, they receive less entropy from electron-positron annihilation, and their final temperature is lower. This result is shown in Figure \ref{fig:alpha_Neff}.

\subsection{Fermi constant, $G_{F}$}
\label{sec:GF}

\begin{figure}[t]
    \centering
    \begin{minipage}{0.49\linewidth}
        \centering
        \includegraphics[width=\linewidth]{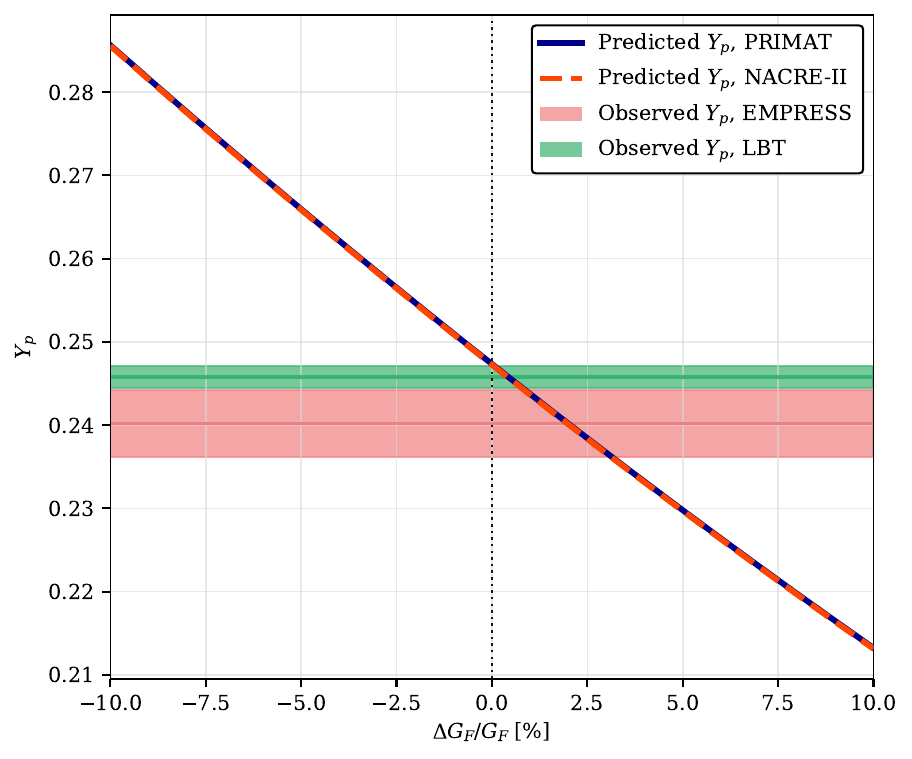}
        \caption{Predicted $Y_p$ as a function of $G_F$/$G_F^{\rm SM}$ calculated using PRIMAT (blue) and NACRE-II (red) reaction rates.}
        \label{fig:GF_Yp}
    \end{minipage}\hfill
    \begin{minipage}{0.49\linewidth}
        \centering
        \includegraphics[width=\linewidth]{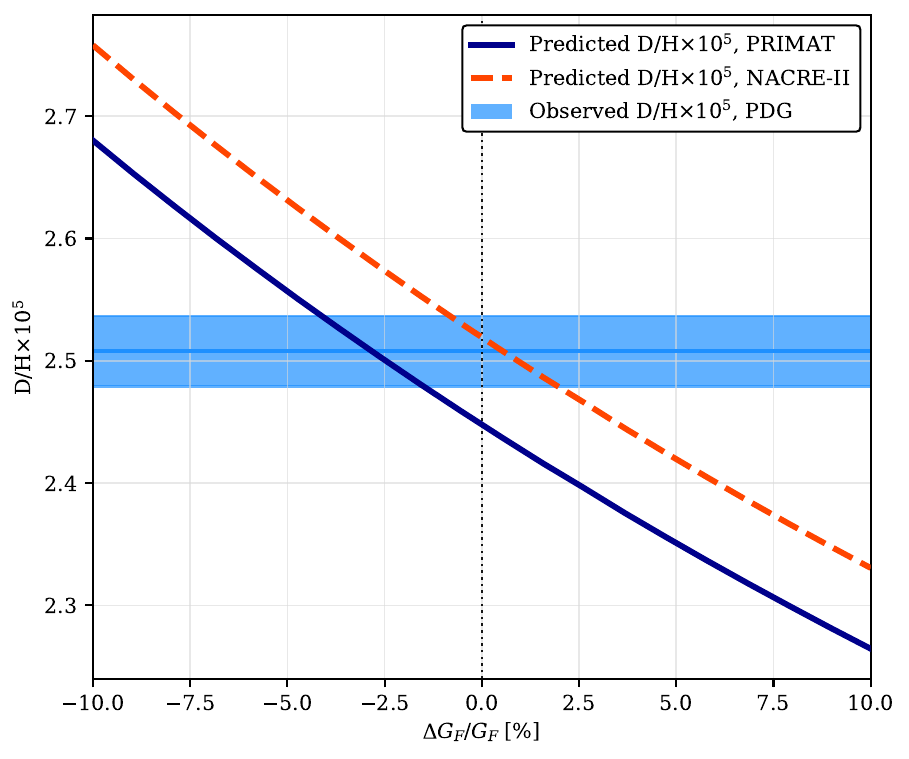}
        \caption{Predicted $D/H \times 10^5$ as a function of $G_F$/$G_F^{\rm SM}$ calculated using PRIMAT (blue) and NACRE-II (red) reaction rates.}
        \label{fig:GF_D}
    \end{minipage}

    \vspace{0.6em} 

    \begin{minipage}{0.49\linewidth}
        \centering
        \includegraphics[width=\linewidth]{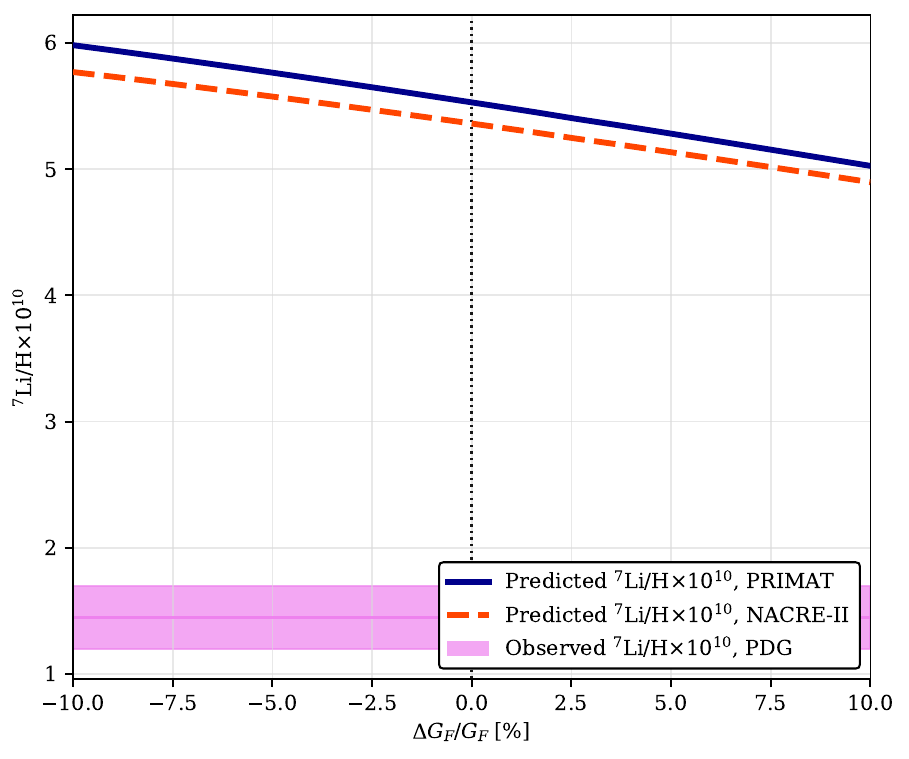}
        \caption{Predicted $^7Li/H \times 10^{10}$ as a function of $G_F$/$G_F^{\rm SM}$ calculated using PRIMAT (blue) and NACRE-II (red) reaction rates.}
        \label{fig:GF_Li7}
    \end{minipage}
\end{figure}

The Fermi constant $G_F$ affects BBN through either one or two channels, depending on the weak rate normalization scheme. When we normalize the weak rates using fundamental parameters, $G_F$ enters directly through the overall rate normalization as shown in Equation \ref{eq:weakratenorm}. When this choice of the weak rate normalization is made, the helium-4 and deuterium final abundance values are significantly impacted, and the lithium-7 final abundance value is modestly impacted, as shown in Figures \ref{fig:GF_Yp}, \ref{fig:GF_D}, and \ref{fig:GF_Li7}. When $G_F$ is increased, the normalization factor increases as well, delaying freeze out and decreasing the neutron to proton ratio, which in turn decreases all primordial abundance values.

Secondarily, in both weak rate normalization schemes, $G_{F}$ enters in  the interaction between electrons, positrons and neutrinos. Here, it enters the neutrino–electron/positron collision terms indirectly through the tree-level determination of sin$^2 \theta_W$, which fixes the neutral-current couplings appearing in the prefactor of the collision terms of the Boltzmann equations for the neutrinos. Due to the fact that sin$^2 \theta_W$ must be real, the smallest allowed value of $G_F$ that we test is 7.8e-12 MeV$^{-2}$, which corresponds to about a 33\% decrease as compared to the SM value. The way in which this impacts the final value of $N_{\rm{eff}}$ is discussed in section \ref{sec:alphaem}.

\subsection{Electron mass, $m_{e}$}

The electron mass, $m_e$ is the parameter with the most pervasive reach in the BBN calculation, entering at every stage of the pipeline shown in Figure \ref{fig:PRyM}. The first place $m_e$ enters is in the calculation of the background thermodynamics, including the temperature of the plasma, and the time of neutrino decoupling due to its dependence on electron-positron annihilation and neutrino scattering with electrons and positrons. In addition, the electron mass plays a role in every part of the calculation of the weak rates, including in the Born rates and their corrections, as well as in the weak rate normalization when the fundamental constant normalization is chosen as shown in Equation \ref{eq:weakratenorm}. Finally, the electron mass also plays a role in the calculation of the normalization of the nuclear reaction rates, as this is calculated using the measured values of atomic mass excess values for neutral atoms from NUBASE2020~\cite{Kondev_2021}.

Regardless of the weak rate normalization chosen, the impact of changing the electron mass on the helium-4 and deuterium abundances is strong, and the impact is modest on $N_{\rm{eff}}$ and the lithium-7 abundance. 

\begin{figure}[t]
    \centering
    \begin{minipage}{0.49\linewidth}
        \centering
        \includegraphics[width=\linewidth]{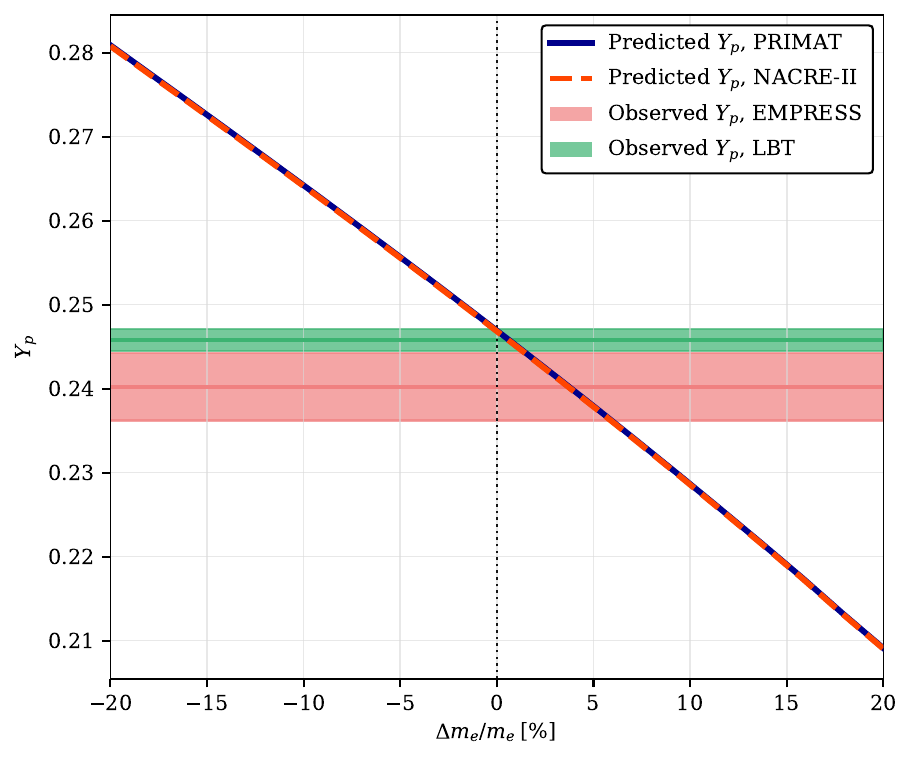}
        \caption{Predicted $Y_p$ as a function of $m_e$/$m_e^{\rm SM}$ calculated using PRIMAT (blue) and NACRE-II (red) reaction rates using the $\tau_n$ weak rate normalization.}
        \label{fig:me_tau_Yp}
    \end{minipage}\hfill
    \begin{minipage}{0.49\linewidth}
        \centering
        \includegraphics[width=\linewidth]{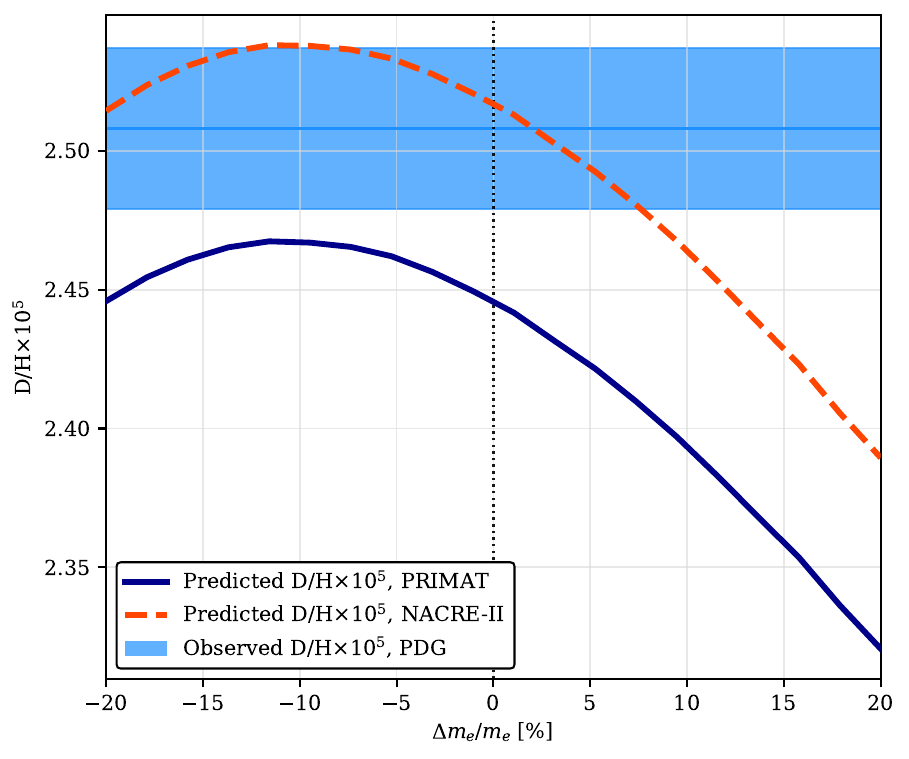}
        \caption{Predicted $D/H \times 10^5$ as a function of $m_e$/$m_e^{\rm SM}$ calculated using PRIMAT (blue) and NACRE-II (red) reaction rates using the $\tau_n$ weak rate normalization.}
        \label{fig:me_tau_D}
    \end{minipage}
\end{figure}

\begin{figure}[t]
    \centering
    \begin{minipage}{0.49\linewidth}
        \centering
        \includegraphics[width=\linewidth]{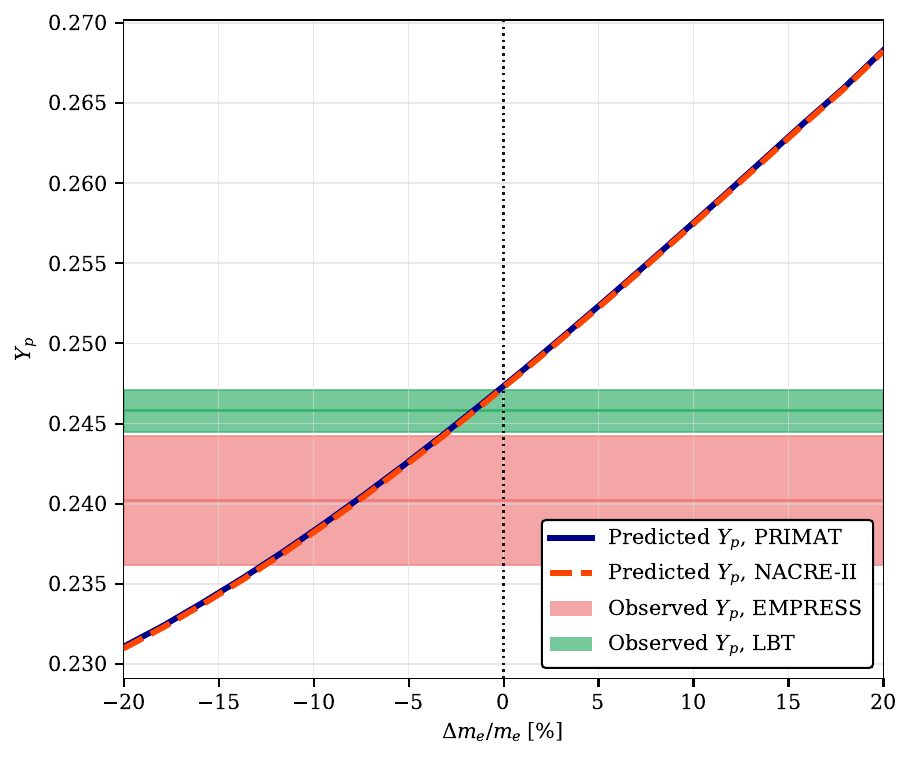}
        \caption{Predicted $Y_p$ as a function of $m_e$/$m_e^{\rm SM}$ calculated using PRIMAT (blue) and NACRE-II (red) reaction rates using the fundamental weak rate normalization.}
        \label{fig:me_fund_Yp}
    \end{minipage}\hfill
    \begin{minipage}{0.49\linewidth}
        \centering
        \includegraphics[width=\linewidth]{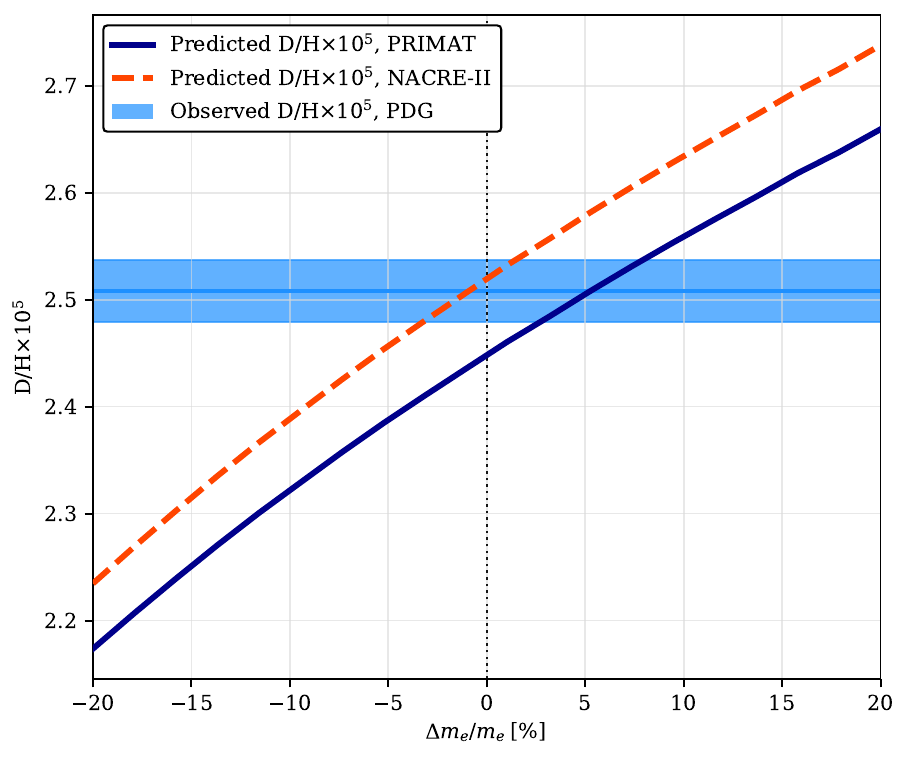}
        \caption{Predicted $D/H \times 10^5$ as a function of $m_e$/$m_e^{\rm SM}$ calculated using PRIMAT (blue) and NACRE-II (red) reaction rates using the fundamental weak rate normalization.}
        \label{fig:me_fund_D}
    \end{minipage}
\end{figure}

The results for the change in the helium-4 and deuterium abundances using weak rate normalization calculated using the neutron lifetime is shown in Figures \ref{fig:me_tau_Yp} and \ref{fig:me_tau_D}. The results for the change in the helium-4 and deuterium abundances using weak rate normalization calculated using fundamental constants is shown in Figures \ref{fig:me_fund_Yp} and \ref{fig:me_fund_D}. 

It should be noted that when the weak rate normalization is calculated using the neutron lifetime, the response function for deuterium is extremely non-linear, and thus the value of $\frac{dlnY}{dlnp}$ reported in Table \ref{tab:sens_taun_nacre_side_by_side} is only valid in a very small range around the fiducial value of $m_e$. This behavior is due to the fact that the two dominant entry points of the electron mass in the calculation, the plasma thermodynamics and the weak rates, produce opposite responses in the calculated value of the deuterium abundance. Increasing $m_e$ shifts $e^+e^-$ annihilation to higher temperatures, so the photon bath is no longer undergoing reheating as it passes through the deuterium-burning window, thus the universe crosses that window more quickly, less deuterium is destroyed, and the final deuterium abundance is increased. In contrast, increasing the electron mass slightly lowers the neutron-to-proton ratio at freeze-out, reducing the final deuterium abundance. 

\begin{figure}[t]
        \centering
        \includegraphics[width=0.6\linewidth]{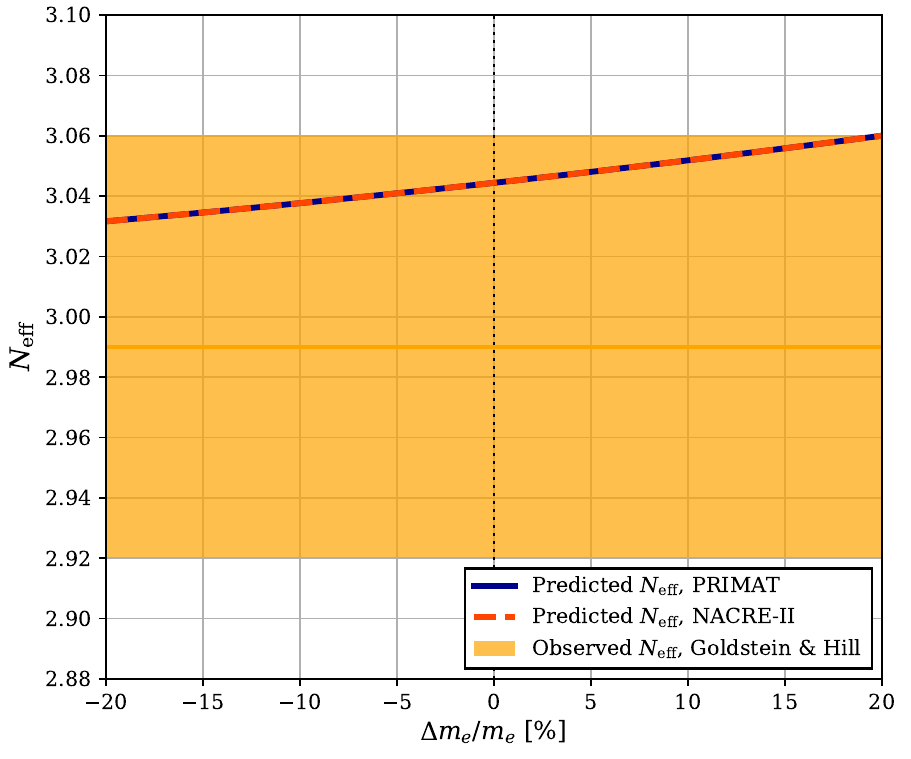}
        \caption{Predicted $N_{\rm{eff}}$ as a function of $m_e$/$m_e^{\rm SM}$ calculated using PRIMAT (blue) and NACRE-II (red) reaction rates using the fundamental weak rate normalization.}
        \label{fig:me_fund_Neff}
\end{figure}

Of all of the parameters studied here, $m_e$ has the most direct effect on the evolution of $N_{\rm eff}$. First, $m_e$ sets the temperature at which electron positron annihilation occurs: heavier electrons causes annihilation to occur earlier, when neutrinos are still partially coupled to the plasma, allowing them to absorb a larger fraction of the entropy released. This raises the final neutrino energy density and thus $N_{\rm eff}$. While $\alpha_{EM}$ and $G_F$ also affect $N_{\rm eff}$ by modifying the strength of the coupling between neutrinos and electrons, as discussed in Sections \ref{sec:alphaem} and \ref{sec:GF}, $m_e$ is unique in that it shifts the timing of the annihilation itself rather than changing the decoupling dynamics. This makes it the most direct handle on the competition between these two timescales that ultimately determines $N_{\rm eff}$. The dependence on $m_e$ of $N_{\rm{eff}}$ is shown in Figure \ref{fig:me_fund_Neff}.

\subsection{Nucleon axial coupling, $g_{A}$}

The nucleon axial coupling, $g_{A}$, which parametrizes the axial-vector contribution to the charged-current weak interaction enters the calculation in two places. When the fundamental weak rate normalization is used, the freeze out temperature and thus the neutron to proton ratio before BBN changes with shifts in the normalization. As shown in Equation \ref{eq:weakratenorm}, the normalization scales like $(1+ 3g_{A}^2)$ and thus higher values of $g_{A}$ delay freeze-out to lower temperature, and reduces the neutron to proton ratio which leads to lower final values in the light element abundances. 

Secondarily, using either weak rate normalization scheme $g_{A}$ is used to calculate the finite nucleon mass effect correction to the Born rates and has a negligible impact on the final abundance values and no impact on the final $N_{\rm{eff}}$ value. 

\begin{figure}[t]
    \centering
    \begin{minipage}{0.49\linewidth}
        \centering
        \includegraphics[width=\linewidth]{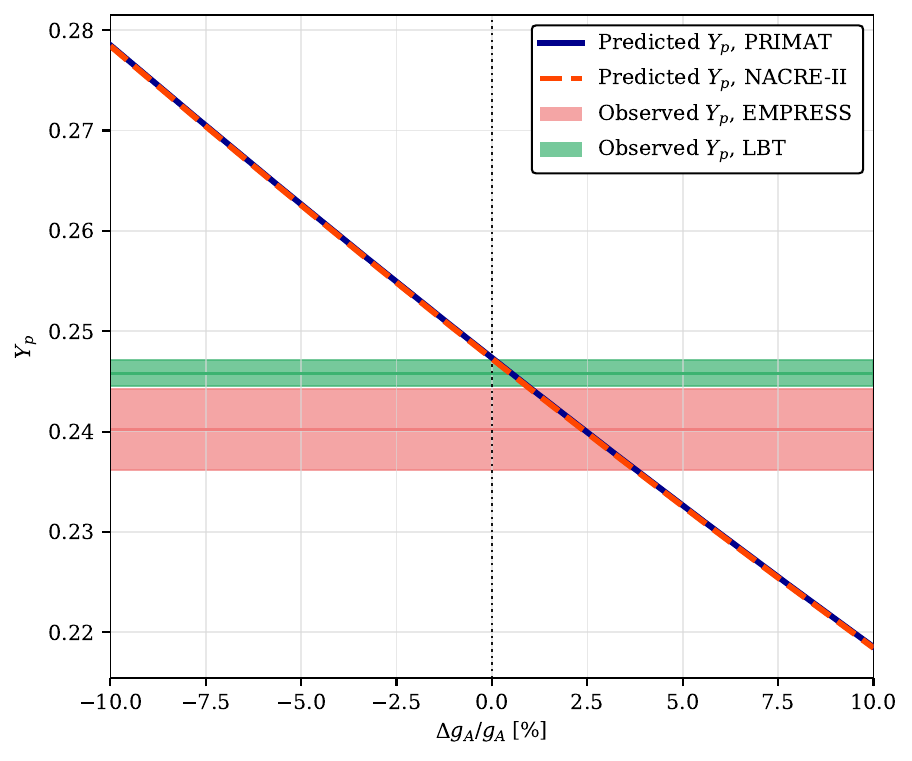}
        \caption{Predicted $Y_p$ as a function of $g_A$/$g_A^{\rm SM}$ calculated using PRIMAT (blue) and NACRE-II (red) reaction rates.}
        \label{fig:gA_Yp}
    \end{minipage}\hfill
    \begin{minipage}{0.49\linewidth}
        \centering
        \includegraphics[width=\linewidth]{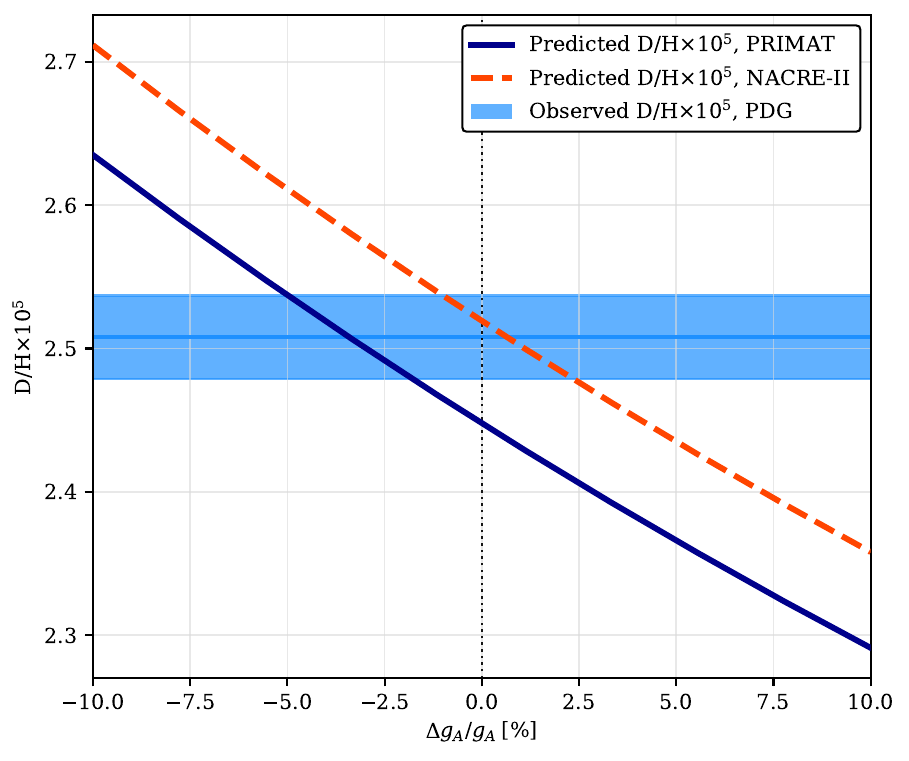}
        \caption{Predicted $D/H \times 10^5$ as a function of $g_A$/$g_A^{\rm SM}$ calculated using PRIMAT (blue) and NACRE-II (red) reaction rates.}
        \label{fig:gA_D}
    \end{minipage}

    \vspace{0.6em} 

    \begin{minipage}{0.49\linewidth}
        \centering
        \includegraphics[width=\linewidth]{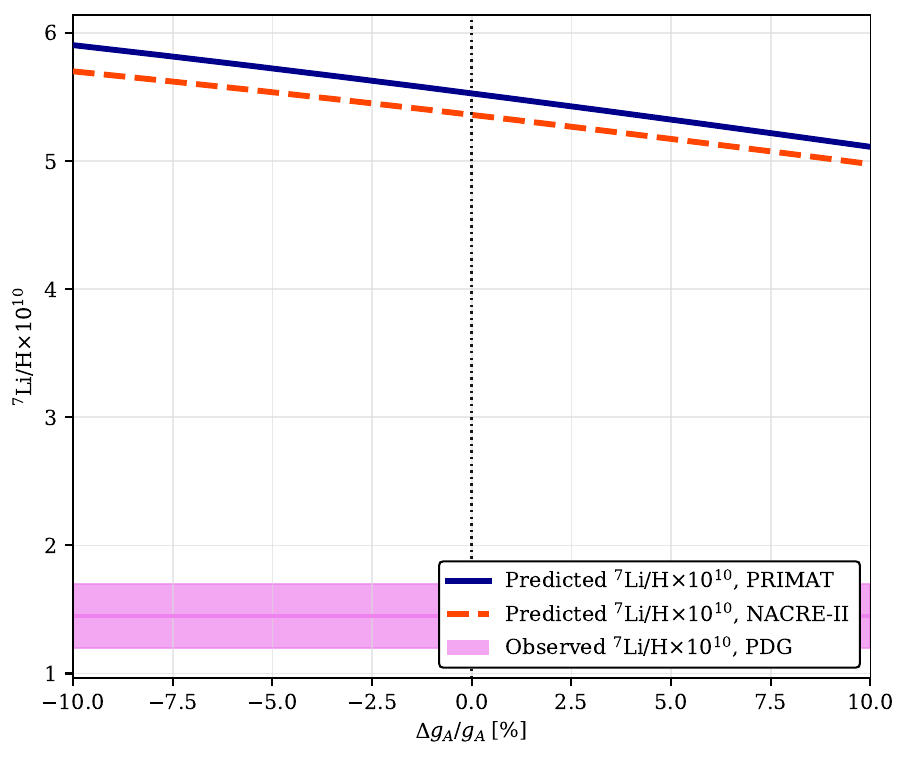}
        \caption{Predicted $^7Li/H \times 10^{10}$as a function of $g_A$/$g_A^{\rm SM}$ calculated using PRIMAT (blue) and NACRE-II (red) reaction rates.}
        \label{fig:gA_Li7}
    \end{minipage} 
\end{figure}

$g_A$ is thus a parameter whose sensitivity is almost entirely controlled by the choice of normalization scheme: under the fundamental parameterization it is among the more influential parameters for $Y_p$, $D/H$, and $^7Li/H$ through its dependence in the weak rate normalization, while under the neutron lifetime parameterization its effect is negligible. The changes in helium-4 deuterium, and lithium-7 are shown in Figures \ref{fig:gA_Yp}, \ref{fig:gA_D}, \ref{fig:gA_Li7}.

\subsection{CKM matrix element, $V_{ud}$}

\begin{figure}[t]
    \centering
    \begin{minipage}{0.49\linewidth}
        \centering
        \includegraphics[width=\linewidth]{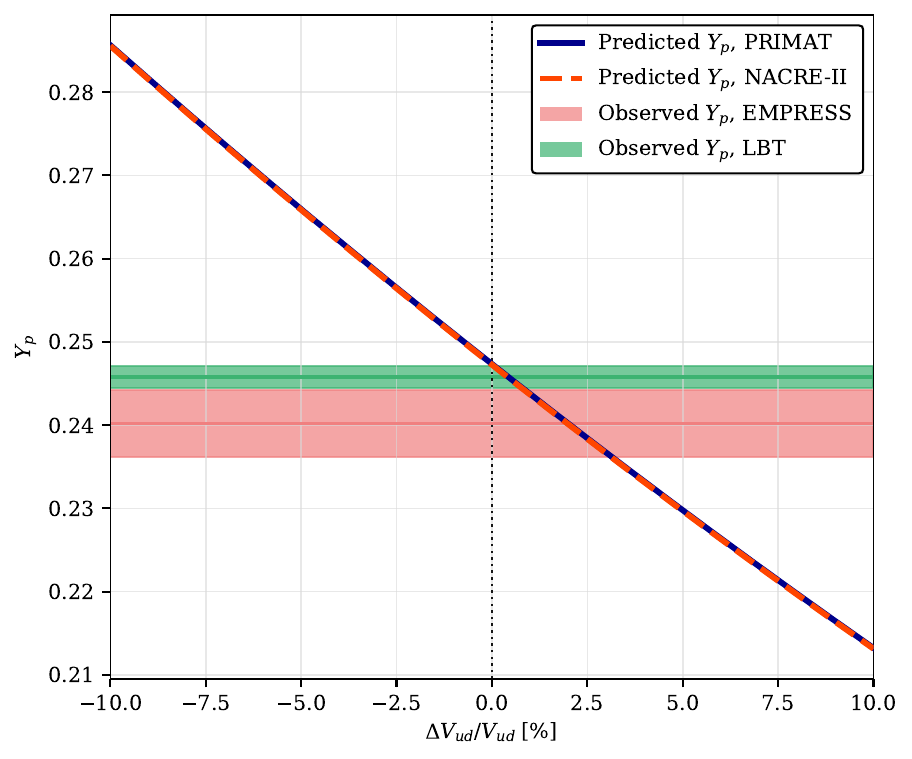}
        \caption{Predicted $Y_p$ as a function of $V_{ud}$/$V_{ud}^{\rm SM}$ calculated using PRIMAT (blue) and NACRE-II (red) reaction rates.}
        \label{fig:Vud_Yp}
    \end{minipage}\hfill
    \begin{minipage}{0.49\linewidth}
        \centering
        \includegraphics[width=\linewidth]{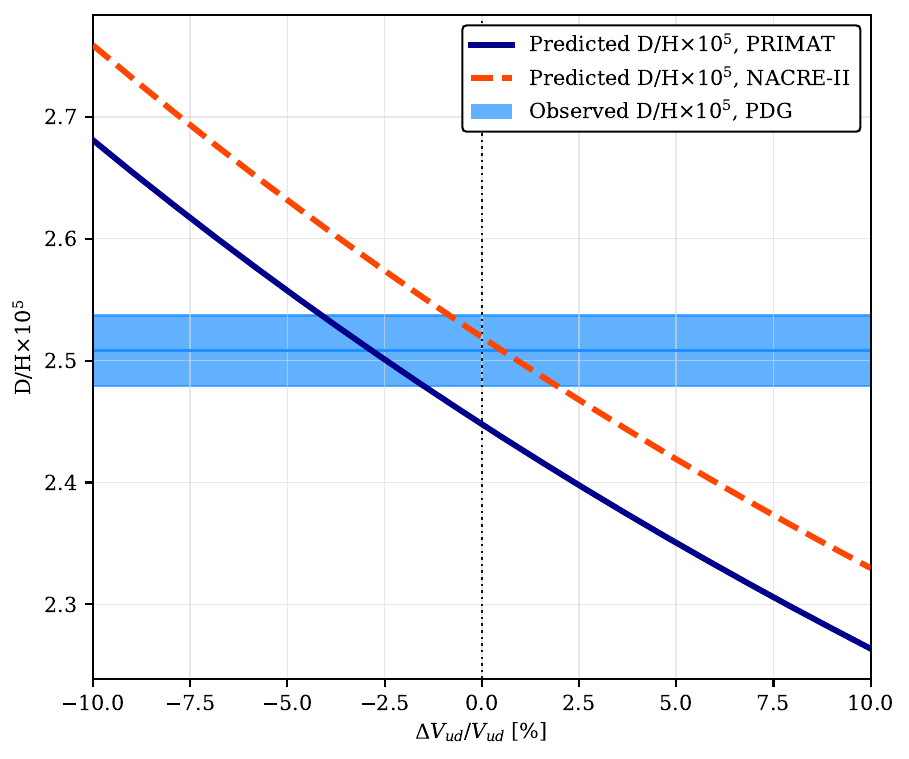}
        \caption{Predicted $D/H \times 10^5$ as a function of $V_{ud}$/$V_{ud}^{\rm SM}$ calculated using PRIMAT (blue) and NACRE-II (red) reaction rates.}
        \label{fig:Vud_D}
    \end{minipage}
\end{figure}

The up-down element of the Cabibbo–Kobayashi–Maskawa (CKM) matrix, $V_{ud}$, enters the calculation exclusively in the normalization of the weak rates when the fundamental parameters are used for the normalization as shown in Equation \ref{eq:weakratenorm}. Increasing $V_{ud}$ increases the normalization factor of the weak rates which scales like $V_{ud}^2$. The changes in helium-4 and deuterium are shown in Figures \ref{fig:Vud_Yp} and \ref{fig:Vud_D}. Changing $V_{ud}$ has no impact on the final value of $N_{\rm{eff}}$. 


\subsection{Weak magnetism constant, $\Delta \kappa$}

The weak magnetism constant, $\Delta \kappa$, is exclusively used to calculate the finite nucleon mass effect correction to the Born rates and has a negligible impact on the final abundance values and no impact on the final $N_{\rm{eff}}$ value. The resulting plots can be seen on GitHub \faGithub \href{https://github.com/Anne-KatherineBurns/bbn-sensitivity-atlas}{\,\texttt{bbn-sensitivity-atlas}}.

\subsection{Proton radius, $r_{p}$}

The proton radius, $r_{p}$, is exclusively used to calculate corrections to the Born rates and has a negligible impact on the final abundance values and no impact on the final $N_{\rm{eff}}$ value. These corrections include corrections for Bremsstrahlung, real photon emission, finite temperature radiative corrections, and finite nucleon mass effects. The resulting plots can be seen on GitHub \faGithub \href{https://github.com/Anne-KatherineBurns/bbn-sensitivity-atlas}{\,\texttt{bbn-sensitivity-atlas}}.

\subsection{Z boson mass, $m_{Z}$}

The mass of the Z boson, $m_{Z}$ is exclusively used to calculate the interaction between electrons, positrons and neutrinos. Here, it enters indirectly through the tree-level determination of sin$^2 \theta_W$, which fixes the neutral-current couplings appearing in the prefactor of the collision terms of the Boltzmann equations for the neutrinos. Due to the fact that sin$^2 \theta_W$ must be real, the smallest allowed value of $m_Z$ that we test is 74.56 GeV, which corresponds to about a 18.2\% decrease as compared to the SM value. Changing $m_{Z}$ has a very small impact on the final abundance values and $N_{\rm{eff}}$. Further discussion on the way in which modifying the interaction between electrons, positrons and neutrinos impacts the final value of $N_{\rm{eff}}$ is discussed in section \ref{sec:alphaem}. The resulting plots can be seen on GitHub \faGithub \href{https://github.com/Anne-KatherineBurns/bbn-sensitivity-atlas}{\,\texttt{bbn-sensitivity-atlas}}.

\subsection{Gravitational constant, $G_{N}$}
\label{sec:GN}

Varying the gravitational constant, $G_N$, changes the Hubble expansion rate, $H(a)\propto \sqrt{G_N\rho_{\rm tot}}$, thereby changing the time-temperature relation and the amount of time during which weak freeze-out and nuclear processing occurs.

In principle, $G_N$ also enters the conversion between the CMB baryon-density parameter $\Omega_b h^2$ and the baryon-to-photon ratio today, $\eta_{0b}$, through the critical density. Since $G_N$ today is constrained to high precision by laboratory and Solar System tests, and BSM scenarios involving varying gravity typically modify $G_N$ only in the early universe, we hold $G_N$ fixed at its SM value in this conversion. 

Increasing $G_N$ increases the expansion rate at fixed energy density. This drives earlier weak freeze-out, corresponding to a larger $n/p$ at freeze-out, and leaves less time for nuclear reactions to proceed before the temperature becomes too low for them to occur. As a result, the final ${}^4{\rm He}$ mass fraction $Y_p$ and the deuterium abundance D/H are increased due to the higher $n/p$ at the start of BBN. In contrast, the final lithium-7 abundance decreases due to suppression of late-time production, largely via Beryllium-7.

\begin{figure}[t]
    \centering
    \begin{minipage}{0.49\linewidth}
        \centering
        \includegraphics[width=\linewidth]{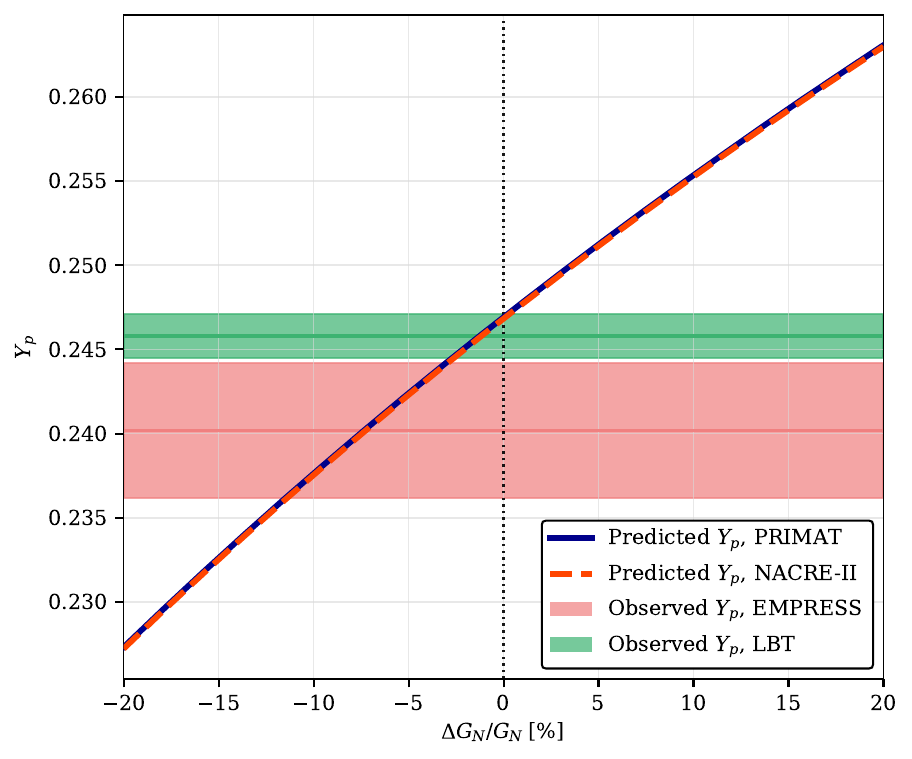}
        \caption{Predicted $Y_p$ as a function of $G_N$/$G_N^{\rm SM}$ calculated using PRIMAT (blue) and NACRE-II (red) reaction rates.}
        \label{fig:GN_Yp}
    \end{minipage}\hfill
    \begin{minipage}{0.49\linewidth}
        \centering
        \includegraphics[width=\linewidth]{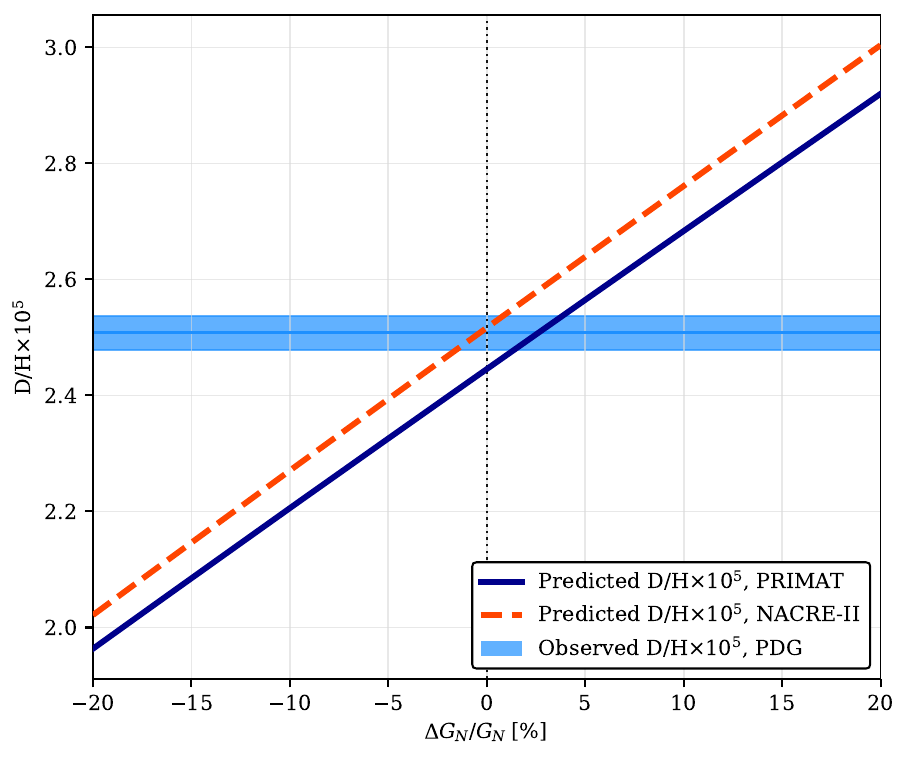}
        \caption{Predicted $D/H \times 10^5$as a function of $G_N$/$G_N^{\rm SM}$ calculated using PRIMAT (blue) and NACRE-II (red) reaction rates.}
        \label{fig:GN_D}
    \end{minipage}

    \vspace{0.6em} 

    \begin{minipage}{0.49\linewidth}
        \centering
        \includegraphics[width=\linewidth]{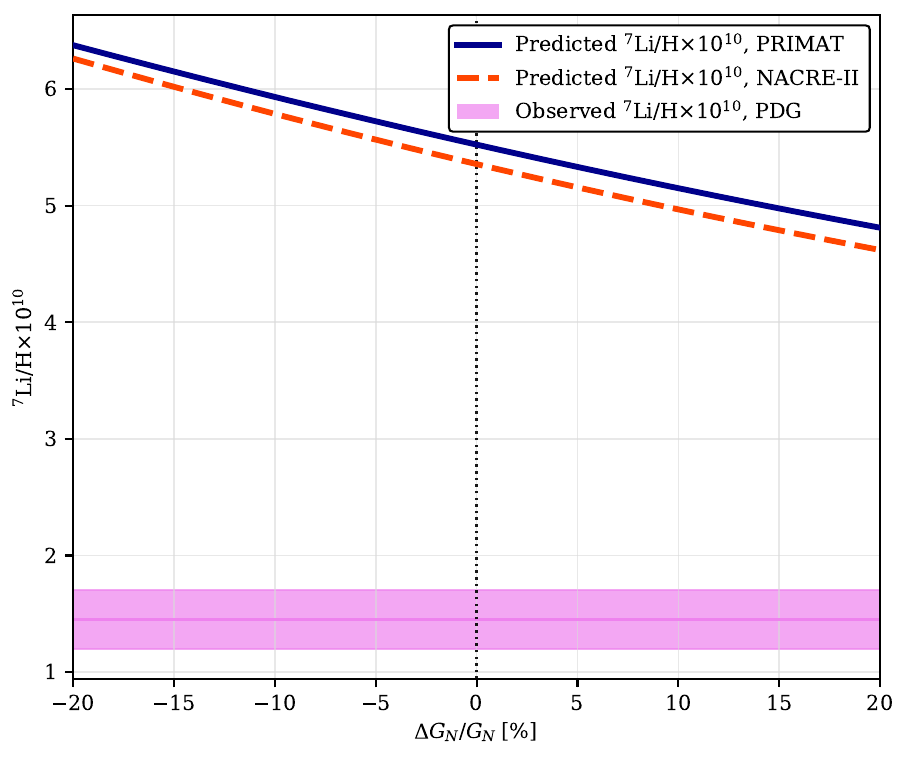}
        \caption{Predicted $^7Li/H \times 10^{10}$as a function of $G_N$/$G_N^{\rm SM}$ calculated using PRIMAT (blue) and NACRE-II (red) reaction rates.}
        \label{fig:GN_Li7}
    \end{minipage}
\end{figure}

Figures~\ref{fig:GN_Yp}--\ref{fig:GN_Li7} show the resulting changes in the primordial abundances of ${}^4{\rm He}$, D, and ${}^7{\rm Li}$. The final value of $N_{\rm eff}$ shifts very slightly with $G_N$ due to the fact that the modified expansion rate slightly changes the timing of non-instantaneous neutrino decoupling relative to $e^+e^-$ annihilation.

\subsection{Baryon abundance, $\Omega_b h^2$}

When the baryon abundance, $\Omega_b h^2$ is held fixed, the code computes the baryon to photon ratio today, $\eta_{0b}$ using the following equation,

\begin{equation}
\eta_{0b}
= \Omega_b h^2\,\frac{\rho_c/h^2}{n_{\gamma,0}\,m_B}
\label{eq:etab0}
\end{equation}

Here, $\rho_c$ is the critical energy density today, $n_{\gamma,0}$ is the present-day photon number density, and $m_B$ is the mean mass per baryon. As shown in the equation, $\eta_{0b}$ is proportional to the baryon abundance, $\Omega_b h^2$, thus increasing $\Omega_b h^2$ increases the inferred baryon density.

When $\Omega_b h^2$ is increased the final value of the helium-4 abundance will increase due to the fact that higher baryon density advances the onset of nuclear reactions and increases their efficiency during which time fewer neutrons decay than in the standard scenario, so a slightly larger fraction of neutrons gets locked into helium-4. For deuterium, increasing the baryon energy density causes a decrease in the final abundance deuterium as it gets burned into helium-4 more efficiently, leaving less residual deuterium. Increasing the baryon energy density causes an increase in the final lithium-7 abundance due to increase in the production of Beryllium-7.

Although $\Omega_b h^2$ is precisely determined by Planck \cite{Planck:2018vyg}, the extreme sensitivity of $D/H$ causes it to be largest contributor to the deuterium uncertainty budget when the PRIMAT nuclear reaction rates are used, as shown in Section \ref{sec:unc}. Further tightening the CMB baryon density constraint would directly sharpen the BBN deuterium prediction.

\begin{figure}[t]
    \centering
    \begin{minipage}{0.49\linewidth}
        \centering
        \includegraphics[width=\linewidth]{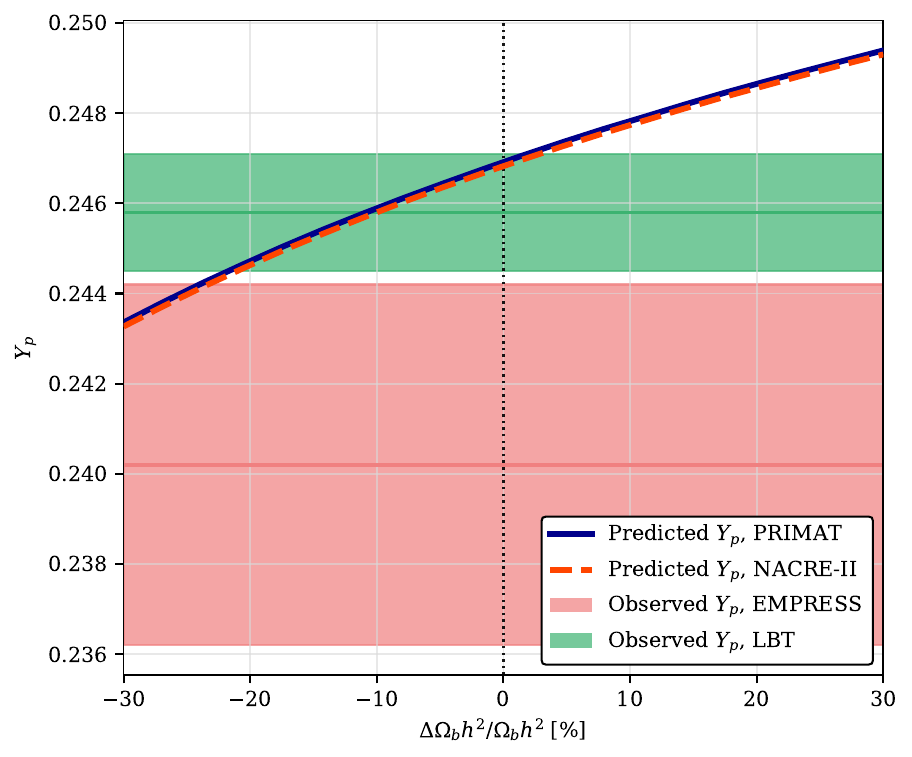}
        \caption{Predicted $Y_p$ as a function of $\Omega_b h^2$/$\Omega_b h^{2}{}^{\rm ,SM}$ calculated using PRIMAT (blue) and NACRE-II (red) reaction rates.}
        \label{fig:O_Yp}
    \end{minipage}\hfill
    \begin{minipage}{0.49\linewidth}
        \centering
        \includegraphics[width=\linewidth]{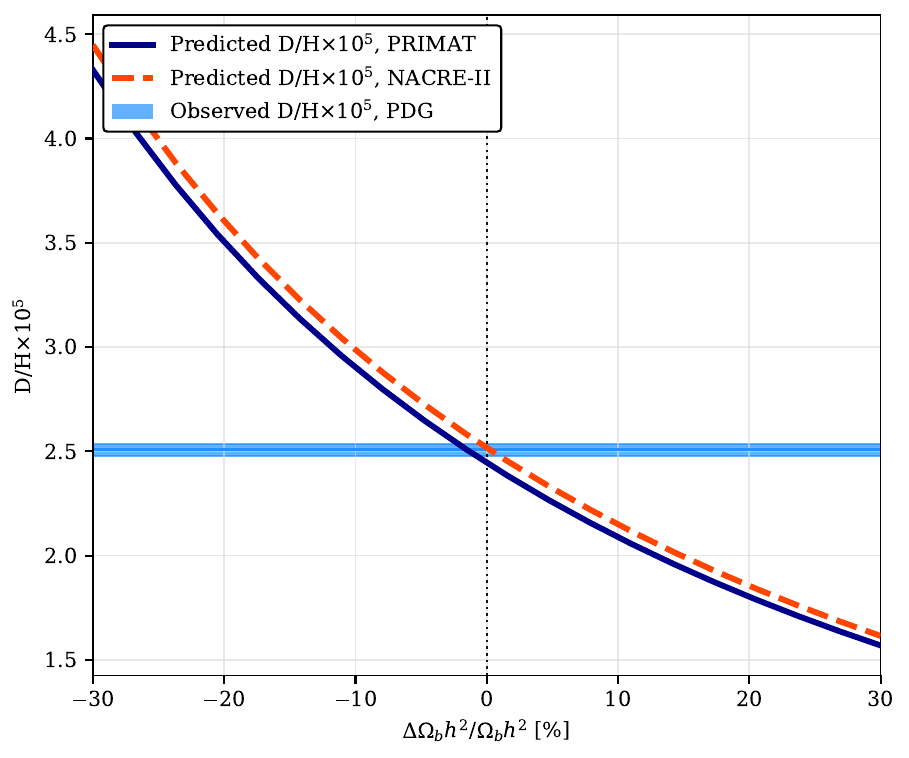}
        \caption{Predicted $D/H \times 10^5$ as a function of $\Omega_b h^2$/$\Omega_b h^{2}{}^{\rm ,SM}$ calculated using PRIMAT (blue) and NACRE-II (red) reaction rates.}
        \label{fig:O_D}
    \end{minipage}

    \vspace{0.6em} 

    \begin{minipage}{0.49\linewidth}
        \centering
        \includegraphics[width=\linewidth]{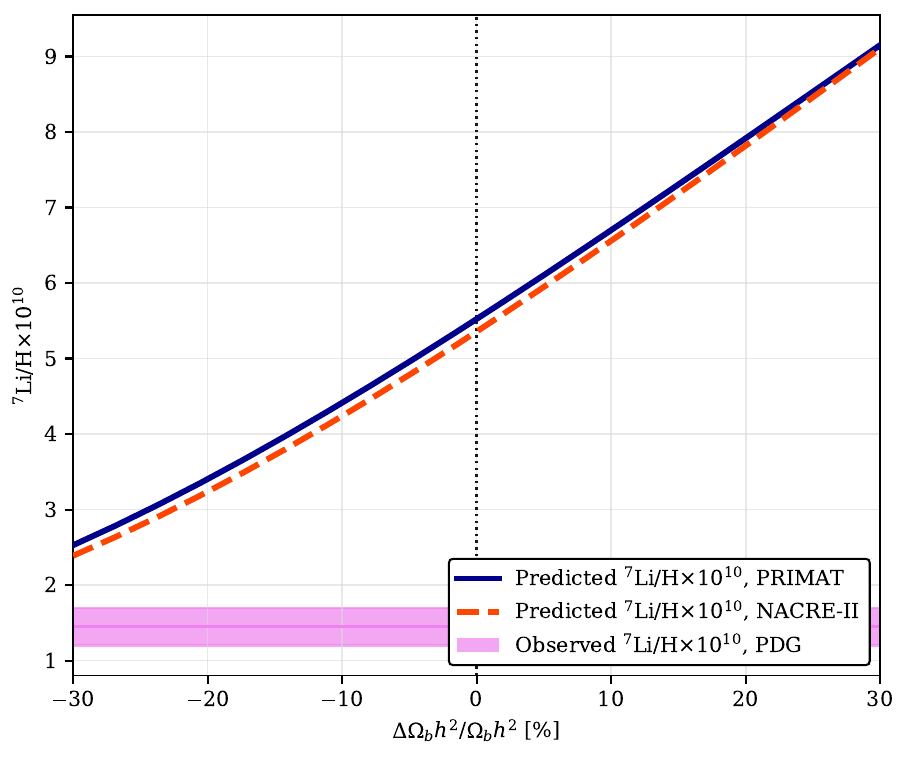}
        \caption{Predicted $^7Li/H \times 10^{10}$ as a function of $\Omega_b h^2$/$\Omega_b h^{2}{}^{\rm ,SM}$ calculated using PRIMAT (blue) and NACRE-II (red) reaction rates.}
        \label{fig:O_Li7}
    \end{minipage}
\end{figure}

$N_{\rm{eff}}$ is not affected by changes in $\Omega_b h^2$. The resulting plots are shown in Figures \ref{fig:O_Yp}, \ref{fig:O_D}, and \ref{fig:O_Li7}.

\subsection{Neutrino degeneracy parameter, $\xi_\nu$}

The neutrino degeneracy parameter, $\xi_\nu$ = $\mu_\nu / T_{\nu}$, which is constrained to be less than 0.71 for all neutrino flavors \cite{Burns:2022hkq}, plays two roles in the calculation: in the determinations of the weak rates and in the computation of the energy density of neutrinos. Most significantly, changing the $\mu_{\nu_e}$ changes the Born Rates, shown in Equation \ref{eq:GammaBorn}. Only the electron neutrino flavor enters these charged-current weak processes directly, however, due to neutrino oscillations, the degeneracy parameters for all three flavors are expected to equilibrate before the onset of BBN \cite{Burns:2022hkq, Escudero:2022okz}, thus we take $\xi_{\nu_e} = \xi_{\nu_\mu} = \xi_{\nu_\tau}$. In addition, in our calculation, neutrinos are assumed to be perfectly Fermi-Dirac and thus, the electron neutrino distribution functions, $f_{\nu_e}(E)$ are given by, 

\begin{equation}
    f_{\nu_e}(E) = \frac{1}{e^{(E - \mu_{\nu_e})/T_{\nu}} + 1}.
\end{equation}

As discussed further in~\cite{Burns:2022hkq}, increasing the neutrino degeneracy parameter reduces the neutron to proton ratio shown in Equation \ref{eq:np_ratio}. This in turn causes all of the final values of the primordial element abundances to decrease, and the helium-4 abundance is impacted especially strongly. $\mu_\nu$ is also used in the calculation of the finite-temperature radiative corrections. This has a secondary effect on the final element abundances is small as compared to the modification of the Born Rates. 

In addition, increasing $\mu_\nu$ increases the neutrino energy density, $\rho_\nu$ in the following way,

\begin{equation}
    \rho_{\nu}(T_{\nu}) = \left(1+\frac{\Delta N_{\rm eff}}{3}\right) \frac{7}{8} \frac{\pi^2}{15} T_{\nu}^4 \left[ 1 + \frac{30}{7} \left(\frac{\xi}{\pi}\right)^2  + \frac{15}{7}\left(\frac{\xi}{\pi}\right)^4 \right]. 
    \label{eq:rho_nu}
\end{equation}

Increasing $\rho_\nu$ increases Hubble and also impacts the calculation of the neutrino temperature evolution. Increasing the expansion rate of the universe has the effect of increasing the final abundances of helium-4 and deuterium and decreasing the final abundance of lithium-7. This effect is discussed in further detail in Section \ref{sec:GN} and is a secondary effect compared to the effect of changing the weak rates.

These two effects compete: the weak-rate modification reduces n/p and thus all abundances, while the increased expansion rate acts to raise $Y_p$ and $D/H$. The weak-rate effect dominates, so the net result is a decrease in all abundances with increasing $\mu_\nu / T_\nu$. A comprehensive analysis of current BBN and CMB constraints on the lepton asymmetry, including the implications of the EMPRESS helium measurement, can be found in \cite{Escudero:2022okz}.

\begin{figure}[t]
    \centering
    \begin{minipage}{0.49\linewidth}
        \centering
        \includegraphics[width=\linewidth]{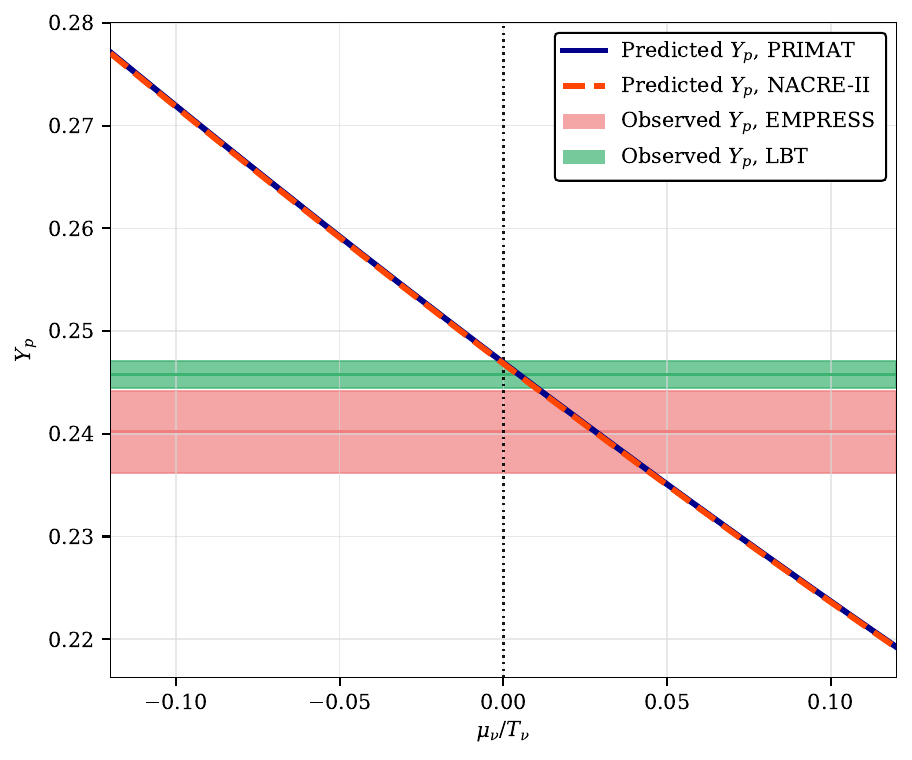}
        \caption{Predicted $Y_p$ as a function of $\mu_\nu / T_\nu$ calculated using PRIMAT (blue) and NACRE-II (red) reaction rates.}
        \label{fig:munu_Yp}
    \end{minipage}\hfill
    \begin{minipage}{0.49\linewidth}
        \centering
        \includegraphics[width=\linewidth]{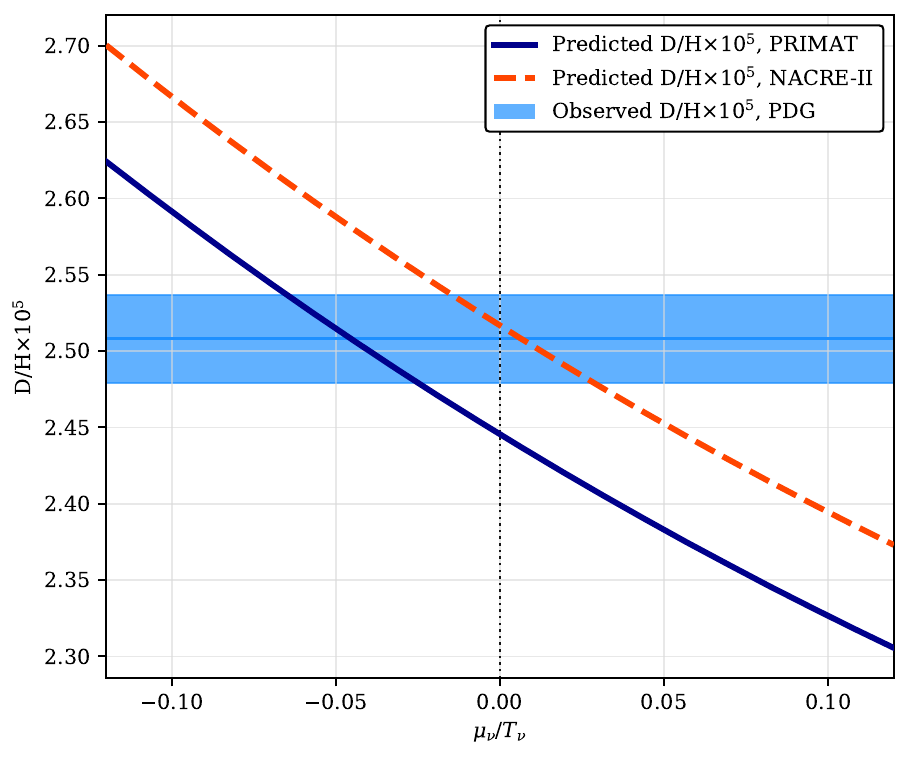}
        \caption{Predicted $D/H \times 10^5$  as a function of $\mu_\nu / T_\nu$ calculated using PRIMAT (blue) and NACRE-II (red) reaction rates.}
        \label{fig:munu_D}
    \end{minipage}
\end{figure}

The changes in helium-4 and deuterium are shown in Figures \ref{fig:munu_Yp} and \ref{fig:munu_D}. 

\subsection{Effective number of relativistic neutrino species, $N_{\rm{eff}}$}

\begin{figure}[t]
    \centering
    \begin{minipage}{0.49\linewidth}
        \centering
        \includegraphics[width=\linewidth]{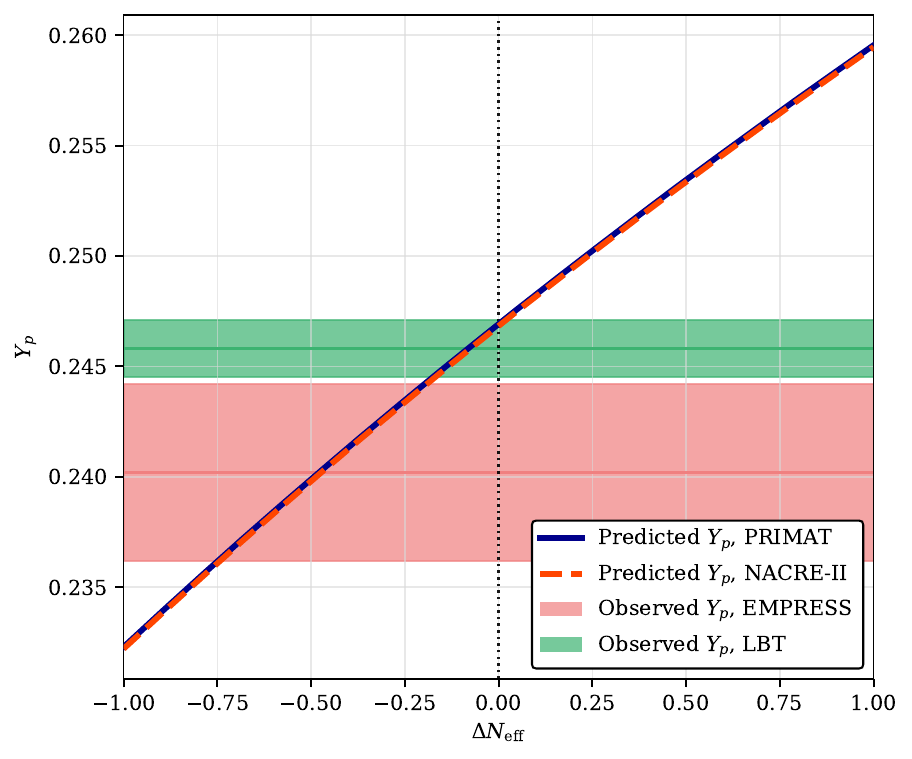}
        \caption{Predicted $Y_p$ as a function of $\Delta N_{\rm eff}$ calculated using PRIMAT (blue) and NACRE-II (red) reaction rates.}
        \label{fig:Neff_Yp}
    \end{minipage}\hfill
    \begin{minipage}{0.49\linewidth}
        \centering
        \includegraphics[width=\linewidth]{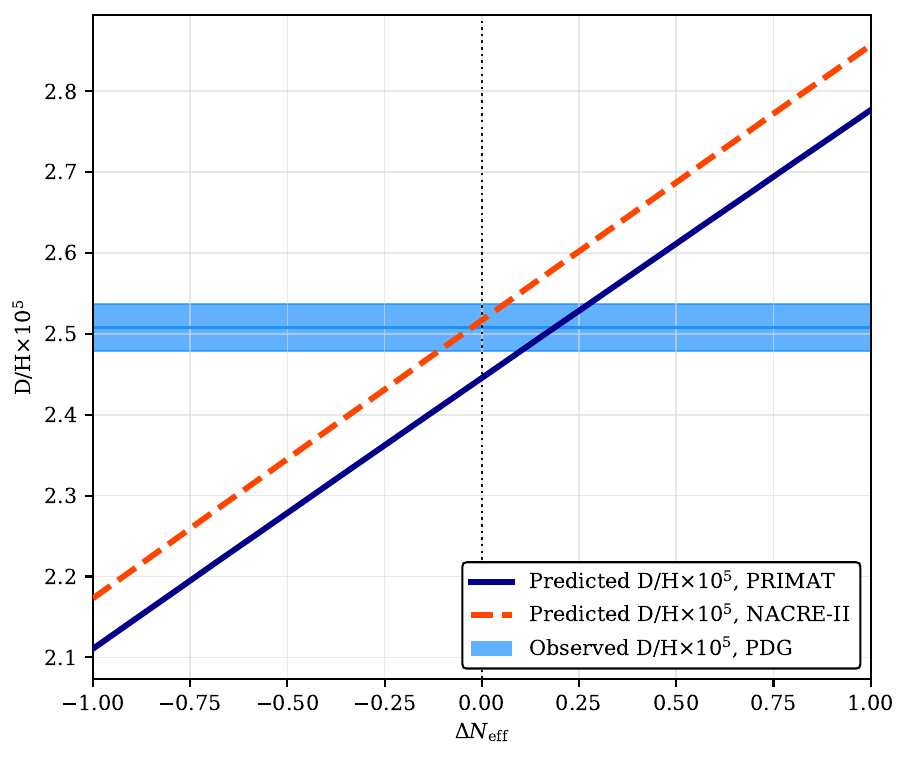}
        \caption{Predicted $D/H \times 10^5$ as a function of $\Delta N_{\rm eff}$ calculated using PRIMAT (blue) and NACRE-II (red) reaction rates.}
        \label{fig:Neff_D}
    \end{minipage}
\end{figure}

Implementing a non-zero $\Delta N_{\rm eff}$ directly modifies the neutrino energy density as shown in Equation \ref{eq:rho_nu}. Increasing $\Delta N_{\rm eff}$ has the effect of increasing the expansion rate of the universe, the impact of which is discussed in Section \ref{sec:GN}. In short, increasing the expansion rate increases the final abundance values of helium-4 and deuterium, as shown in Figures \ref{fig:Neff_Yp} and \ref{fig:Neff_D} and decreases the final abundance of lithium-7.

The combined CMB+BAO+BBN value of $N_{\rm eff} = 2.990 \pm 0.070$ \cite{Goldstein:2026iuu} corresponds to $\Delta N_{\rm eff} \approx -0.054$, which would lower the predicted $Y_p$ by approximately 0.001, well within the current LBT observational uncertainty \cite{Aver:2026dxv}. Moreover, the 95\% C.L.\ upper bound of $\Delta N_{\rm eff} < 0.107$ \cite{Goldstein:2026iuu} directly limits the room for additional light relics that could raise $Y_p$ or alter D/H, illustrating the power of combining precise $Y_p$ measurements with CMB and BAO data to jointly constrain light relic scenarios.

\section{Sensitivity Analysis: Thermonuclear Reaction Rates}
\label{sec:p_var_nuc}

While we will leave detailed study of each thermonuclear reaction rate to future work, we summarize here the main trends in the sensitivities of the final abundance values to variations in the nuclear rates. 

The final value of the helium-4 abundance is extremely insensitive to the thermonuclear reaction rates. This is due to the fact that nearly all available neutrons are processed into helium-4 regardless of the exact values of the reaction rates, so the final abundance is primarily set by the neutron-to-proton ratio at the onset of nucleosynthesis rather than by the details of the nuclear network. It is most sensitive to the reaction $d(d,n)^3He$, for which the slope of the sensitivity curve, d$Y_p$/dp, is equal to $\sim 1.6 \times 10^{-5}$ for the PRIMAT \cite{Iliadis16} rate and $\sim 7.9 \times 10^{-5}$ for the NACRE-II rate \cite{Xu:2013fha}.

In contrast, the final value of the deuterium abundance is much more sensitive to the values of the thermonuclear reaction rates, in particular to $d(p,\gamma)^3He$ \cite{Mossa:2020gjc}, $d(d,n)^3He$ \cite{Iliadis16,Xu:2013fha}, and $d(d,p)t$ \cite{Iliadis16,Xu:2013fha}, the three reactions targeted in recent data-driven analyses of the primordial deuterium prediction~\cite{Launders:2026ciu}. These three reactions are the dominant reactions responsible for the destruction of deuterium during BBN, so changing their rates changes the efficiency of deuterium burning. Reactions such as $t(d,n)^4He$ \cite{Descouvemont04, Xu:2013fha} also destroy deuterium but rely on the presence of heavier elements. 

\begin{figure}[t]
    \centering
    \includegraphics[width=1.0\linewidth]{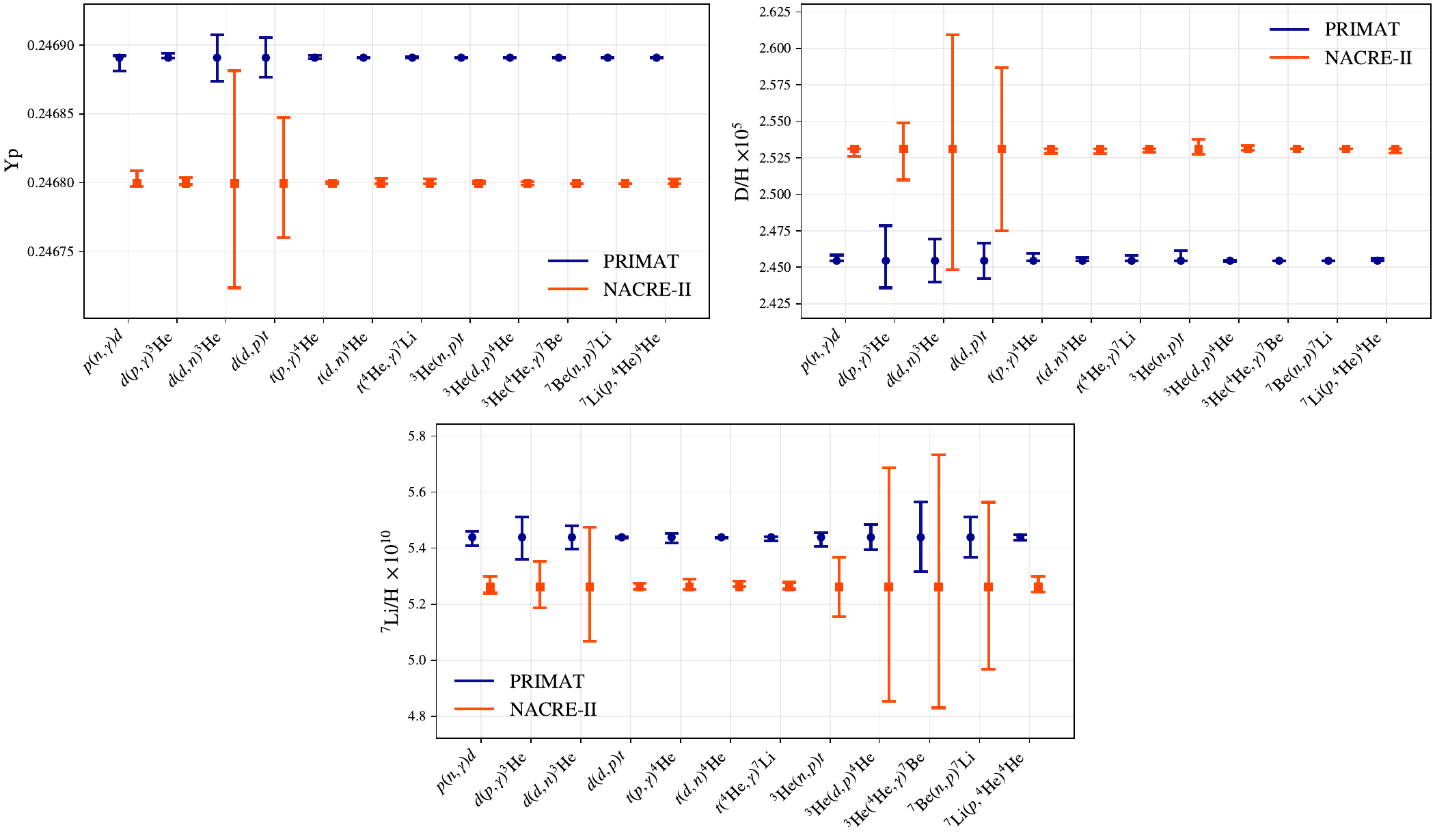}
    \caption{Sensitivity of standard BBN abundance predictions to $\pm 1\sigma$ variations of individual nuclear reaction rates. For each reaction, the vertical bars show the range of predicted $Y_p$, $\mathrm{D/H}\times 10^5$, and ${}^7\mathrm{Li}/\mathrm{H}\times 10^{10}$ obtained by setting that rate to its lower and upper $1\sigma$ values while keeping all other rates fixed; markers indicate the median prediction. Blue circles correspond to the PRIMAT rate compilation, and orange squares to NACRE-II.}
    \label{fig:RxnRates}
\end{figure}

Among the three reactions to which the deuterium abundance is most sensitive, the dominant source of the PRIMAT--NACRE-II discrepancy in D/H is the radiative capture reaction $d(p,\gamma)^{3}\mathrm{He}$. A direct comparison of the tabulated rates in the BBN-relevant temperature range, detailed in Table~\ref{table:rate_comparison}, shows that the NACRE-II rate for $d(p,\gamma)^{3}\mathrm{He}$ deviates from the PRIMAT rate by up to $\sim 30\%$ with a mean absolute deviation of $\sim 15\%$, while the two compilations' $1\sigma$ uncertainty bands fail to overlap over much of this range. By contrast, the $d(d,n)^{3}\mathrm{He}$ and $d(d,p)t$ rates agree at the $\sim 5\%$ level, with maximum deviations of $5.0\%$ and $4.6\%$, respectively, with overlapping uncertainty bands. Given the local sensitivity $\partial Y_{\rm D/H}/\partial p \simeq -0.02$ for this reaction (see Table~\ref{tab:sens_tau_n_rates_first12_side_by_side}), where $p$ is measured in units of the quoted $1\sigma$ NACRE-II rate uncertainty, the $\sim 20$--$30\%$ central-value rate offset between the two compilations produces a D/H shift of order a few percent, more than sufficient to account for the $\sim 2.9\%$ offset between the PRIMAT and NACRE-II central predictions for D/H visible in Fig.~\ref{fig:RxnRates}. The smaller rate differences in the $d+d$ channels contribute at the sub-percent to percent level to D/H and cannot account for the bulk of the offset. We therefore conclude that the tension between the PRIMAT-based SM prediction for D/H and its observed value, and conversely the agreement found when NACRE-II rates are employed, can be traced primarily to the treatment of the $d(p,\gamma)^{3}\mathrm{He}$ cross section in the two databases. A recent data-driven analysis using Gaussian process regression on the experimental cross-section data for $d(p,\gamma)^{3}\mathrm{He}$, $d(d,n)^{3}\mathrm{He}$, and $d(d,p)t$~\cite{Launders:2026ciu} obtains a D/H prediction in closer agreement with the PRIMAT-based result, lending support to the interpretation that the NACRE-II treatment of $d(p,\gamma)^{3}\mathrm{He}$ is the source of the offset.

\begin{table}[t]
\centering
\setlength{\heavyrulewidth}{0.08em}
\setlength{\lightrulewidth}{0.05em}
\setlength{\cmidrulewidth}{0.04em}
\setlength{\tabcolsep}{6pt}
\small
\begin{tabular}{l @{\hspace{1.5em}} || @{\hspace{1.5em}} c @{\hspace{1.0em}} c}
\textbf{Nuclear Reaction} & \textbf{max\,$|\Delta|$} & \textbf{mean\,$|\Delta|$} \\
\hline
$\mathrm{D}+p \to {}^{3}\mathrm{He}+\gamma$                        & $30.38\%$ & $15.06\%$ \\
${}^{7}\mathrm{Be}+n \to {}^{7}\mathrm{Li}+p$                      & $6.88\%$  & $4.74\%$  \\
${}^{3}\mathrm{He}+\mathrm{D} \to {}^{4}\mathrm{He}+p$             & $6.67\%$  & $5.42\%$  \\
${}^{3}\mathrm{He}+{}^{4}\mathrm{He} \to {}^{7}\mathrm{Be}+\gamma$ & $5.83\%$  & $4.73\%$  \\
${}^{7}\mathrm{Li}+p \to {}^{4}\mathrm{He}+{}^{4}\mathrm{He}$      & $5.56\%$  & $2.16\%$  \\
$\mathrm{D}+\mathrm{D} \to {}^{3}\mathrm{He}+n$                    & $5.05\%$  & $4.07\%$  \\
${}^{3}\mathrm{H}+{}^{4}\mathrm{He} \to {}^{7}\mathrm{Li}+\gamma$  & $4.79\%$  & $4.25\%$  \\
$\mathrm{D}+\mathrm{D} \to {}^{3}\mathrm{H}+p$                     & $4.62\%$  & $2.42\%$  \\
${}^{3}\mathrm{He}+n \to {}^{3}\mathrm{H}+p$                       & $2.59\%$  & $2.17\%$  \\
${}^{3}\mathrm{H}+\mathrm{D} \to {}^{4}\mathrm{He}+n$              & $1.42\%$  & $0.58\%$  \\
${}^{3}\mathrm{H}+p \to {}^{4}\mathrm{He}+\gamma$                  & $0.05\%$  & $0.01\%$  \\
$n+p \to \mathrm{D}+\gamma$                                        & $0.01\%$  & $0.00\%$  \\
\end{tabular}
\caption{Comparison of the NACRE-II and PRIMAT thermo-nuclear reaction rate compilations in the BBN-relevant temperature window $T \in [0.01, 0.1]\,\mathrm{MeV}$. For each of the 12 key reactions, we report the maximum absolute deviation from unity ($|\Delta| \equiv |R_{\mathrm{NACRE\text{-}II}}/R_{\mathrm{PRIMAT}} - 1|$) of the pointwise rate ratio, along with its mean across the window. Reactions are sorted in order of decreasing maximum deviation. The $d(p,\gamma){}^{3}\mathrm{He}$ rate shows by far the largest disagreement between the two compilations, consistent with it being the primary driver of the PRIMAT--NACRE-II discrepancy in the predicted D/H abundance.}
\label{table:rate_comparison}
\end{table}

The final value of the lithium-7 abundance is also sensitive to several of the nuclear reaction rates. The majority of lithium-7 that we observe today is created as Beryllium-7 during BBN and then is converted to lithium-7 via electron capture long after BBN \cite{Fields:2022mpw}. As a result, the final lithium-7 abundance is most sensitive to the following types of reactions: 

\begin{itemize}

\item Reactions producing Helium-3, which is the primary nuclei that later produces Beryllium-7 such as, $d(p,\gamma)^3He$ and $d(d,n)^3He$
\item Reactions destroying Helium-3 such as, ${}^3He(d,p)^4He$
\item Reactions producing either Beryllium-7 or lithium-7 such as, ${}^3He(^4He,\gamma)^7Be$ and ${}^7Be(n,p)^7Li$

\end{itemize}

Figure \ref{fig:RxnRates} is a summary plot of the sensitivity of all of the nuclear reaction rates, illustrating the discussion above for both the PRIMAT \cite{Pitrou:2018cgg} and NACRE-II \cite{Xu:2013fha} compilations. The qualitative trends are similar in the two compilations, but the sizes of the uncertainty spans differ for several reactions, reflecting differences in the adopted nuclear rate uncertainties.  In all cases, the rates are varied one at a time while holding the others fixed. Observational bands are not shown in this figure: the variation in $Y_p$ is so small that the observational $1\sigma$ band would span most of the plotted range, while the observed lithium-7 abundance lies outside the plotted $y$-axis range used here. Numerical sensitivity results are reported in Table \ref{tab:sens_tau_n_rates_first12_side_by_side}. 

These trends motivate prioritizing improved determinations of the deuterium destruction and mass-7 production and destruction rates in future studies, especially when propagating nuclear uncertainties in non-standard BBN scenarios.

\section{Tabulated Sensitivity Coefficients}
\label{sec:sens}

Table  \ref{tab:sens_taun_nacre_side_by_side}, \ref{tab:sens_fund_side_by_side}, \ref{tab:sens_taun_side_by_side_neff}, and \ref{tab:sens_tau_n_rates_first12_side_by_side}  collect the local linear sensitivity coefficients for all parameters and nuclear reaction rates studied in this work. For the fundamental physics parameters (Tables~\ref{tab:sens_taun_nacre_side_by_side} and  \ref{tab:sens_fund_side_by_side}), we report the logarithmic derivative $\frac{d\ln Y}{d\ln p}$ calculated at the fiducial value, which gives the fractional response of observable $Y$ to a fractional change in parameter $p$. In other words, a 1\% shift in $p$ changes $Y$ by approximately $\frac{d\ln Y}{d\ln p}$\%. For the non-standard parameters $\Delta N_{\rm eff}$ and $\mu_\nu / T_\nu$ (Table~\ref{tab:sens_taun_side_by_side_neff}), we instead report $\frac{dY}{dp}$ calculated at the fiducial value, as the fiducial values are zero. For the nuclear reaction rates (Table~\ref{tab:sens_tau_n_rates_first12_side_by_side}), we report $\frac{dY}{dp}$ calculated at the fiducial value where $p$ is the lognormal rate-shift parameter measured in units of the quoted $1\sigma$ uncertainty.

Table~\ref{tab:sens_taun_nacre_side_by_side} uses the $\tau_n$ weak rate normalization and includes all SM and $\Lambda$CDM parameters except $V_{ud}$. Table~\ref{tab:sens_fund_side_by_side} uses the fundamental constant weak rate normalization, and reports results for the subset of parameters whose sensitivities depend on the normalization scheme. In all tables, results are shown for both the PRIMAT and NACRE-II nuclear reaction rate compilations. The $R^2$ column reports the quality of the linear fit, with values significantly below unity indicating that the response is nonlinear over the scanned range. Here, 

\begin{equation}
R^2_{\rm lin} = 1 - \frac{\sum_i (y_i - \hat{y}_i)^2}{\sum_i (y_i - \bar{y})^2}\,,
\end{equation}

where $y_i$ are the numerically computed values of the observable at each scan point, $\hat{y}_i$ are the corresponding values from the best-fit linear model, and $\bar{y}$ is the mean of the $y_i$. For parameter--observable pairs exhibiting very small variation, the standard definition of $R^2$ can be misleading, since the total variance becomes very small. In those cases, we instead quote a modified goodness-of-fit statistic that measures consistency with a horizontal line at a fixed tolerance level. All other parameter--observable pairs use $R^2_{\rm lin}$. The physical interpretation of each parameter's sensitivity is discussed in Sections~\ref{sec:PVar} and~\ref{sec:p_var_nuc}.

Previous BBN sensitivity analyses have been published, but have either focused on nuclear rates and cosmological parameters \cite{Cyburt:2015mya}, or have studied fundamental constant variations within specific unification frameworks that correlate changes between parameters \cite{Clara:2020efx, Deal:2021kjs, 
Baryakhtar:2024rky, Baryakhtar:2025uxs}. Here we extend these approaches by computing the independent response of each observable to each input parameter individually, providing a model-independent atlas that can be applied to arbitrary BSM scenarios. As a concrete illustration of how to use our results within a specific model, consider the sensitivity of $Y_p$ to the fine-structure constant $\alpha_{EM}$. While our tables report the independent responses $\frac{d\ln Y_p}{d\ln \alpha_{EM}}$ and $\frac{d\ln Y_p}{d\ln Q}$ separately, the total effective sensitivity to $\alpha_{EM}$ including the electromagnetic contribution to the neutron-proton mass difference can be recovered by combining these two entries, weighted by $d\ln Q^{\rm QED}/d\ln \alpha_{EM}$. More generally, any model predicting correlations among our 14 parameters can construct its effective sensitivity coefficients as a linear combination of the entries in our atlas.

\begin{table}[t]
\centering
\vspace*{0pt}

\setlength{\heavyrulewidth}{0.08em}%
\setlength{\lightrulewidth}{0.05em}%
\setlength{\cmidrulewidth}{0.04em}%

\begin{minipage}[t]{0.49\textwidth}
\centering
\tiny   
{\footnotesize\bfseries PRIMAT}\par\vspace{0.5ex}
\vspace{2.0ex}
\setlength{\tabcolsep}{4pt}

\begin{tabular}{l@{\hspace{-15pt}}c l c c}

Parameter & $p_{\rm fid}$ & Observable & $\frac{d\ln Y}{d\ln p}$ & $R^2$ \\
\addlinespace[0.8ex]
\midrule
\addlinespace[0.8ex]
 
\multirow[t]{4}{*}{$\alpha_{EM}$} & \multirow[t]{4}{*}{0.00729735} & Y$_p$            & +0.013 & 0.9999 \\
                            &                                 & D/H              & +0.008 & 0.9856 \\
                            &                                 & $^7$Li/H         & +0.007 & 0.9782 \\
                            &                                 & N$_{\text{eff}}$ & +0.004 & 0.8363 \\
\addlinespace[0.8ex]\cmidrule(lr){1-5}\addlinespace[0.8ex]
 
\multirow[t]{4}{*}{$G_F$}      & \multirow[t]{4}{*}{1.16638e-11 MeV$^{-2}$} & Y$_p$            & +0.000 & 0.9991 \\
                            &                                 & D/H              & +0.003 & 0.9810 \\
                            &                                 & $^7$Li/H         & -0.004 & 0.9819 \\
                            &                                 & N$_{\text{eff}}$ & +0.010 & 0.9817 \\
\addlinespace[0.8ex]\cmidrule(lr){1-5}\addlinespace[0.8ex]
 
\multirow[t]{4}{*}{$m_e$}      & \multirow[t]{4}{*}{0.510999 MeV}    & Y$_p$            & -0.734 & 0.9991 \\
                            &                                 & D/H              & -0.150 & 0.7976 \\
                            &                                 & $^7$Li/H         & -0.849 & 0.9873 \\
                            &                                 & N$_{\text{eff}}$ & +0.023 & 0.9970 \\
\addlinespace[0.8ex]\cmidrule(lr){1-5}\addlinespace[0.8ex]
 
\multirow[t]{4}{*}{$\tau_n$}  & \multirow[t]{4}{*}{878.4 s}       & Y$_p$            & +0.731 & 0.9975 \\
                            &                                 & D/H              & +0.420 & 0.9997 \\
                            &                                 & $^7$Li/H         & +0.437 & 0.9884 \\
                            &                                 & N$_{\text{eff}}$ & +0.000 & 1.0000 \\
\addlinespace[0.8ex]\cmidrule(lr){1-5}\addlinespace[0.8ex]
 
\multirow[t]{4}{*}{$G_N$ }      & \multirow[t]{4}{*}{6.70883e-45 MeV$^{-2}$}  & Y$_p$            & +0.354 & 0.9973 \\
                            &                                 & D/H              & +0.976 & 1.0000 \\
                            &                                 & $^7$Li/H         & -0.702 & 0.9976 \\
                            &                                 & N$_{\text{eff}}$ & -0.004 & 0.9949 \\
\addlinespace[0.8ex]\cmidrule(lr){1-5}\addlinespace[0.8ex]
 
\multirow[t]{4}{*}{$g_A$}      & \multirow[t]{4}{*}{1.2753}      & Y$_p$            & +0.000 & 0.9758 \\
                            &                                 & D/H              & +0.000 & 0.9926 \\
                            &                                 & $^7$Li/H         & +0.000 & 0.9897 \\
                            &                                 & N$_{\text{eff}}$ & +0.000 & 1.0000 \\
 
\addlinespace[0.8ex]\cmidrule(lr){1-5}\addlinespace[0.8ex]
 
\multirow[t]{4}{*}{$m_{Z}$}      & \multirow[t]{4}{*}{91188.0 MeV}       & Y$_p$            & +0.000 & 0.9999 \\
                            &                                 & D/H              & -0.004 & 0.9846 \\
                            &                                 & $^7$Li/H         & +0.004 & 0.9823 \\
                            &                                 & N$_{\text{eff}}$ & -0.011 & 0.8648 \\
\addlinespace[0.8ex]\cmidrule(lr){1-5}\addlinespace[0.8ex]
 
\multirow[t]{4}{*}{$\Delta \kappa$} & \multirow[t]{4}{*}{3.70589}  & Y$_p$            & +0.000 & 0.9993 \\
                               &                              & D/H              & +0.000 & 0.9967 \\
                               &                              & $^7$Li/H         & +0.000 & 0.9990 \\
                               &                              & N$_{\text{eff}}$ & +0.000 & 1.0000 \\
\addlinespace[0.8ex]\cmidrule(lr){1-5}\addlinespace[0.8ex]
 
\multirow[t]{4}{*}{$r_p$}  & \multirow[t]{4}{*}{8.409e-14 cm} & Y$_p$            & +0.000 & 1.0000 \\
                               &                              & D/H              & +0.000 & 1.0000 \\
                               &                              & $^7$Li/H         & +0.000 & 1.0000 \\
                               &                              & N$_{\text{eff}}$ & +0.000 & 1.0000 \\
\addlinespace[0.8ex]\cmidrule(lr){1-5}\addlinespace[0.8ex]
 
\multirow[t]{4}{*}{Q}       & \multirow[t]{4}{*}{1.29333 MeV}     & Y$_p$            & +1.567 & 0.9976 \\
                            &                                 & D/H              & +0.863 & 0.9998 \\
                            &                                 & $^7$Li/H         & +1.012 & 0.9893 \\
                            &                                 & N$_{\text{eff}}$ & +0.000 & 1.0000 \\
\addlinespace[0.8ex]\cmidrule(lr){1-5}\addlinespace[0.8ex]
 
\multirow[t]{4}{*}{$\Omega_b h^2$} & \multirow[t]{4}{*}{0.02237}    & Y$_p$            & +0.039 & 0.9903 \\
                             &                                & D/H              & -1.639 & 0.9582 \\
                             &                                & $^7$Li/H         & +2.079 & 0.9973 \\
                             &                                & N$_{\text{eff}}$ & +0.000 & 1.0000 \\
 
\end{tabular}
 
\label{tab:sens_taun_primat_side_by_side}
 
\end{minipage}
\hfill
\begin{minipage}[t]{0.49\textwidth}
\centering
\tiny   
{\footnotesize\bfseries NACRE-II}\par\vspace{0.5ex}
\vspace{2.0ex}
\setlength{\tabcolsep}{4pt}

\begin{tabular}{l@{\hspace{-15pt}}c l c c}
 
Parameter & $p_{\rm fid}$ & Observable & $\frac{d\ln Y}{d\ln p}$ & $R^2$ \\
\addlinespace[0.8ex]
\midrule
\addlinespace[0.8ex]
 
\multirow[t]{4}{*}{$\alpha_{EM}$} & \multirow[t]{4}{*}{0.00729735} & Y$_p$            & +0.013 & 0.9999 \\
                            &                                 & D/H              & +0.008 & 0.9854 \\
                            &                                 & $^7$Li/H         & +0.006 & 0.9725 \\
                            &                                 & N$_{\text{eff}}$ & +0.004 & 0.8363 \\
\addlinespace[0.8ex]\cmidrule(lr){1-5}\addlinespace[0.8ex]
 
\multirow[t]{4}{*}{$G_F$}      & \multirow[t]{4}{*}{1.16638e-11 MeV$^{-2}$} & Y$_p$            & +0.000 & 0.9991 \\
                            &                                 & D/H              & +0.003 & 0.9813 \\
                            &                                 & $^7$Li/H         & -0.004 & 0.9819 \\
                            &                                 & N$_{\text{eff}}$ & +0.010 & 0.9817 \\
\addlinespace[0.8ex]\cmidrule(lr){1-5}\addlinespace[0.8ex]
 
\multirow[t]{4}{*}{$m_e$}      & \multirow[t]{4}{*}{0.510999 MeV}    & Y$_p$            & -0.718 & 0.9991 \\
                            &                                 & D/H              & -0.145 & 0.7849 \\
                            &                                 & $^7$Li/H         & -0.850 & 0.9856 \\
                            &                                 & N$_{\text{eff}}$ & +0.023 & 0.9970 \\
\addlinespace[0.8ex]\cmidrule(lr){1-5}\addlinespace[0.8ex]
 
\multirow[t]{4}{*}{$\tau_n$}  & \multirow[t]{4}{*}{878.4 s}       & Y$_p$            & +0.731 & 0.9975 \\
                            &                                 & D/H              & +0.420 & 0.9997 \\
                            &                                 & $^7$Li/H         & +0.411 & 0.9869 \\
                            &                                 & N$_{\text{eff}}$ & +0.000 & 1.0000 \\
\addlinespace[0.8ex]\cmidrule(lr){1-5}\addlinespace[0.8ex]
 
\multirow[t]{4}{*}{$G_N$}      & \multirow[t]{4}{*}{6.70883e-45 MeV$^{-2}$}  & Y$_p$            & +0.354 & 0.9973 \\
                            &                                 & D/H              & +0.974 & 1.0000 \\
                            &                                 & $^7$Li/H         & -0.756 & 0.9969 \\
                            &                                 & N$_{\text{eff}}$ & -0.004 & 0.9949 \\
\addlinespace[0.8ex]\cmidrule(lr){1-5}\addlinespace[0.8ex]
 
\multirow[t]{4}{*}{$g_A$}      & \multirow[t]{4}{*}{1.2753}      & Y$_p$            & +0.000 & 0.9759 \\
                            &                                 & D/H              & +0.000 & 0.9926 \\
                            &                                 & $^7$Li/H         & +0.000 & 0.9909 \\
                            &                                 & N$_{\text{eff}}$ & +0.000 & 1.0000 \\
\addlinespace[0.8ex]\cmidrule(lr){1-5}\addlinespace[0.8ex]
 
\multirow[t]{4}{*}{$m_{Z}$}      & \multirow[t]{4}{*}{91188.0 MeV}       & Y$_p$            & +0.000 & 0.9999 \\
                            &                                 & D/H              & -0.004 & 0.9849 \\
                            &                                 & $^7$Li/H         & +0.004 & 0.9804 \\
                            &                                 & N$_{\text{eff}}$ & -0.011 & 0.8648 \\
\addlinespace[0.8ex]\cmidrule(lr){1-5}\addlinespace[0.8ex]
 
\multirow[t]{4}{*}{$\Delta \kappa$} & \multirow[t]{4}{*}{3.70589}  & Y$_p$            & +0.000 & 0.9994 \\
                               &                              & D/H              & +0.000 & 0.9953 \\
                               &                              & $^7$Li/H         & +0.000 & 0.9959 \\
                               &                              & N$_{\text{eff}}$ & +0.000 & 1.0000 \\
\addlinespace[0.8ex]\cmidrule(lr){1-5}\addlinespace[0.8ex]
 
\multirow[t]{4}{*}{$r_p$}  & \multirow[t]{4}{*}{8.409e-14 cm} & Y$_p$            & +0.000 & 1.0000 \\
                               &                              & D/H              & +0.000 & 1.0000 \\
                               &                              & $^7$Li/H         & +0.000 & 1.0000 \\
                               &                              & N$_{\text{eff}}$ & +0.000 & 1.0000 \\
\addlinespace[0.8ex]\cmidrule(lr){1-5}\addlinespace[0.8ex]
 
\multirow[t]{4}{*}{Q}       & \multirow[t]{4}{*}{1.29333 MeV}     & Y$_p$            & +1.566 & 0.9976 \\
                            &                                 & D/H              & +0.862 & 0.9998 \\
                            &                                 & $^7$Li/H         & +0.959 & 0.9880 \\
                            &                                 & N$_{\text{eff}}$ & +0.000 & 1.0000 \\
\addlinespace[0.8ex]\cmidrule(lr){1-5}\addlinespace[0.8ex]
 
\multirow[t]{4}{*}{$\Omega_b h^2$} & \multirow[t]{4}{*}{0.02237}    & Y$_p$            & +0.039 & 0.9903 \\
                             &                                & D/H              & -1.638 & 0.9586 \\
                             &                                & $^7$Li/H         & +2.176 & 0.9960 \\
                             &                                & N$_{\text{eff}}$ & +0.000 & 1.0000 \\
 
\end{tabular}

\end{minipage}

\caption{Local linear sensitivities, $\frac{dlnY}{dlnp}$, and linear-fit quality, 
$R^2$, computed using weak rate normalization calculated from the neutron lifetime. All SM and $\Lambda$CDM parameters are included, except for $V_{ud}$ which is excluded due to the fact that the only role it plays in the calculation is in the weak rate normalization from fundamental parameters. The left table shows results using the PRIMAT nuclear reaction rates, and the right table shows the results using the NACRE-II nuclear reaction rates.}
\label{tab:sens_taun_nacre_side_by_side}
\end{table}

\begin{table}[t]
\centering
\vspace*{0pt}

\begin{minipage}[t]{0.49\textwidth}
\centering
\tiny   
{\footnotesize\bfseries PRIMAT}\par\vspace{0.5ex}
\vspace{2.0ex}
\setlength\tabcolsep{4pt}

\setlength{\heavyrulewidth}{0.08em}%
\setlength{\lightrulewidth}{0.05em}%
\setlength{\cmidrulewidth}{0.04em}%

\begin{tabular}{l@{\hspace{-12pt}}c l c c}
Parameter & $p_{\rm fid}$ & Observable & $\frac{d\ln Y}{d\ln p}$ & $R^2$ \\
\addlinespace[1ex]
\midrule
\addlinespace[1ex]
\multirow[t]{4}{*}{$G_F$}  & \multirow[t]{4}{*}{1.16638e-11 MeV$^{-2}$} & Y$_p$         & -1.452 & 0.9990 \\
                        &                                  & D/H           & -0.827 & 0.9959 \\
                        &                                  & $^7$Li/H      & -0.880 & 0.9992 \\
                        &                                  & N$_{\text{eff}}$ & +0.010 & 0.9987 \\
\addlinespace[1ex]
\cmidrule(lr){1-5}
\addlinespace[1ex]

\multirow[t]{4}{*}{$m_e$}  & \multirow[t]{4}{*}{0.510999 MeV}    & Y$_p$         & +0.394 & 0.9954 \\
                        &                                  & D/H           & +0.489 & 0.9948 \\
                        &                                  & $^7$Li/H      & -0.185 & 0.8256 \\
                        &                                  & N$_{\text{eff}}$ & +0.023 & 0.9970 \\
\addlinespace[1ex]
\cmidrule(lr){1-5}
\addlinespace[1ex]

\multirow[t]{4}{*}{$g_A$}  & \multirow[t]{4}{*}{1.2753}      & Y$_p$         & -1.214 & 0.9995 \\
                        &                                  & D/H           & -0.699 & 0.9975 \\
                        &                                  & $^7$Li/H      & -0.725 & 0.9992 \\
                        &                                  & N$_{\text{eff}}$ & +0.000 & 1.0000 \\
\addlinespace[1ex]
\cmidrule(lr){1-5}
\addlinespace[1ex]

\multirow[t]{4}{*}{$V_{ud}$} & \multirow[t]{4}{*}{0.97367}     & Y$_p$         & -1.460 & 0.9990 \\
                        &                                  & D/H           & -0.843 & 0.9960 \\
                        &                                  & $^7$Li/H      & -0.872 & 0.9992 \\
                        &                                  & N$_{\text{eff}}$ & +0.000 & 1.0000 \\

\addlinespace[1ex]
\cmidrule(lr){1-5}
\addlinespace[1ex]

\multirow[t]{4}{*}{$Q$} & \multirow[t]{4}{*}{1.29333 MeV}     & Y$_p$         & -3.198 & 1.0000 \\
                        &                                  & D/H           & -1.878 & 0.9941 \\
                        &                                  & $^7$Li/H      & -1.835 & 0.9820 \\
                        &                                  & N$_{\text{eff}}$ & +0.000 & 1.0000 \\

\addlinespace[1ex]
\cmidrule(lr){1-5}
\addlinespace[1ex]

\multirow[t]{4}{*}{$\Delta \kappa$} & \multirow[t]{4}{*}{3.70589}  & Y$_p$            & +0.000 & 0.9995 \\
                               &                              & D/H              & +0.000 & 0.9942 \\
                               &                              & $^7$Li/H         & +0.000 & 0.9985 \\
                               &                              & N$_{\text{eff}}$ & +0.000 & 1.0000 \\
\addlinespace[0.8ex]\cmidrule(lr){1-5}\addlinespace[0.8ex]
\multirow[t]{4}{*}{$\alpha_{EM}$} & \multirow[t]{4}{*}{0.00729735} & Y$_p$            & -0.009 & 1.0000 \\
                            &                                 & D/H              & -0.004 & 0.9621 \\
                            &                                 & $^7$Li/H         & -0.006 & 0.9708 \\
                            &                                 & N$_{\text{eff}}$ & +0.004 & 0.8363 \\
\addlinespace[0.8ex]\cmidrule(lr){1-5}\addlinespace[0.8ex]
\multirow[t]{4}{*}{$r_p$}  & \multirow[t]{4}{*}{8.409e-14 cm} & Y$_p$            & +0.000 & 1.0000 \\
                               &                              & D/H              & +0.000 & 1.0000 \\
                               &                              & $^7$Li/H         & +0.000 & 1.0000 \\
                               &                              & N$_{\text{eff}}$ & +0.000 & 1.0000 \\

\end{tabular}

\par\smallskip

\end{minipage}%
\hfill
\begin{minipage}[t]{0.49\textwidth}
\centering
\tiny   
{\footnotesize\bfseries NACRE-II}\par\vspace{0.5ex}
\vspace{2.0ex}
\setlength\tabcolsep{4pt}

\setlength{\heavyrulewidth}{0.08em}%
\setlength{\lightrulewidth}{0.05em}%
\setlength{\cmidrulewidth}{0.04em}%

\begin{tabular}{l@{\hspace{-12pt}}c l c c}
Parameter & $p_{\rm fid}$ & Observable & $\frac{d\ln Y}{d\ln p}$ & $R^2$ \\
\addlinespace[1ex]
\midrule
\addlinespace[1ex]

\multirow[t]{4}{*}{$G_F$}  & \multirow[t]{4}{*}{1.16638e-11 MeV$^{-2}$} & Y$_p$   & -1.452 & 0.9990 \\
                     &                               & D/H   & -0.827 & 0.9960 \\
                     &                               & $^7$Li/H & -0.829 & 0.9988\\
                     &                               & N$_{\text{eff}}$ & +0.010 & 0.9987 \\
\addlinespace[1ex]
\cmidrule(lr){1-5}
\addlinespace[1ex]

\multirow[t]{4}{*}{$m_e$}  & \multirow[t]{4}{*}{0.510999 MeV}    & Y$_p$   & +0.394 & 0.9954 \\
                     &                               & D/H   & +0.493 & 0.9946 \\
                     &                               & $^7$Li/H & -0.225 & 0.8482 \\
                     &                               & N$_{\text{eff}}$ & +0.023 & 0.9970 \\
\addlinespace[1ex]
\cmidrule(lr){1-5}
\addlinespace[1ex]

\multirow[t]{4}{*}{$g_A$}  & \multirow[t]{4}{*}{1.2753}      & Y$_p$   & -1.215 & 0.9995 \\
                     &                               & D/H   & -0.698 & 0.9975 \\
                     &                               & $^7$Li/H & -0.682 & 0.9988 \\
                     &                               & N$_{\text{eff}}$ & +0.000 & 1.0000 \\
\addlinespace[1ex]
\cmidrule(lr){1-5}
\addlinespace[1ex]

\multirow[t]{4}{*}{$V_{ud}$} & \multirow[t]{4}{*}{0.97367}     & Y$_p$   & -1.465 & 0.9991 \\
                     &                               & D/H   & -0.860 & 0.9962 \\
                     &                               & $^7$Li/H & -0.823 & 0.9988 \\
                     &                               & N$_{\text{eff}}$ & +0.000 & 1.0000  \\

\addlinespace[1ex]
\cmidrule(lr){1-5}
\addlinespace[1ex]

\multirow[t]{4}{*}{$Q$} & \multirow[t]{4}{*}{1.29333 MeV}     & Y$_p$         & -3.199 & 1.0000 \\
                        &                                  & D/H           & -1.878 & 0.9942 \\
                        &                                  & $^7$Li/H      & -1.720 & 0.9776 \\
                        &                                  & N$_{\text{eff}}$ & +0.000 & 1.0000 \\

\addlinespace[1ex]
\cmidrule(lr){1-5}
\addlinespace[1ex]

\multirow[t]{4}{*}{$\Delta \kappa$} & \multirow[t]{4}{*}{3.70589}  & Y$_p$            & +0.000 & 0.9994 \\
                               &                              & D/H              & +0.000 & 0.9939 \\
                               &                              & $^7$Li/H         & +0.000 & 0.9969 \\
                               &                              & N$_{\text{eff}}$ & +0.000 & 1.0000 \\
\addlinespace[0.8ex]\cmidrule(lr){1-5}\addlinespace[0.8ex]
\multirow[t]{4}{*}{$\alpha_{EM}$} & \multirow[t]{4}{*}{0.00729735} & Y$_p$            & -0.010 & 1.0000 \\
                            &                                 & D/H              & -0.005 & 0.9671 \\
                            &                                 & $^7$Li/H         & -0.006 & 0.9650 \\
                            &                                 & N$_{\text{eff}}$ & +0.004 & 0.8363 \\
\addlinespace[0.8ex]\cmidrule(lr){1-5}\addlinespace[0.8ex]
\multirow[t]{4}{*}{$r_p$}  & \multirow[t]{4}{*}{8.409e-14 cm} & Y$_p$            & +0.000 & 1.0000 \\
                               &                              & D/H              & +0.000 & 1.0000 \\
                               &                              & $^7$Li/H         & +0.000 & 1.0000 \\
                               &                              & N$_{\text{eff}}$ & +0.000 & 1.0000 \\

\end{tabular}

\par\smallskip

\end{minipage}

\caption{Local linear sensitivities, $\frac{dlnY}{dlnp}$, and linear-fit quality, 
$R^2$, computed using weak rate normalization calculated from fundamental parameters. Only the parameters involved in calculating the weak rate normalization are included as the response functions of all other parameters do not change based on the choice of weak rate normalization. The left table shows results using the PRIMAT nuclear reaction rates, and the right table shows the results using the NACRE-II nuclear reaction rates.}

\label{tab:sens_fund_side_by_side}
\end{table}

\begin{table}[t]
\centering
\vspace*{0pt}

\setlength{\heavyrulewidth}{0.08em}%
\setlength{\lightrulewidth}{0.05em}%
\setlength{\cmidrulewidth}{0.04em}%

\begin{minipage}[t]{0.49\textwidth}
\centering
\scriptsize
{\footnotesize\bfseries PRIMAT}\par\vspace{0.5ex}
\vspace{2.0ex}
\setlength{\tabcolsep}{4pt}

\begin{tabular}{l l c c}
Parameter & Observable & $\frac{dY}{dp}$ & $R^2$ \\
\addlinespace[0.8ex]
\midrule
\addlinespace[0.8ex]

\multirow[t]{3}{*}{$\Delta \text{N}_{\text{eff}}$} & Y$_p$    & +0.014 & 0.9984 \\
                                                  & D/H      & +0.333 & 1.0000 \\
                                                  & $^7$Li/H & -0.509 & 0.9993 \\

\addlinespace[0.8ex]\cmidrule(lr){1-4}\addlinespace[0.8ex]

\multirow[t]{4}{*}{$\xi_\nu$} & Y$_p$            & -0.241 & 0.9994 \\
                                       & D/H              & -1.316 & 0.9956 \\
                                       & $^7$Li/H         & -3.488 & 0.9988 \\
                                       & N$_{\text{eff}}$ & +0.000 & 0.8331 \\

\end{tabular}

\end{minipage}
\hfill
\begin{minipage}[t]{0.49\textwidth}
\centering
\scriptsize
{\footnotesize\bfseries NACRE-II}\par\vspace{0.5ex}
\vspace{2.0ex}
\setlength{\tabcolsep}{4pt}

\begin{tabular}{l l c c}
Parameter & Observable & $\frac{dY}{dp}$ & $R^2$ \\
\addlinespace[0.8ex]
\midrule
\addlinespace[0.8ex]

\multirow[t]{3}{*}{$\Delta \text{N}_{\text{eff}}$} & Y$_p$    & +0.014 & 0.9984 \\
                                                  & D/H      & +0.342 & 1.0000 \\
                                                  & $^7$Li/H & -0.535 & 0.9990 \\

\addlinespace[0.8ex]\cmidrule(lr){1-4}\addlinespace[0.8ex]

\multirow[t]{4}{*}{$\xi_\nu$} & Y$_p$            & -0.242 & 0.9994 \\
                                       & D/H              & -1.354 & 0.9994 \\
                                       & $^7$Li/H         & -3.205 & 0.9984 \\
                                       & N$_{\text{eff}}$ & +0.000 & 0.8331 \\

\end{tabular}

\end{minipage}

\caption{Local linear sensitivities, $\frac{dY}{dp}$, and linear-fit quality, $R^2$, for variations in the non-standard parameters $\Delta N_{\rm eff}$ and $\mu_\nu/T_\nu$. The derivative $\frac{dY}{dp}$ gives the change in the observable $Y$ per unit change in the parameter $p$ evaluated at $p_{\rm fid}$ = 0. The left table uses the PRIMAT nuclear reaction rates and the right table uses the NACRE-II nuclear reaction rates. The linear sensitivity, $dN_{\rm eff}/d\xi_\nu = 0$ at $\xi_\nu = 0$ because the neutrino energy density receives equal contributions from neutrinos and anti-neutrinos, whose responses to $\xi_\nu$ cancel at first order. The leading correction is quadratic and small: $\Delta N_{\rm eff} = 0.013$ for $\xi_\nu = 0.1$, and does not appear in our linear sensitivity analysis.}
\label{tab:sens_taun_side_by_side_neff}
\end{table}

\begin{table}[t]
\centering
\vspace*{0pt}

\setlength{\heavyrulewidth}{0.08em}%
\setlength{\lightrulewidth}{0.05em}%
\setlength{\cmidrulewidth}{0.04em}%

\begin{minipage}[t]{0.49\textwidth}
\centering
\tiny   
{\footnotesize\bfseries PRIMAT}\par\vspace{0.5ex}
\vspace{2.0ex}
\setlength{\tabcolsep}{4pt}

\begin{tabular}{l l c c}

Reaction & Observable & $\frac{dY}{dp}$ & $R^2_{\rm lin}$ \\
\addlinespace[0.8ex]
\midrule
\addlinespace[0.8ex]

\multirow[t]{3}{*}{$n + p \to d + \gamma$}  & Y$_p$ & +0.000 & 1.0000 \\
 &  D/H & -0.002 & 0.9999 \\
 &  $^7$Li/H & +0.032 & 1.0000 \\

\addlinespace[0.8ex]\cmidrule(lr){1-4}\addlinespace[0.8ex]

\multirow[t]{3}{*}{$d + p \to {}^{3}{\rm He} + \gamma$}  & Y$_p$ & +0.000 & 1.0000 \\
 &  D/H & -0.018 & 0.9999 \\
 &  $^7$Li/H & +0.072 & 0.9998 \\

\addlinespace[0.8ex]\cmidrule(lr){1-4}\addlinespace[0.8ex]

\multirow[t]{3}{*}{$d + d \to {}^{3}{\rm He} + n$}  & Y$_p$ & +0.000 & 1.0000 \\
 &  D/H & -0.015 & 1.0000 \\
 &  $^7$Li/H & +0.041 & 1.0000 \\

\addlinespace[0.8ex]\cmidrule(lr){1-4}\addlinespace[0.8ex]

\multirow[t]{3}{*}{$d + d \to t + p$}  & Y$_p$ & +0.000 & 1.0000 \\
& D/H & -0.012 & 1.0000 \\
& $^7$Li/H & +0.003 & 0.9997 \\

\addlinespace[0.8ex]\cmidrule(lr){1-4}\addlinespace[0.8ex]

\multirow[t]{3}{*}{$t + p \to {}^4\text{He} + \gamma$}  & Y$_p$ & +0.000 & 1.0000 \\
& D/H & +0.000 & 0.9993 \\
& $^7$Li/H & +0.021 & 0.9912 \\

\addlinespace[0.8ex]\cmidrule(lr){1-4}\addlinespace[0.8ex]

\multirow[t]{3}{*}{$t + d \to {}^4\text{He} + n$}  & Y$_p$ & +0.000 & 1.0000 \\
& D/H & +0.000 & 1.0000 \\
& $^7$Li/H & +0.000 & 1.0000 \\

\addlinespace[0.8ex]\cmidrule(lr){1-4}\addlinespace[0.8ex]

\multirow[t]{3}{*}{$t + {}^4\text{He} \to {}^{7}{\rm Li} + \gamma$}  & Y$_p$ & +0.000 & 1.0000 \\
& D/H & +0.000 & 1.0000 \\
& $^7$Li/H & +0.006 & 0.9994 \\

\addlinespace[0.8ex]\cmidrule(lr){1-4}\addlinespace[0.8ex]

\multirow[t]{3}{*}{${}^{3}{\rm He} + n \to t + p$}  & Y$_p$ & +0.000 & 1.0000 \\
& D/H & +0.001 & 0.9993 \\
& $^7$Li/H & -0.020 & 1.0000 \\

\addlinespace[0.8ex]\cmidrule(lr){1-4}\addlinespace[0.8ex]

\multirow[t]{3}{*}{${}^{3}{\rm He} + d \to {}^4\text{He}+ p$}  & Y$_p$ & +0.000 & 1.0000 \\
& D/H & +0.000 & 0.9965 \\
& $^7$Li/H & -0.046 & 1.0000 \\

\addlinespace[0.8ex]\cmidrule(lr){1-4}\addlinespace[0.8ex]

\multirow[t]{3}{*}{${}^{3}{\rm He} + {}^4\text{He} \to {}^{7}{\rm Be} + \gamma$}  & Y$_p$ & +0.000 & 1.0000 \\
& D/H & +0.000 & 1.0000 \\
& $^7$Li/H & +0.127 & 0.9998 \\

\addlinespace[0.8ex]\cmidrule(lr){1-4}\addlinespace[0.8ex]

\multirow[t]{3}{*}{${}^{7}{\rm Li} + p \to {}^4\text{He} + {}^4\text{He}$}  & Y$_p$ & +0.000 & 1.0000 \\
& D/H & +0.000 & 1.0000 \\
& $^7$Li/H & -0.009 & 0.9997 \\

\addlinespace[0.8ex]\cmidrule(lr){1-4}\addlinespace[0.8ex]

\multirow[t]{3}{*}{${}^{7}{\rm Be} + n \to {}^{7}{\rm Li} + p$}  & Y$_p$ & +0.000 & 1.0000 \\
& D/H & +0.000 & 1.0000 \\
& $^7$Li/H & -0.074 & 1.0000 \\

\end{tabular}

\end{minipage}
\hfill
\begin{minipage}[t]{0.49\textwidth}
\centering
\tiny   
{\footnotesize\bfseries NACRE-II}\par\vspace{0.5ex}
\vspace{2.0ex}
\setlength{\tabcolsep}{4pt}

\begin{tabular}{l l c c}

Reaction & Observable & $\frac{dY}{dp}$ & $R^2_{\rm lin}$ \\
\addlinespace[0.8ex]
\midrule
\addlinespace[0.8ex]

\multirow[t]{3}{*}{$n + p \to d + \gamma$}  & Y$_p$ & +0.000 & 1.0000 \\
& D/H & -0.002 & 0.9999 \\
& $^7$Li/H & +0.032 & 1.0000 \\

\addlinespace[0.8ex]\cmidrule(lr){1-4}\addlinespace[0.8ex]

\multirow[t]{3}{*}{$d + p \to {}^{3}{\rm He} + \gamma$}  & Y$_p$ & +0.000 & 0.9989 \\
& D/H & -0.020 & 0.9999 \\
& $^7$Li/H & +0.085 & 0.9997 \\

\addlinespace[0.8ex]\cmidrule(lr){1-4}\addlinespace[0.8ex]

\multirow[t]{3}{*}{$d + d \to {}^{3}{\rm He} + n$}  & Y$_p$ & +0.000 & 0.9998 \\
& D/H & -0.082 & 1.0000 \\
& $^7$Li/H & +0.207 & 0.9994 \\

\addlinespace[0.8ex]\cmidrule(lr){1-4}\addlinespace[0.8ex]

\multirow[t]{3}{*}{$d + d \to t + p$}  & Y$_p$ & +0.000 & 0.9999 \\
& D/H & -0.055 & 1.0000 \\
& $^7$Li/H & +0.008 & 0.9997 \\

\addlinespace[0.8ex]\cmidrule(lr){1-4}\addlinespace[0.8ex]

\multirow[t]{3}{*}{$t + p \to {}^4\text{He} + \gamma$}  & Y$_p$ & +0.000 & 0.9910 \\
& D/H & -0.001 & 0.9993 \\
& $^7$Li/H & +0.021 & 0.9912 \\

\addlinespace[0.8ex]\cmidrule(lr){1-4}\addlinespace[0.8ex]

\multirow[t]{3}{*}{$t + d \to {}^4\text{He} + n$}  & Y$_p$ & +0.000 & 1.0000 \\
& D/H & +0.000 & 1.0000 \\
& $^7$Li/H & -0.009 & 0.9981 \\

\addlinespace[0.8ex]\cmidrule(lr){1-4}\addlinespace[0.8ex]

\multirow[t]{3}{*}{$t + {}^4\text{He} \to {}^{7}{\rm Li} + \gamma$}  & Y$_p$ & +0.000 & 1.0000 \\
& D/H & +0.000 & 1.0000 \\
& $^7$Li/H & +0.012 & 0.9979 \\

\addlinespace[0.8ex]\cmidrule(lr){1-4}\addlinespace[0.8ex]

\multirow[t]{3}{*}{${}^{3}{\rm He} + n \to t + p$}  & Y$_p$ & +0.000 & 0.9997 \\
& D/H & +0.004 & 1.0000 \\
& $^7$Li/H & -0.107 & 1.0000 \\

\addlinespace[0.8ex]\cmidrule(lr){1-4}\addlinespace[0.8ex]

\multirow[t]{3}{*}{${}^{3}{\rm He} + d \to {}^4\text{He} + p$}  & Y$_p$ & +0.000 & 0.9984 \\
& D/H & -0.002 & 0.9673 \\
& $^7$Li/H & -0.426 & 0.9996 \\

\addlinespace[0.8ex]\cmidrule(lr){1-4}\addlinespace[0.8ex]

\multirow[t]{3}{*}{${}^{3}{\rm He} + {}^4\text{He} \to {}^{7}{\rm Be} + \gamma$}  & Y$_p$ & +0.000 & 1.0000 \\
& D/H & +0.000 & 1.0000 \\
& $^7$Li/H & +0.460 & 0.9978 \\

\addlinespace[0.8ex]\cmidrule(lr){1-4}\addlinespace[0.8ex]

\multirow[t]{3}{*}{${}^{7}{\rm Li} + p \to {}^4\text{He} + {}^4\text{He}$}  & Y$_p$ & +0.000 & 1.0000 \\
& D/H & +0.000 & 1.0000 \\
& $^7$Li/H & -0.025 & 0.9982 \\

\addlinespace[0.8ex]\cmidrule(lr){1-4}\addlinespace[0.8ex]

\multirow[t]{3}{*}{${}^{7}{\rm Be} + n \to {}^{7}{\rm Li} + p$}  & Y$_p$ & +0.000 & 1.0000 \\
& D/H & +0.000 & 1.0000 \\
& $^7$Li/H & -0.305 & 0.9998 \\

\end{tabular}

\end{minipage}

\caption{Local linear sensitivities, $\frac{dY}{dp}$, and linear-fit quality, $R^2$, for variations in the twelve most impactful nuclear reaction rates. Here $p$ is the lognormal rate-shift parameter measured in units of the quoted $1\sigma$ uncertainty (i.e.\ $p=1$ corresponds to a $+1\sigma$ shift of the reaction rate, and $p=-1$ to a $-1\sigma$ shift), so $\frac{dY}{dp}$ gives the change in $Y$ per $1\sigma$ rate shift evaluated at $p_{\rm fid}=0$. The left table shows results for the PRIMAT nuclear reaction rates and the right table shows results for the NACRE-II nuclear reaction rates.}

\label{tab:sens_tau_n_rates_first12_side_by_side}
\end{table}

\section{Uncertainty Budget: Standard Model BBN}
\label{sec:unc}

We present a detailed uncertainty budget for the BBN abundance predictions
obtained via linear error propagation. The contribution of each input
parameter is computed as $\sigma_{Y_i} = |dY/dp_i|\,\sigma_{p_i}$ and the total
input uncertainty is given by $\sigma_Y = \sqrt{\sum_i \sigma^2_{Y_i}}$, where
the derivatives $dY/dp$ are computed numerically as centered finite
differences from the parameter scans described in Section~\ref{sec:PVar}.
Off-diagonal covariance terms are neglected, so the propagated variance is
approximated as a quadrature sum of the individual parameter contributions. This choice is discussed in detail in Section \ref{sec:covariance}. Users applying these results to models with correlated parameters should instead propagate the full covariance matrix using the response functions provided here.

This decomposition allows us to identify and rank the dominant sources of uncertainty for each observable, separately for the PRIMAT and NACRE-II nuclear rate compilations and each choice of weak rate normalization. We verify that the linear approximation is well-justified by confirming that the curvature corrections are negligible for all dominant parameters, as indicated by the $R^2$ values reported in Section~\ref{sec:sens}. 

All fundamental physics input uncertainties are taken to be symmetric 1$\sigma$ errors adopted from recommended literature values quoted in symmetric form. The nuclear reaction rates are treated separately, as their uncertainties are generally temperature dependent and are not well described by a single symmetric linear error bar. For each reaction, we use the PRyMordial nuisance-parameter prescription, in which a Gaussian-distributed parameter in units of 1$\sigma$ rescales the tabulated median rate through its temperature-dependent lognormal uncertainty band. The resulting response derivatives are extracted at the fiducial point and propagated into the total error budget in quadrature, identically to the fundamental parameters. 

Two parameters warrant further scrutiny. First, the electron mass sensitivity for D/H under the neutron lifetime normalization is highly nonlinear, $R^2_{\rm lin} \approx 0.8$; however, because the experimental uncertainty on $m_e$ is extremely small, this parameter contributes negligibly to the total deuterium error budget and the nonlinearity has no practical impact. Second, and more consequentially, the D/H response to $\Omega_b h^2$ is only approximately linear, $R^2_{\rm lin} \approx 0.96$, and $\Omega_b h^2$ is the dominant contributor to the deuterium uncertainty budget when the PRIMAT rates are used. The linear propagation therefore slightly underestimates the true deuterium uncertainty from this source, and the corresponding entry in Table~\ref{tab:budget_comparison_taun} should be interpreted with this caveat in mind.

The input parameter uncertainties are taken from the Particle Data Group~\cite{PDG2025} and Ivanov, et al. \cite{Ivanov:2012qe} for the fundamental constants,
from Planck~\cite{Planck:2018vyg} for $\Omega_b h^2$, and from the PRIMAT~\cite{Pitrou:2018cgg} and NACRE-II~\cite{Xu:2013fha} compilations for the nuclear reaction rates. We present the uncertainty budget separately for both rate compilations and for each choice of weak rate normalization. The spread between the two sets of central nuclear reaction rate values provides an additional measure of systematic uncertainty associated with the choice of nuclear rate evaluation, which is not captured within either compilation's individual error budget.

The total theory uncertainties incorporate additional sources of error beyond the scope of our linear analysis, including higher-order effects, numerical integration uncertainties, and marginalization over correlated nuclear rate uncertainties. The total theoretical uncertainties are the following:

$Y_p = 0.2467 \pm 0.0002$ ~\cite{Yeh:2022heq} 

$\text{D/H}  \times 10^{5} = 2.439 \pm 0.037$~\cite{Pitrou:2020etk} 

${}^7\text{Li/H} \times 10^{10} = 5.464 \pm 0.220$~\cite{Pitrou:2020etk}.

Our linear error budget recovers these total uncertainties to within approximately 30\%, with the residual difference attributable to the additional systematic effects included in the full analysis. Here, we present our uncertainty budgets for both the PRIMAT and NACRE-II compilations. The primary contribution of this work is therefore not the total uncertainty itself, but the decomposition of the error budget into its constituent parts, identifying which parameters most urgently require improved experimental or theoretical input. The predicted theory values and uncertainty coming from linear error propagation of parameter uncertainties are shown in Table \ref{tab:th_vals}.

\textbf{\begin{table*}[t]
\centering
\begin{tabular}{l l c c}
\noalign{\vspace{4pt}}
Observable & Compilation & $\tau_n$ normalization & Fundamental normalization \\[6pt]
\hline
\noalign{\vspace{6pt}}
\multirow{2}{*}{$Y_p$}
 & PRIMAT   & $0.24691 \pm 0.00012$ & $0.24734 \pm 0.00034$ \\
 & NACRE-II & $0.24683 \pm 0.00015$ & $0.24725 \pm 0.00035$ \\[6pt]
\hline
\noalign{\vspace{6pt}}
\multirow{2}{*}{$\text{D/H} \times 10^5$}
 & PRIMAT   & $2.445 \pm 0.038$ & $2.448 \pm 0.038$ \\
 & NACRE-II & $2.517 \pm 0.104$ & $2.519 \pm 0.104$ \\[6pt]
\hline
\noalign{\vspace{6pt}}
\multirow{2}{*}{${}^7\text{Li/H} \times 10^{10}$}
 & PRIMAT   & $5.523 \pm 0.196$ & $5.528 \pm 0.196$ \\
 & NACRE-II & $5.356 \pm 0.745$ & $5.361 \pm 0.745$ \\[6pt]
\hline
\end{tabular}
\caption{Predicted BBN abundances comparing the $\tau_n$ and fundamental weak rate normalizations for both PRIMAT and NACRE-II reaction rate compilations.}
\label{tab:th_vals}
\end{table*}}
These uncertainties assume $\Delta N_{\text{eff}} = 0$, and thus represent the SM predictions. We report the uncertainties when $\Delta N_{\text{eff}}$ is allowed to vary freely in the following section.

Table \ref{tab:budget_comparison_taun} is the ranked uncertainty budget for all of the parameters from most influential to least for each of the abundances using the $\tau_n$ weak rate normalization. Table \ref{tab:budget_comparison_fund} is the ranked uncertainty budget for all of the parameters from most influential to least for each of the abundances using the fundamental weak rate normalization. Parameters contributing less than $0.1\%$ of the total uncertainty are omitted for brevity. The full results are shown on GitHub \faGithub \href{https://github.com/Anne-KatherineBurns/bbn-sensitivity-atlas}{\,\texttt{bbn-sensitivity-atlas}}. 

\subsection{Impact of neglected covariances}
\label{sec:covariance}

Aside from the weak-sector parameters and nuclear reaction rates, the dominant contributors to each budget are determined by mutually independent measurements, so their
cross-terms vanish by construction. The principal weak-sector correlations, in
turn, are either inactivated by the choice of normalization scheme or act on
sub-dominant parameter pairs. In order to justify the treatment of both the weak-sector parameters and the nuclear reaction rates as independent, which we explicitly bound the maximal contribution of correlations to the error budget of each abundance value. 

Among the fundamental parameters that contribute significantly to the
uncertainty budget, the weak-sector inputs $g_A$, $V_{ud}$, and $\tau_n$ are physically correlated through their joint role in the charged-current rates. 
The cross-terms involving $\tau_n$ are absent from both budgets, however,
because $\tau_n$ and the fundamental parameters never enter the same
normalization scheme. The one meaningful weak-sector covariance that remains active is that between $g_A$ and $V_{ud}$ within the fundamental-normalization budget. Treating this cross-term in the worst case, even maximal correlation between
$g_A$ and $V_{ud}$ inflates the fundamental-normalization $Y_p$ uncertainty by
at most $\sim$25\%, and the true inflation is smaller since their correlation
is far from maximal. For D/H and ${}^7Li/H$ the same cross-term is entirely negligible, as $g_A$ and $V_{ud}$ contribute only marginally to the budgets of those abundances.

The remaining assumption to examine is the treatment of the nuclear reaction
rates as mutually independent. Correlations among them arising from shared
experimental normalizations or joint R-matrix analyses are not provided by
the rate compilations, so rather than adopt a specific correlation model, we
bound their possible impact. For any of the predicted abundances, we can write the nuclear contributions to the error budget as $a_i = (dY/dp_i)\,\sigma_{p_i}$. Then, for any valid correlation matrix, the propagated uncertainty from the nuclear block is
bounded between the uncorrelated value
$\sigma_{\rm diag} = \sqrt{\sum_i a_i^2}$ and the fully-correlated value
$\sigma_{\rm max} = \sum_i |a_i|$, the latter being attained
in the extreme limit in which the reaction-rate contributions combine fully
coherently rather than in quadrature. Holding
the particle physics and cosmology contributions fixed at their quadrature
values, this provides a rigorous upper bound on the inflation of the total
quoted uncertainty for each observable that can come from correlations between  nuclear reaction rates.

Applying this bound over the full reaction network, we find that the impact on
$Y_p$ is negligible: even under maximal correlation of all 63 reaction rates, the total $Y_p$ uncertainty
inflates by at most a factor of $1.01$, reflecting the fact that nuclear rates
account for less than $1\%$ of its variance. For the nuclear-dominated D/H and
${}^7$Li/H predictions the bound is necessarily looser. The D/H uncertainty
inflates by at most a factor of $\sim$1.4 (PRIMAT) or $\sim$1.6 (NACRE-II), the
larger value reflecting the greater nuclear-physics share of the NACRE-II D/H
budget. The uncertainty of the ${}^7$Li/H prediction, whose budget is almost entirely nuclear, inflates by
at most a factor of $\sim$2.3 in both compilations. We emphasize that these are upper limits: they assume perfect correlation among all 63 reaction rates,
whereas the true correlations are considerably smaller, so the actual impact on
the quoted uncertainties is well below these ceilings. 

\subsection{Dominant contributors to the uncertainty budget}

Comparing the ranked uncertainty budgets across compilations and
normalization schemes reveals several robust patterns as well as
important compilation-dependent differences. For $Y_p$ under the
$\tau_n$ normalization, the neutron lifetime is the leading
contributor in both compilations, accounting for 67.1\% of the
variance with PRIMAT and 45.2\% with NACRE-II\@. However, the
sub-leading contributions differ markedly: with PRIMAT, the baryon
abundance $\Omega_b h^2$ is second (27.1\%), whereas with NACRE-II
the nuclear rate $d(d,n){}^3\text{He}$ rises to second place
(26.6\%), indicating that the helium-4 uncertainty is not entirely
insensitive to nuclear physics when the NACRE-II rate uncertainties
are used. Under the fundamental normalization, $g_A$ dominates
overwhelmingly in both compilations ($\sim$78--83\% of the variance),
with $V_{ud}$ a distant second ($\sim$12\%). The ranking beyond this, however is still 
compilation-dependent.

For D/H, the ranking is strongly compilation-dependent.  With PRIMAT,
$\Omega_b h^2$ leads (51\%), followed by $d(p,\gamma){}^3\text{He}$
($\sim$22\%) and $d(d,n){}^3\text{He}$ ($\sim$15\%).  With NACRE-II,
by contrast, $d(d,n){}^3\text{He}$ dominates (62\%), followed by
$d(d,p)t$ (27\%), with $\Omega_b h^2$ reduced to just 7\%.
Deuterium is thus cosmology-limited under PRIMAT but
nuclear-physics-limited under NACRE-II, and this pattern is
independent of the weak-rate normalization scheme.

For ${}^7\text{Li/H}$, the budget is nuclear-dominated in both
compilations, with ${}^3\text{He}({}^4\text{He},\gamma){}^7\text{Be}$
consistently the leading contributor ($\sim$38--42\%).  The
sub-leading structure differs: with PRIMAT, $\Omega_b h^2$,
${}^7\text{Be}(n,p){}^7\text{Li}$, and
$d(p,\gamma){}^3\text{He}$ each contribute 13--16\%, whereas with
NACRE-II, ${}^3\text{He}(d,p){}^4\text{He}$ rises to 33\% and
${}^7\text{Be}(n,p){}^7\text{Li}$ contributes 17\%, together
accounting for the majority of the variance.

These patterns identify concrete targets for future improvement:
tightening $Y_p$ predictions requires better weak-interaction inputs
(principally $\tau_n$ or $g_A$ depending on the normalization scheme),
while sharpening the D/H prediction requires either improved
determinations of $\Omega_b h^2$ or resolution of the discrepancies
between nuclear rate evaluations---particularly for the
deuterium-burning reactions $d(d,n){}^3\text{He}$, $d(d,p)t$, and
$d(p,\gamma){}^3\text{He}$.

\begin{table*}[t]
\centering
\footnotesize

\begin{minipage}[t]{0.45\textwidth}
\centering

\textbf{PRIMAT} \\[6pt]
\tiny
\begin{tabular}{l @{\hspace{10pt}} lc @{\hspace{8pt}} c}

Observable & Parameter & $\sigma_{Y_i}$ & \% var. \\ [6pt]
\hline
\noalign{\vspace{6pt}}
\multirow{7}{*}{$Y_p$}  
 & $\tau_n$                                              & $1.03 \times 10^{-4}$ & 67.1 \\
 & $\Omega_b h^2$                                        & $6.53 \times 10^{-5}$ & 27.1 \\
  & $n(p,\gamma)d$                                        & $1.89 \times 10^{-5}$ &  2.3 \\
& $d(d,p)t$                                             & $1.69 \times 10^{-5}$ &  1.8 \\
 & $d(d,n){}^3\text{He}$                                 & $1.63 \times 10^{-5}$ &  1.7 \\ [6pt]
\hline
\noalign{\vspace{6pt}}
\multirow{5}{*}{$\text{D/H} \times 10^5$}
 & $\Omega_b h^2$                                        & $2.69 \times 10^{-2}$ & 51.2 \\
 & $d(p,\gamma){}^3\text{He}$                            & $1.79 \times 10^{-2}$ & 22.7 \\
 & $d(d,n){}^3\text{He}$                                 & $1.48 \times 10^{-2}$ & 15.4 \\
 & $d(d,p)t$                                             & $1.20 \times 10^{-2}$ & 10.3 \\
 & $n(p,\gamma)d$                                        & $1.94 \times 10^{-3}$ &  0.3 \\ [6pt]
\hline
\noalign{\vspace{6pt}}
\multirow{10}{*}{${}^7\text{Li/H} \times 10^{10}$}
 & ${}^3\text{He}(^4\text{He},\gamma){}^7\text{Be}$           & $1.27 \times 10^{-1}$ & 41.9 \\
 & $\Omega_b h^2$                                        & $7.70 \times 10^{-2}$ & 15.4 \\
 & ${}^7\text{Be}(n,p){}^7\text{Li}$                     & $7.41 \times 10^{-2}$ & 14.3 \\
 & $d(p,\gamma){}^3\text{He}$                            & $7.18 \times 10^{-2}$ & 13.4 \\
 & ${}^3\text{He}(d,p){}^4\text{He}$                     & $4.58 \times 10^{-2}$ &  5.5 \\
 & $d(d,n){}^3\text{He}$                                 & $4.09 \times 10^{-2}$ &  4.4 \\
 & $n(p,\gamma)d$                                        & $3.10 \times 10^{-2}$ &  2.6 \\
 & $t(p,\gamma){}^4\text{He}$                            & $2.01 \times 10^{-2}$ &  1.1 \\
 & ${}^3\text{He}(n,p)t$                                 & $2.00 \times 10^{-2}$ &  1.0 \\
 & ${}^7\text{Li}(p,^4\text{He}){}^4\text{He}$                & $9.42 \times 10^{-3}$ &  0.2 \\
  & $t(^4\text{He},\gamma){}^7\text{Li}$                & $5.34 \times 10^{-3}$ &  0.1 \\

\end{tabular}
\end{minipage}
\hfill
\begin{minipage}[t]{0.45\textwidth}
\centering

\textbf{NACRE~II} \\[6pt]
\tiny
\begin{tabular}{l @{\hspace{10pt}} lc @{\hspace{8pt}} c}

Observable & Parameter & $\sigma_{Y_i}$ & \% var. \\
\noalign{\vspace{6pt}}
\hline
\noalign{\vspace{6pt}}
\multirow{5}{*}{$Y_p$}
 & $\tau_n$                                              & $1.03 \times 10^{-4}$ & 45.2 \\
 & $d(d,n){}^3\text{He}$                                 & $7.89 \times 10^{-5}$ & 26.6 \\
 & $\Omega_b h^2$                                        & $6.14 \times 10^{-5}$ & 17.6 \\
 & $d(d,p)t$                                             & $4.23 \times 10^{-5}$ &  7.7 \\
 & $n(p,\gamma)d$                                        & $2.63 \times 10^{-5}$ &  3.0 \\
 \noalign{\vspace{6pt}}
\hline
\noalign{\vspace{6pt}}
\multirow{6}{*}{$\text{D/H} \times 10^5$}
 & $d(d,n){}^3\text{He}$                                 & $8.20 \times 10^{-2}$ & 61.9 \\
 & $d(d,p)t$                                             & $5.42 \times 10^{-2}$ & 27.1 \\
 & $\Omega_b h^2$                                        & $2.77 \times 10^{-2}$ &  7.1 \\
 & $d(p,\gamma){}^3\text{He}$                            & $2.01 \times 10^{-2}$ &  3.7 \\
 & ${}^3\text{He}(n,p)t$                                 & $4.75 \times 10^{-3}$ &  0.2 \\

 \noalign{\vspace{6pt}}
\hline
\noalign{\vspace{6pt}}
\multirow{8}{*}{${}^7\text{Li/H} \times 10^{10}$}
 & ${}^3\text{He}(^4\text{He},\gamma){}^7\text{Be}$           & $4.59 \times 10^{-1}$ & 38.0 \\
 & ${}^3\text{He}(d,p){}^4\text{He}$                     & $4.25 \times 10^{-1}$ & 32.5 \\
 & ${}^7\text{Be}(n,p){}^7\text{Li}$                     & $3.05 \times 10^{-1}$ & 16.8 \\
 & $d(d,n){}^3\text{He}$                                 & $2.08 \times 10^{-1}$ &  7.8 \\
 & ${}^3\text{He}(n,p)t$                                 & $1.07 \times 10^{-1}$ &  2.1 \\
 & $d(p,\gamma){}^3\text{He}$                            & $8.54 \times 10^{-2}$ &  1.3 \\
 & $\Omega_b h^2$                                        & $7.81 \times 10^{-2}$ &  1.1 \\
 & $n(p,\gamma)d$                                        & $3.20 \times 10^{-2}$ &  0.2 \\
& ${}^7\text{Li}(p,^4\text{He}){}^4\text{He}$                & $2.60 \times 10^{-2}$ &  0.1 \\
 & $t(p,\gamma){}^4\text{He}$                            & $2.12 \times 10^{-2}$ &  0.1 \\

\end{tabular}
\end{minipage}
\caption{Uncertainty budget for BBN abundances, comparing PRIMAT (left) and NACRE~II (right) compilations using the $\tau_n$ weak rate normalization. 
Parameters are ranked by $\sigma_{Y_i}$ within each observable. 
Only contributions with fractional variance $>0.1\%$ are shown.}
\label{tab:budget_comparison_taun}
\end{table*}

\begin{table*}[t]
\centering
\begin{minipage}[t]{0.45\textwidth}
\centering

\textbf{\footnotesize\bfseries PRIMAT} \\[6pt]
\tiny
\begin{tabular}{l @{\hspace{10pt}} lc @{\hspace{8pt}} c}

Observable & Parameter & $\sigma_{Y_i}$ & \% var. \\ [6pt]
\hline
\noalign{\vspace{6pt}}
\multirow{6}{*}{$Y_p$}
 & $g_A$                                              & $3.06 \times 10^{-4}$ &  83.2 \\
 & $V_{ud}$                                           & $1.19 \times 10^{-4}$ &  12.5 \\
 & $\Omega_b h^2$                                     & $6.47 \times 10^{-5}$ &   3.7 \\
 & $d(d,n){}^3\text{He}$                              & $1.62 \times 10^{-5}$ &   0.2 \\
 & $n(p,\gamma)d$                                     & $1.44 \times 10^{-5}$ &   0.2 \\
 & $d(d,p)t$                                          & $1.40 \times 10^{-5}$ &   0.2 \\ [6pt]
\hline
\noalign{\vspace{6pt}}
\multirow{7}{*}{$\text{D/H} \times 10^5$}
 & $\Omega_b h^2$                                     & $2.69 \times 10^{-2}$ &  51.4 \\
 & $d(p,\gamma){}^3\text{He}$                         & $1.79 \times 10^{-2}$ &  22.6 \\
 & $d(d,n){}^3\text{He}$                              & $1.46 \times 10^{-2}$ &  15.1 \\
 & $d(d,p)t$                                          & $1.20 \times 10^{-2}$ &  10.2 \\
 & $n(p,\gamma)d$                                     & $2.07 \times 10^{-3}$ &   0.3 \\
 & $g_A$                                              & $1.74 \times 10^{-3}$ &   0.2 \\
 & ${}^3\text{He}(n,p)t$                              & $1.19 \times 10^{-3}$ &   0.1 \\ [6pt]
\hline
\noalign{\vspace{6pt}}
\multirow{10}{*}{${}^7\text{Li/H} \times 10^{10}$}
 & ${}^3\text{He}(^4He,\gamma){}^7\text{Be}$       & $1.27 \times 10^{-1}$ &  42.0 \\
 & $\Omega_b h^2$                                     & $7.71 \times 10^{-2}$ &  15.5 \\
 & ${}^7\text{Be}(n,p){}^7\text{Li}$                  & $7.37 \times 10^{-2}$ &  14.1 \\
 & $d(p,\gamma){}^3\text{He}$                         & $7.16 \times 10^{-2}$ &  13.3 \\
 & ${}^3\text{He}(d,p){}^4\text{He}$                  & $4.62 \times 10^{-2}$ &   5.6 \\
 & $d(d,n){}^3\text{He}$                              & $4.06 \times 10^{-2}$ &   4.3 \\
 & $n(p,\gamma)d$                                     & $3.19 \times 10^{-2}$ &   2.6 \\
 & $t(p,\gamma){}^4\text{He}$                         & $2.07 \times 10^{-2}$ &   1.1 \\
 & ${}^3\text{He}(n,p)t$                              & $1.97 \times 10^{-2}$ &   1.0 \\
 & ${}^7\text{Li}(p,{}^4\text{He}){}^4\text{He}$            & $9.42 \times 10^{-3}$ &   0.2 \\
  & $t({}^4\text{He},\gamma){}^7\text{Li}$            & $5.62 \times 10^{-3}$ &   0.1 \\ [6pt]

\end{tabular}
\end{minipage}
\hfill
\begin{minipage}[t]{0.45\textwidth}
\centering

\textbf{\footnotesize\bfseries NACRE~II} \\[6pt]
\tiny
\begin{tabular}{l @{\hspace{10pt}} lc @{\hspace{8pt}} c}

Observable & Parameter & $\sigma_{Y_i}$ & \% var. \\ [6pt]
\hline
\noalign{\vspace{6pt}}
\multirow{6}{*}{$Y_p$}
 & $g_A$                                              & $3.06 \times 10^{-4}$ &  77.8 \\
 & $V_{ud}$                                           & $1.19 \times 10^{-4}$ &  11.7 \\
 & $d(d,n){}^3\text{He}$                              & $7.96 \times 10^{-5}$ &   5.3 \\
 & $\Omega_b h^2$                                     & $6.40 \times 10^{-5}$ &   3.4 \\
 & $d(d,p)t$                                          & $4.55 \times 10^{-5}$ &   1.7 \\
 & $n(p,\gamma)d$                                     & $1.29 \times 10^{-5}$ &   0.1 \\ [6pt]
\hline
\noalign{\vspace{6pt}}
\multirow{5}{*}{$\text{D/H} \times 10^5$}
 & $d(d,n){}^3\text{He}$                              & $8.20 \times 10^{-2}$ &  61.6 \\
 & $d(d,p)t$                                          & $5.47 \times 10^{-2}$ &  27.4 \\
 & $\Omega_b h^2$                                     & $2.77 \times 10^{-2}$ &   7.0 \\
 & $d(p,\gamma){}^3\text{He}$                         & $2.00 \times 10^{-2}$ &   3.7 \\
 & ${}^3\text{He}(n,p)t$                              & $4.41 \times 10^{-3}$ &   0.2 \\ [6pt]
\hline
\noalign{\vspace{6pt}}
\multirow{9}{*}{${}^7\text{Li/H} \times 10^{10}$}
 & ${}^3\text{He}({}^4\text{He},\gamma){}^7\text{Be}$       & $4.60 \times 10^{-1}$ &  38.0 \\
 & ${}^3\text{He}(d,p){}^4\text{He}$                  & $4.26 \times 10^{-1}$ &  32.6 \\
 & ${}^7\text{Be}(n,p){}^7\text{Li}$                  & $3.06 \times 10^{-1}$ &  16.8 \\
 & $d(d,n){}^3\text{He}$                              & $2.07 \times 10^{-1}$ &   7.7 \\
 & ${}^3\text{He}(n,p)t$                              & $1.07 \times 10^{-1}$ &   2.0 \\
 & $d(p,\gamma){}^3\text{He}$                         & $8.50 \times 10^{-2}$ &   1.3 \\
 & $\Omega_b h^2$                                     & $7.83 \times 10^{-2}$ &   1.1 \\
 & $n(p,\gamma)d$                                     & $3.22 \times 10^{-2}$ &   0.2 \\
 & ${}^7\text{Li}(p,{}^4\text{He}){}^4\text{He}$            & $2.50 \times 10^{-2}$ &   0.1 \\ 
  & $t(p,\gamma){}^4\text{He}$            & $2.10 \times 10^{-2}$ &   0.1 \\ [6pt]

\end{tabular}
\end{minipage}
\caption{Uncertainty budget for BBN abundances, comparing PRIMAT (left) and NACRE~II (right) compilations using the fundamental weak rate normalization. 
Parameters are ranked by $\sigma_{Y_i}$ within each observable. 
Only contributions with fractional variance $>0.1\%$ are shown.}
\label{tab:budget_comparison_fund}
\end{table*}

\section{Uncertainty Budget: $N_{\rm{eff}}$ Dependence}
\label{sec:unc_neff}

When $\Delta N_\text{eff}$ is promoted to be a free parameter with $\sigma_{\Delta N_\text{eff}} = 0.070$ corresponding to the current combined CMB+BAO+BBN constraint~\cite{Goldstein:2026iuu}, it dominates the $Y_p$ uncertainty budget, accounting for between 88\% and 98\% of the total variance, depending on the weak rate normalization and choice of nuclear network compilation. This is expected: $Y_p$ is set primarily by the neutron-to-proton ratio at freeze-out, which depends sensitively on the competition between the weak interaction rate and the Hubble expansion rate. Since $N_\text{eff}$ directly controls the radiation energy density and hence the expansion rate, even a modest uncertainty on $\Delta N_\text{eff}$ propagates strongly into $Y_p$. The resulting uncertainty increases by roughly a factor of three relative to the standard case, from $\sigma(Y_p) \sim 0.0003$ to $\sim 0.0010$, and in doing so diminishes the distinction between the $\tau_n$ and fundamental weak rate normalizations. Both now yield similar uncertainties since neither contributes as significantly to the total error budget. The impact on D/H is more moderate, with $\Delta N_\text{eff}$ contributing $\sim$5--28\% of the variance depending on the compilation, while for ${}^7\text{Li/H}$ the effect is minor ($\lesssim 3\%$), as the uncertainty remains dominated by the ${}^3\text{He}({}^4\text{He},\gamma){}^7\text{Be}$ reaction rate and other nuclear inputs. Table \ref{tab:th_vals_neff} gives the theoretical values of the abundances with $\Delta N_\text{eff}$ as a free parameter. Tables \ref{tab:unc_neff_taun} and \ref{tab:unc_neff_fund} show the ranked uncertainty budget for all of the parameters from most influential to least for each of the abundances. Parameters contributing less than 0.1\% of the total uncertainty are omitted for brevity.

We note one important covariance not captured by the quadrature treatment in this budget: when $\Delta N_{\rm eff}$ is promoted to a free parameter using the combined CMB+BAO+BBN fit, $\Omega_b h^2$ and $\Delta N_{\rm eff}$ are jointly constrained and therefore correlated~\cite{DESI:2025zgx}. However, its impact on $Y_p$ is
strongly suppressed, since $\Delta N_{\rm eff}$ alone already accounts for 88--98\% of the $Y_p$ variance. On the other hand, the cross-term is least suppressed for D/H, where $\Delta N_{\rm eff}$ and $\Omega_b h^2$
contribute comparably. Applying the same worst-case bound used above, even
maximal correlation between them inflates the D/H uncertainty by at most
$\sim$28\% (PRIMAT) or $\sim$6\% (NACRE-II). The larger PRIMAT value reflects
that D/H is cosmology-limited there, so $\Omega_b h^2$ and $\Delta N_{\rm eff}$
are both major contributors (37\% and 28\% of the variance), whereas under
NACRE-II the nuclear rates dominate and both fall below 7\%. As the true
correlation is smaller than unity, the actual effect lies below these ceilings.

\begin{figure*}[t]
    \centering
    \includegraphics[width=\textwidth]{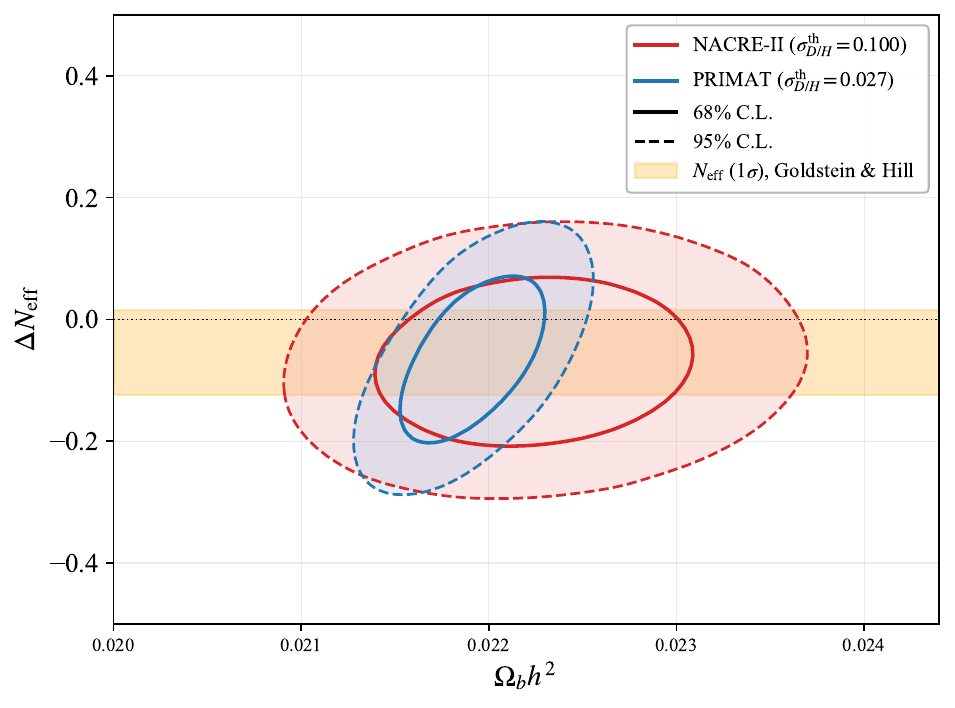}
    \caption{Joint constraints on $(\Omega_b h^2,\,\Delta N_{\rm eff})$ from the
    observed primordial D/H~\cite{PDG2025} and $Y_p$~\cite{Aver:2026dxv},
    computed with the NACRE-II (red) and PRIMAT (blue) reaction rate
    compilations using the $\tau_n$ weak-rate normalization. Filled regions show
    the 68.3\% and 95.4\% C.L.. The
    dotted horizontal line marks the Standard Model value
    $\Delta N_{\rm eff} = 0$, and the yellow band is the combined
    CMB+BAO+BBN determination $N_{\rm eff} = 2.990 \pm 0.070$~\cite{Goldstein:2026iuu},
    shown for context. The two contours share a similar extent in $\Delta N_{\rm eff}$, set by the
common $Y_p$ measurement, and differ mainly in their $\Omega_b h^2$ width,
which tracks the D/H theoretical uncertainty.}
    \label{fig:omega_neff_2d}
\end{figure*}

\begin{figure}[t]
    \centering
    \includegraphics[width=\columnwidth]{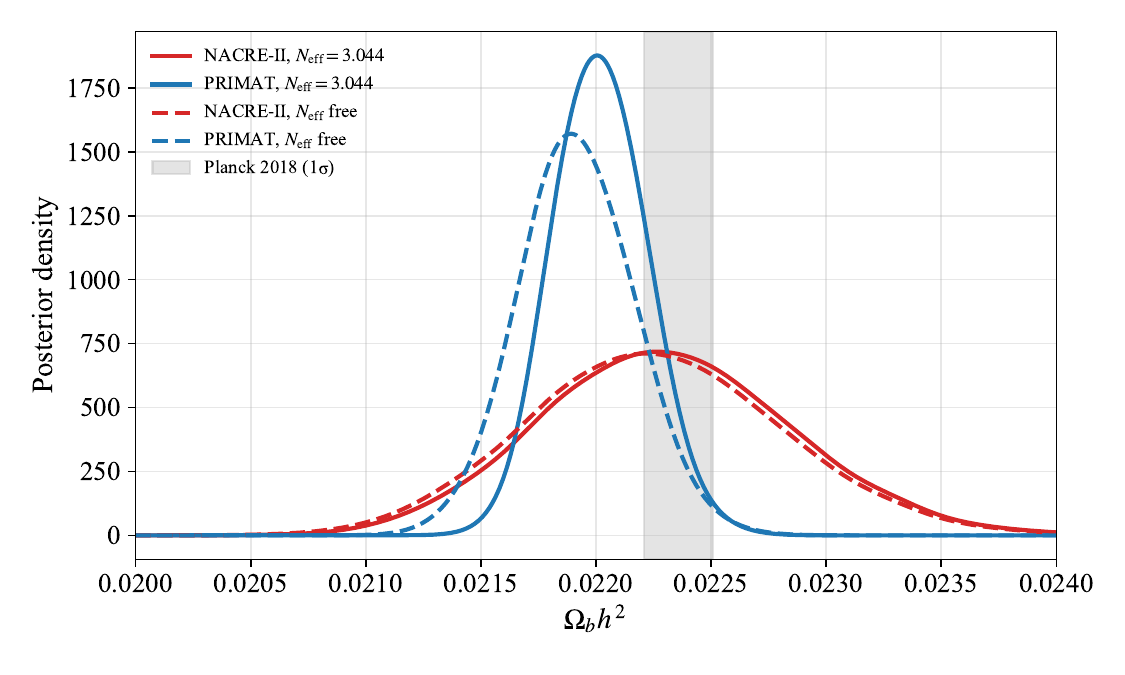}
    \caption{Marginal posteriors on $\Omega_b h^2$ inferred from the observed
    D/H~\cite{PDG2025} and $Y_p$~\cite{Aver:2026dxv}. Solid curves fix
    $\Delta N_{\rm eff} = 0$ and dashed curves marginalize over $\Delta N_{\rm eff}$
    with a flat prior. Red denotes NACRE-II and blue PRIMAT. The gray band is the
    Planck 2018 value $\Omega_b h^2 = 0.02236 \pm
    0.00015$~\cite{Planck:2018vyg}. Freeing $\Delta N_{\rm eff}$ leaves each posterior nearly unchanged, the additional freedom being absorbed by $Y_p$ rather than D/H.}
    \label{fig:omega_posteriors}
\end{figure}

This degeneracy between $\Omega_b h^2$ and $\Delta N_{\rm eff}$ can be used to
determine what the primordial abundances alone imply for the baryon density once
the radiation content is allowed to vary. We invert the BBN predictions for the
observed D/H~\cite{PDG2025} and $Y_p$~\cite{Aver:2026dxv}, using the theoretical
errors of Sec.~\ref{sec:unc} with the $\Omega_b h^2$ and $\Delta N_{\rm eff}$
contributions removed. The $\Omega_b h^2$ term is excluded because it is the
quantity being inferred, so its uncertainty must not also enter the likelihood
width. The $\Delta N_{\rm eff}$ term is excluded because it is varied explicitly
here, as the vertical axis of Fig.~\ref{fig:omega_neff_2d} and the marginalization
variable of Fig.~\ref{fig:omega_posteriors}. The remaining D/H theory
uncertainty is therefore $0.027$ (PRIMAT) and $0.100$ (NACRE-II), used in both the fixed and free $\Delta N_{\rm eff}$ cases. The resulting constraints are shown in Figs.~\ref{fig:omega_neff_2d}
and~\ref{fig:omega_posteriors}. Fixing $\Delta N_{\rm eff} = 0$, deuterium pins
the baryon density to $\Omega_b h^2 = 0.02202 \pm 0.00021$ (PRIMAT) and
$0.02231 \pm 0.00056$ (NACRE-II), the latter looser by roughly a factor of
$2.7$ owing to the larger NACRE-II D/H rate uncertainty. Promoting
$\Delta N_{\rm eff}$ to a free parameter and marginalizing over it with a flat
prior leaves these constraints almost unchanged, $0.02192 \pm 0.00026$ (PRIMAT)
and $0.02226 \pm 0.00056$ (NACRE-II): the inferred baryon density is largely
insensitive to the radiation content, since the freedom in $\Delta N_{\rm eff}$
is absorbed almost entirely by $Y_p$ rather than by D/H, which sets
$\Omega_b h^2$. The contour is narrower in the $\Omega_b h^2$ direction for PRIMAT than for NACRE-II, owing to the smaller PRIMAT D/H theoretical uncertainty. The external
$N_{\rm eff} = 2.990 \pm 0.070$ band~\cite{Goldstein:2026iuu} is shown for
context, but is not strictly independent of the abundance data, as it was
derived using the primordial $Y_p$ abundance. All determinations remain within
$\sim2\sigma$ of the Planck value $\Omega_b h^2 = 0.02236 \pm
0.00015$~\cite{Planck:2018vyg}.

The key implication here is that the precision of the $Y_p$ prediction is currently limited almost entirely by the uncertainty on $N_{\rm{eff}}$ when this parameter is allowed to vary. Improved CMB measurements from the Simons Observatory, which are projected to achieve $\sigma(N_\text{eff}) \sim 0.045$ \cite{SimonsObservatory:2025wwn}, would reduce the $Y_p$ uncertainty by roughly a factor of 1.6, at which point the neutron lifetime and nuclear rate uncertainties would again become relevant. 

\begin{table*}[t]
\centering
\begin{tabular}{l l c c}
\noalign{\vspace{4pt}}
Observable & Compilation & $\tau_n$ normalization & Fundamental normalization \\[6pt]
\hline
\noalign{\vspace{6pt}}
\multirow{2}{*}{$Y_p$}
 & PRIMAT   & $0.2469 \pm 0.0010$ & $0.2473 \pm 0.0010$ \\
 & NACRE~II & $0.2468 \pm 0.0010$ & $0.2473 \pm 0.0010$ \\[6pt]
\hline
\noalign{\vspace{6pt}}
\multirow{2}{*}{$\text{D/H} \times 10^5$}
 & PRIMAT   & $2.445 \pm 0.044$ & $2.448 \pm 0.044$ \\
 & NACRE~II & $2.517 \pm 0.107$ & $2.519 \pm 0.107$ \\[6pt]
\hline
\noalign{\vspace{6pt}}
\multirow{2}{*}{${}^7\text{Li/H} \times 10^{10}$}
 & PRIMAT   & $5.523 \pm 0.199$ & $5.528 \pm 0.199$ \\
 & NACRE~II & $5.356 \pm 0.746$ & $5.361 \pm 0.746$ \\[6pt]
\hline
\end{tabular}
\caption{Predicted BBN abundances with $\Delta N_\text{eff}$ as a free parameter ($\sigma_{\Delta N_\text{eff}} = 0.070$), comparing the $\tau_n$ and fundamental ($g_A$, $V_{ud}$) weak rate normalizations for both PRIMAT and NACRE~II reaction rate compilations.}
\label{tab:th_vals_neff}
\end{table*}

\begin{table*}[t]
\centering
\footnotesize

\label{tab:budget_comparison_neff}
\begin{minipage}[t]{0.45\textwidth}
\centering
\textbf{PRIMAT} \\[6pt]
\tiny
\begin{tabular}{l @{\hspace{10pt}} lc @{\hspace{8pt}} c}

Observable & Parameter & $\sigma_{Y_i}$ & \% var. \\ [6pt]
\hline
\noalign{\vspace{6pt}}
\multirow{3}{*}{$Y_p$}
 & $\Delta N_{\text{eff}}$                               & $9.50 \times 10^{-4}$ & 98.3 \\
 & $\tau_n$                                              & $1.03 \times 10^{-4}$ &  1.1 \\
 & $\Omega_b h^2$                                        & $6.53 \times 10^{-5}$ &  0.5 \\ [6pt]
\hline
\noalign{\vspace{6pt}}
\multirow{6}{*}{$\text{D/H} \times 10^5$}
 & $\Omega_b h^2$                                        & $2.69 \times 10^{-2}$ & 37.0 \\
 & $\Delta N_{\text{eff}}$                               & $2.33 \times 10^{-2}$ & 27.8 \\
 & $d(p,\gamma){}^3\text{He}$                            & $1.79 \times 10^{-2}$ &  16.4 \\
 & $d(d,n){}^3\text{He}$                                 & $1.48 \times 10^{-2}$ &  11.1 \\
 & $d(d,p)t$                                             & $1.20 \times 10^{-2}$ &  7.4 \\
 & $n(p,\gamma)d$                                        & $1.93 \times 10^{-3}$ &  0.2 \\ [6pt]
\hline
\noalign{\vspace{6pt}}
\multirow{12}{*}{${}^7\text{Li/H} \times 10^{10}$}
 & ${}^3\text{He}({}^4\text{He},\gamma){}^7\text{Be}$           & $1.27 \times 10^{-1}$ & 40.6 \\
 & $\Omega_b h^2$                                        & $7.70 \times 10^{-2}$ & 15.0 \\
 & ${}^7\text{Be}(n,p){}^7\text{Li}$                     & $7.41 \times 10^{-2}$ & 13.8 \\
 & $d(p,\gamma){}^3\text{He}$                            & $7.18 \times 10^{-2}$ & 13.0 \\
 & ${}^3\text{He}(d,p){}^4\text{He}$                     & $4.58 \times 10^{-2}$ &  5.3 \\
 & $d(d,n){}^3\text{He}$                                 & $4.09 \times 10^{-2}$ &  4.2 \\
  & $\Delta N_{\text{eff}}$                               & $3.56 \times 10^{-2}$ &  3.2 \\
 & $n(p,\gamma)d$                                        & $3.10 \times 10^{-2}$ &  2.5 \\
 & $t(p,\gamma){}^4\text{He}$                            & $2.09 \times 10^{-2}$ &  1.1 \\
 & ${}^3\text{He}(n,p)t$                                 & $2.00 \times 10^{-2}$ &  1.0 \\
 & ${}^7\text{Li}(p,{}^4\text{He}){}^4\text{He}$                & $9.42 \times 10^{-3}$ &  0.2 \\
 & $t({}^4\text{He},\gamma){}^7\text{Li}$                       & $5.33 \times 10^{-3}$ &  0.1 \\

\end{tabular}
\end{minipage}
\hfill
\begin{minipage}[t]{0.45\textwidth}
\centering
\textbf{NACRE~II} \\[6pt]
\tiny
\begin{tabular}{l @{\hspace{10pt}} lc @{\hspace{8pt}} c}

Observable & Parameter & $\sigma_{Y_i}$ & \% var. \\
\noalign{\vspace{6pt}}
\hline
\noalign{\vspace{6pt}}
\multirow{4}{*}{$Y_p$}
 & $\Delta N_{\text{eff}}$                               & $9.53 \times 10^{-4}$ & 97.5 \\
 & $\tau_n$                                              & $1.03 \times 10^{-4}$ &  1.1 \\
 & $d(d,n){}^3\text{He}$                                 & $7.89 \times 10^{-5}$ &  0.7 \\
 & $\Omega_b h^2$                                        & $6.14 \times 10^{-5}$ &  0.4 \\
  & $d(d,p)t$                                       & $4.23 \times 10^{-5}$ &  0.2 \\
    & $n(p,\gamma)d$                                       & $2.63 \times 10^{-5}$ &  0.1 \\
\noalign{\vspace{6pt}}
\hline
\noalign{\vspace{6pt}}
\multirow{6}{*}{$\text{D/H} \times 10^5$}
 & $d(d,n){}^3\text{He}$                                 & $8.12 \times 10^{-2}$ & 58.8 \\
 & $d(d,p)t$                                             & $5.41 \times 10^{-2}$ & 25.7 \\
  & $\Omega_b h^2$                                        & $2.77 \times 10^{-2}$ &  6.7 \\
 & $\Delta N_{\text{eff}}$                               & $2.40 \times 10^{-2}$ & 5.0 \\
 & $d(p,\gamma){}^3\text{He}$                            & $2.00 \times 10^{-2}$ &  3.5 \\
 & ${}^3\text{He}(n,p)t$                                 & $4.75 \times 10^{-3}$ &  0.2 \\
\noalign{\vspace{6pt}}
\hline
\noalign{\vspace{6pt}}
\multirow{10}{*}{${}^7\text{Li/H} \times 10^{10}$}
 & ${}^3\text{He}({}^4\text{He},\gamma){}^7\text{Be}$           & $4.59 \times 10^{-1}$ & 37.8 \\
 & ${}^3\text{He}(d,p){}^4\text{He}$                     & $4.25 \times 10^{-1}$ & 32.5 \\
 & ${}^7\text{Be}(n,p){}^7\text{Li}$                     & $3.05 \times 10^{-1}$ & 16.7 \\
 & $d(d,n){}^3\text{He}$                                 & $2.07 \times 10^{-1}$ &  7.8 \\
 & ${}^3\text{He}(n,p)t$                                 & $1.07 \times 10^{-1}$ &  2.1 \\
 & $d(p,\gamma){}^3\text{He}$                            & $8.54 \times 10^{-2}$ &  1.3 \\
 & $\Omega_b h^2$                                        & $7.81 \times 10^{-2}$ &  1.1 \\
 & $\Delta N_{\text{eff}}$                               & $3.74 \times 10^{-2}$ &  0.3 \\
 & $n(p,\gamma)d$                                        & $3.19 \times 10^{-2}$ &  0.2 \\
 & ${}^7\text{Li}(p,{}^4\text{He}){}^4\text{He}$                & $2.60 \times 10^{-2}$ &  0.1 \\
 & $t(p,\gamma){}^4\text{He}$                & $2.12 \times 10^{-2}$ &  0.1 \\

\end{tabular}
\end{minipage}
\caption{Same as Table~\ref{tab:budget_comparison_taun}, but with $\Delta N_{\text{eff}}$ allowed to vary freely with $\sigma(\Delta N_{\text{eff}}) = 0.070$, computed using the $\tau_n$ weak rate normalization.}
\label{tab:unc_neff_taun}
\end{table*}

\begin{table*}[t]
\centering
\begin{minipage}[t]{0.45\textwidth}
\centering

\textbf{\footnotesize\bfseries PRIMAT} \\[6pt]
\tiny
\begin{tabular}{l @{\hspace{10pt}} lc @{\hspace{8pt}} c}

Observable & Parameter & $\sigma_{Y_i}$ & \% var. \\ [6pt]
\hline
\noalign{\vspace{6pt}}
\multirow{4}{*}{$Y_p$}
 & $\Delta N_\text{eff}$                              & $9.47 \times 10^{-4}$ &  88.8 \\
 & $g_A$                                              & $3.06 \times 10^{-4}$ &   9.3 \\
 & $V_{ud}$                                           & $1.19 \times 10^{-4}$ &   1.4 \\
 & $\Omega_b h^2$                                     & $6.47 \times 10^{-5}$ &   0.4 \\ [6pt]
\hline
\noalign{\vspace{6pt}}
\multirow{6}{*}{$\text{D/H} \times 10^5$}
 & $\Omega_b h^2$                                     & $2.69 \times 10^{-2}$ &  37.1 \\
 & $\Delta N_\text{eff}$                              & $2.33 \times 10^{-2}$ &  27.8 \\
 & $d(p,\gamma){}^3\text{He}$                         & $1.79 \times 10^{-2}$ &   16.3 \\
 & $d(d,n){}^3\text{He}$                              & $1.46 \times 10^{-2}$ &   10.9 \\
 & $d(d,p)t$                                          & $1.20 \times 10^{-2}$ &   7.4 \\
 & $n(p,\gamma)d$                                     & $2.07 \times 10^{-3}$ &   0.2 \\ 
 & $g_A$                                     & 
 $1.74 \times 10^{-3}$ &   0.2 \\ 
 & $^3\text{He}(n,p)t$                                     & $1.19 \times 10^{-3}$ &   0.1 \\ [6pt]
\hline
\noalign{\vspace{6pt}}
\multirow{11}{*}{${}^7\text{Li/H} \times 10^{10}$}
 & ${}^3\text{He}({}^4\text{He},\gamma){}^7\text{Be}$       & $1.27 \times 10^{-1}$ &  40.6 \\
 & $\Omega_b h^2$                                     & $7.71 \times 10^{-2}$ &  15.0 \\
 & ${}^7\text{Be}(n,p){}^7\text{Li}$                  & $7.37 \times 10^{-2}$ &  13.7 \\
 & $d(p,\gamma){}^3\text{He}$                         & $7.16 \times 10^{-2}$ &  12.9 \\
  & ${}^3\text{He}(d,p){}^4\text{He}$                  & $4.62 \times 10^{-2}$ &   5.4 \\
 & $d(d,n){}^3\text{He}$                              & $4.06 \times 10^{-2}$ &   4.2 \\
 & $\Delta N_\text{eff}$                              & $3.58 \times 10^{-2}$ &  3.2 \\
 & $n(p,\gamma)d$                                     & $3.19 \times 10^{-2}$ &   2.6 \\
 & $t(p,\gamma){}^4\text{He}$                         & $2.07 \times 10^{-2}$ &   1.1 \\
 & ${}^3\text{He}(n,p)t$                              & $1.97 \times 10^{-2}$ &   1.0 \\
 & ${}^7\text{Li}(p,{}^4\text{He}){}^4\text{He}$            & $9.42 \times 10^{-3}$ &   0.2 \\ 
  & $t({}^4\text{He},\gamma){}^7\text{Li}$            & $5.62 \times 10^{-3}$ &   0.1 \\ [6pt]

\end{tabular}
\end{minipage}
\hfill
\begin{minipage}[t]{0.45\textwidth}
\centering

\textbf{\footnotesize\bfseries NACRE~II} \\[6pt]
\tiny
\begin{tabular}{l @{\hspace{10pt}} lc @{\hspace{8pt}} c}

Observable & Parameter & $\sigma_{Y_i}$ & \% var. \\ [6pt]
\hline
\noalign{\vspace{6pt}}
\multirow{5}{*}{$Y_p$}
 & $\Delta N_\text{eff}$                              & $9.50 \times 10^{-4}$ &  88.2 \\
 & $g_A$                                              & $3.06 \times 10^{-4}$ &   9.2 \\
 & $V_{ud}$                                           & $1.19 \times 10^{-4}$ &   1.4 \\
 & $d(d,n){}^3\text{He}$                              & $7.96 \times 10^{-5}$ &   0.6 \\
 & $\Omega_b h^2$                                     & $6.40 \times 10^{-5}$ &   0.4 \\ 
& $d(d,p)t$                              & $4.55 \times 10^{-5}$ &   0.2 \\ [6pt]
\hline
\noalign{\vspace{6pt}}
\multirow{6}{*}{$\text{D/H} \times 10^5$}
 & $d(d,n){}^3\text{He}$                              & $8.20 \times 10^{-2}$ &  58.6 \\
 & $d(d,p)t$                                          & $5.47 \times 10^{-2}$ &  26.0 \\
& $\Omega_b h^2$                                     & $2.77 \times 10^{-2}$ &   6.7 \\
 & $\Delta N_\text{eff}$                              & $2.40 \times 10^{-2}$ &  5.0 \\
 & $d(p,\gamma){}^3\text{He}$                         & $2.00 \times 10^{-2}$ &   3.5 \\
 & ${}^3\text{He}(n,p)t$                              & $4.41 \times 10^{-3}$ &   0.2 \\ [6pt]
\hline
\noalign{\vspace{6pt}}
\multirow{10}{*}{${}^7\text{Li/H} \times 10^{10}$}
 & ${}^3\text{He}({}^4\text{He},\gamma){}^7\text{Be}$       & $4.60 \times 10^{-1}$ &  37.9 \\
 & ${}^3\text{He}(d,p){}^4\text{He}$                  & $4.26 \times 10^{-1}$ &  32.5 \\
 & ${}^7\text{Be}(n,p){}^7\text{Li}$                  & $3.06 \times 10^{-1}$ &  16.8 \\
 & $d(d,n){}^3\text{He}$                              & $2.07 \times 10^{-1}$ &   7.7 \\
 & ${}^3\text{He}(n,p)t$                              & $1.07 \times 10^{-1}$ &   2.0 \\
 & $d(p,\gamma){}^3\text{He}$                         & $8.50 \times 10^{-2}$ &   1.3 \\
 & $\Omega_b h^2$                                     & $7.83 \times 10^{-2}$ &   1.1 \\
 & $\Delta N_\text{eff}$                              & $3.75 \times 10^{-2}$ &   0.3 \\
 & $n(p,\gamma)d$                                     & $3.22 \times 10^{-2}$ &   0.2 \\
 & ${}^7\text{Li}(p,{}^4\text{He}){}^4\text{He}$            & $2.50 \times 10^{-2}$ &   0.1 \\ 
 & $t(p,\gamma){}^4\text{He}$            & $2.10 \times 10^{-2}$ &   0.1 \\ [6pt]

\end{tabular}
\end{minipage}
\caption{Same as Table~\ref{tab:budget_comparison_fund}, but with $\Delta N_{\text{eff}}$ allowed to vary freely with $\sigma(\Delta N_{\text{eff}}) = 0.070$), computed using the fundamental weak rate normalization.}
\label{tab:unc_neff_fund}
\end{table*}

\section{Illustrative Applications}
\label{sec:app}

\subsection{Resolving the deuterium tension}

When the PRIMAT compilation of nuclear reaction rates is used, there is a slight almost $2\sigma$ tension between the calculated and observed primordial deuterium abundance \cite{PDG2025}. The atlas presented here allows us to quickly identify ways to alleviate this tension without significantly affecting the prediction for the helium-4 abundance, which is in good agreement with the LBT result. Using the $\tau_n$ normalization for the weak rates, Table \ref{tab:budget_comparison_taun} shows that the parameter contributing most strongly to the uncertainty in the deuterium prediction is the baryon abundance, $\Omega_b h^2$, with $\sigma_{Y_i} = 2.69 \times 10^{-2}$. From the left column of Table \ref{tab:sens_taun_nacre_side_by_side}, we find that $\frac{d\ln Y_p}{d\ln \Omega_b h^2} = +0.039,$
so a 1\% change in $\Omega_b h^2$ changes $Y_p$ by only 0.039\%, while changing the final value of D/H by $-1.639\%$. This is a well-known result: the final deuterium abundance is highly sensitive to $\Omega_b h^2$. Reducing $\Omega_b h^2$ by just $0.8\sigma$ from its measured value yields a primordial deuterium abundance of $D/H = 2.483$ and a primordial helium-4 abundance of $Y_p = 0.247$, both within $1\sigma$ of their observed values. 

In practice, however, this resolution is disfavored by current CMB data. The Planck 2018 determination of $\Omega_b h^2 = 0.02236 \pm 0.00015$~\cite{Planck:2018vyg} is obtained from a global fit to the temperature and polarization power spectra, so even a modest downward shift would worsen the fit to the acoustic peak structure. More recent ACT DR6 analyses likewise find $\Lambda$CDM parameters consistent with Planck-based results and do not indicate the lower baryon density required to substantially alleviate the deuterium tension~\cite{AtacamaCosmologyTelescope:2025blo}. Thus, accommodating such a shift would likely require a nonstandard reinterpretation of the CMB data that is not supported by present observations.

As an alternative, we can carry out the same exercise using the nuclear reaction rates. Again referring to Table \ref{tab:budget_comparison_taun}, we see that the two parameters contributing next most strongly to the uncertainty in deuterium are the deuterium-burning reactions $d(p,\gamma)^3\mathrm{He}$ and $d(d,n)^3\mathrm{He}$. From Table \ref{tab:sens_tau_n_rates_first12_side_by_side}, we find that $\frac{d(D/H)}{d\bigl(d(p,\gamma)^3\mathrm{He}\bigr)} = -0.018$ and $
\frac{d(D/H)}{d\bigl(d(d,n)^3\mathrm{He}\bigr)} = -0.015,
$
while $\frac{dY_p}{d(d(p,\gamma)^3\mathrm{He})}$ and $\frac{dY_p}{d(d(d,n)^3\mathrm{He})}$ are both approximately zero, meaning that a 1$\sigma$ shift in either of these reactions leaves $Y_p$ essentially unchanged. Reducing $d(p,\gamma)^3\mathrm{He}$ by $0.7\sigma$ and $d(d,n)^3\mathrm{He}$ by $0.5\sigma$ from their measured values over temperature brings both the primordial helium-4 and deuterium abundances into agreement with observation.

\subsection{Addressing the Cosmological Lithium Problem}

Over the years, many solutions have been proposed to address the longstanding cosmological Lithium Problem \cite{Fields:2011zzb}. Here, we illustrate how the atlas presented in this work can be used to explore potential resolutions, with more detailed analysis being left to future work. Using the NACRE-II compilation of nuclear reaction rates, which represents a more conservative assessment of nuclear rate uncertainties, together with the $\tau_n$ normalization for the weak rates, we see from Table \ref{tab:budget_comparison_taun} that most of the uncertainty in the predicted lithium-7 abundance arises from nuclear reaction rates. The dominant contributions come from ${}^3\text{He}(^4\text{He},\gamma)^7\text{Be}$, ${}^3\text{He}(d,p)^4\text{He}$, ${}^7\text{Be}(n,p)^7\text{Li}$, $d(d,n)^3\text{He}$, and ${}^3\text{He}(n,p)t$. Turning to Table \ref{tab:sens_tau_n_rates_first12_side_by_side}, we can determine the impact of each of these reactions on the helium-4 and deuterium abundances. The predicted value of $Y_p$ is essentially unchanged for a 1$\sigma$ shift in any of these five reactions. Deuterium is likewise insensitive to all of them except $d(d,n)^3\text{He}$, for which Table~\ref{tab:sens_tau_n_rates_first12_side_by_side} gives $dY/dp = -0.082$, so a +1$\sigma$ increase in the rate lowers $D/H \times 10^5$ by 0.082.

This indicates that ${}^3\text{He}(^4\text{He},\gamma)^7\text{Be}$, ${}^3\text{He}(d,p)^4\text{He}$, ${}^7\text{Be}(n,p)^7\text{Li}$, and ${}^3\text{He}(n,p)t$ can be varied with essentially no impact on the helium-4 and deuterium abundances, whereas somewhat greater care is required for $d(d,n)^3\text{He}$. Since the predicted lithium-7 abundance exceeds the observed primordial value by roughly a factor of four, fully resolving the Lithium Problem would require substantial shifts in the relevant input parameters. One combination of parameter shifts that resolves the tension without significantly affecting the predicted helium-4 and deuterium abundances is obtained by decreasing the rate of ${}^3\text{He}(^4\text{He},\gamma)^7\text{Be}$ by $5\sigma$ across the relevant temperature range, while increasing the rate of ${}^3\text{He}(d,p)^4\text{He}$ by $5\sigma$, and those of ${}^7\text{Be}(n,p)^7\text{Li}$ and ${}^3\text{He}(n,p)t$ by $4\sigma$. Treating the four rate uncertainties as independent, the combined shift corresponds to $\chi^2 = 82$ for four degrees of freedom, or approximately 8.2$\sigma$ in an equivalent single-parameter test.

The magnitude of the required rate shifts confirms that the Lithium Problem is unlikely to be resolved by nuclear physics alone, consistent with the conclusions of \cite{Fields:2011zzb, Cyburt:2015mya}. This motivates continued investigation of astrophysical solutions, such as Lithium destruction in metal-poor stellar atmospheres~\cite{Fields:2022mpw}, as well as new physics scenarios involving light electrically-neutral particles that have substantial interaction with nuclei~\cite{Goudelis:2015wpa}.

\section{Conclusions}
\label{sec:conclusion}

In this study we have presented a comprehensive, single-code sensitivity atlas for BBN, quantifying the dependence of the primordial abundances of helium-4, deuterium, and lithium-7 as well as $N_{\rm{eff}}$ on variations in 14 fundamental particle physics and cosmological parameters and 63 thermonuclear reaction rates. All calculations were performed using \faGithub \href{https://github.com/vallima/PRyMordial}{\,\texttt{PRyMordial}}, which incorporates QED radiative corrections, finite nucleon-mass effects, weak magnetism, finite-temperature corrections, and non-instantaneous neutrino decoupling. By computing each sensitivity within a consistent framework and repeating the full analysis for two nuclear reaction rate compilations and two weak-rate normalization schemes we provide a model-independent reference that can be applied to arbitrary BSM scenarios without the ambiguity introduced by comparing results across different codes and input assumptions.

The principal results of this work are the sensitivity matrices presented in Tables \ref{tab:sens_taun_nacre_side_by_side}-\ref{tab:sens_tau_n_rates_first12_side_by_side} and the ranked uncertainty budgets in Tables \ref{tab:budget_comparison_taun}, \ref{tab:budget_comparison_fund}, \ref{tab:unc_neff_taun}, and \ref{tab:unc_neff_fund}. For the helium-4 mass fraction, we find that the dominant sensitivities are to the neutron-proton mass difference, the electron mass, and the parameters controlling the weak-rate normalization, with the neutron lifetime and the baryon abundance being the leading contributors to the uncertainty budget under the neutron lifetime normalization, and the nucleon axial coupling dominating when the fundamental constant normalization is used. 

For deuterium, the baryon abundance and the deuterium-burning reaction rates $d(p,\gamma)^3He$ and $d(d,n)^3He$ are the primary sources of uncertainty when the PRIMAT rates are used, while the $d(d,n)^3He$ and $d(d,p)t$ rates dominate when the NACRE-II rates are used. The lithium-7 uncertainty budget is dominated by nuclear reaction rates in both compilations, along with the baryon abundance when the PRIMAT rates are used. 

When $\Delta N_{\rm eff}$ is promoted to be a free parameter using the combined CMB+BAO+BBN uncertainty, it dominates the $Y_p$ error budget by between 88\% and 98\% of the total variance, increasing the helium-4 uncertainty by roughly a factor of three. This dominance is the result of the direct impact that shifting $N_{\rm eff}$ has on the expansion rate at the time of weak rate freeze-out, and thereby the neutron-to-proton ratio at the onset of BBN. Upcoming CMB measurements from the Simons Observatory, projected to achieve an uncertainty in $\Delta N_{\rm eff}$ of 0.045 which will reduce this uncertainty by approximately a factor of 1.6, at which point the neutron lifetime, the baryon abundance, and nuclear rate uncertainties would again become the limiting factors.

As illustrative applications, we have shown that the mild deuterium tension present when the PRIMAT rates are used can be alleviated by modest shifts in the deuterium-burning reaction rates $d(p,\gamma)^3He$ and $d(d,n)^3He$ without disturbing the helium-4 prediction, and that resolving the Lithium Problem through nuclear rate adjustments alone requires shifts of 4-5$\sigma$ in multiple reactions, reinforcing the need for astrophysical or new-physics explanations.

Looking ahead, the atlas and uncertainty budgets presented here identify concrete targets for future improvement. On the nuclear physics side, reducing the uncertainties in the $d(p,\gamma)^3He$, $d(d,n)^3He$, $d(d,p)t$, and $^3He(^4He,\gamma)^7Be$ reaction rates would have the greatest impact on tightening the deuterium and lithium-7 predictions. On the cosmological side, improved determinations of $N_{\rm{eff}}$ from the CMB and continued refinement of the baryon density would sharpen the BBN predictions for all observables. On the observational side, the new LBT measurement of $Y_p$ has already reduced the helium-4 observational uncertainty by a factor of two. Further progress in this direction will bring the observational precision closer to the theoretical precision and enhance the power of BBN as a probe of new physics at MeV scales.

The sensitivity coefficients, response functions, and uncertainty budgets presented in this work are intended to serve as a practical reference for the community. The full set of numerical results is publicly available on GitHub \faGithub \href{https://github.com/Anne-KatherineBurns/bbn-sensitivity-atlas}{\,\texttt{bbn-sensitivity-atlas}}.

\section*{Acknowledgments}

Thank you very much to Jordi Salvadó, Tim Tait, Gabriela Barenboim, Marc Sher, and Helena García Escudero for the illuminating discussions and valuable feedback on our work. AKB acknowledges support from the ``Unit of Excellence Maria de Maeztu 2020-2023'' award to the ICC-UB CEX2019-000918-M and grant PID2022-136224NB-C21 funded by MCIN / MINECO / MCOC. AKB is also supported by the European Union’s Horizon 2020 research and innovation program under the Marie Sklodowska-Curie grant agreement No 860881-HIDDeN, and Horizon Europe research and innovation programme under the Marie Sk lodowska-Curie Staff Exchange grant agreement No 101086085 – ASYMMETRY. 

\vspace{2em} 

\noindent \textbf{Statement of AI Usage}
\vspace{1em} 

\noindent In the preparation of this manuscript, the author made limited use of the AI assistant Claude (Anthropic). Its use was confined to two purposes: occasional assistance with the manuscript prose (improving wording, clarity, and grammar), and assistance with writing and debugging code used in the analysis. All scientific content, methodology, results, and conclusions are the author's own. The AI tool was not used to generate research ideas or interpret results, and all output was reviewed and verified by the author, who takes full responsibility for the content of this work.

\bibliographystyle{unsrt} 
\bibliography{bib}

\end{document}